\documentclass[12pt,a4paper,leqno]{article}
\usepackage[utf8]{inputenc}
\usepackage{geometry}
\usepackage{xcolor}
\usepackage[round]{natbib}
\usepackage{graphicx}
\usepackage{pdflscape}  
\usepackage{appendix}
\usepackage{threeparttable}
\usepackage{booktabs}
\usepackage{placeins}  % \FloatBarrier: keep floats within their (sub)sections

\usepackage{amsthm}
\newtheorem{example}{Example}

\usepackage{amssymb}
\usepackage{amsbsy}
\usepackage{bm}
\usepackage{amsmath}
\usepackage{amsthm}
\usepackage{graphicx}
\usepackage{mathtools}
\usepackage{rotating}
\usepackage{setspace}
\usepackage{footmisc}
\usepackage{color, colortbl}
\definecolor{ashgrey}{rgb}{0.7, 0.75, 0.71}
\definecolor{columbiablue}{rgb}{0.61, 0.87, 1.0}
\definecolor{coral}{rgb}{1.0, 0.5, 0.31}

  \definecolor{mcsorange}{HTML}{CC5500}

\definecolor{colBVAR}{HTML}{bababa}
\definecolor{colBART}{HTML}{d7191c}
\definecolor{colmixBART}{HTML}{fdae61}
\definecolor{colerrorBART}{HTML}{abd9e9}
\definecolor{colfullBART}{HTML}{2c7bb6}
  \definecolor{mcsblue}{HTML}{CC5500}
  \definecolor{irfblue}{HTML}{1F4E79}      % + shock (contractionary) line, matches plot
\definecolor{irfred}{HTML}{9E1B1B}       % - shock (expansionary, mirrored) line, matches plot

\definecolor{colcons}{HTML}{e31a1c}
\definecolor{colSV}{HTML}{a6cee3}
\definecolor{colhBART}{HTML}{1f78b4}

\usepackage{xcolor}
\usepackage{tikz}
\usepackage{enumitem}
\newlist{steps}{enumerate}{1}
\setlist[steps,1]{label = Step \arabic*:}

\usepackage{adjustbox}
\usepackage{dcolumn}
\newcolumntype{d}[1]{D..{#1}} % for alignment of numbers on decimal marker

\usepackage{caption}
\usepackage{subcaption}
\definecolor{nblue}{HTML}{000660}
\usepackage[colorlinks=true,urlcolor=nblue,linkcolor=nblue,citecolor=nblue]{hyperref}

\newcommand*{\myeqref}[2][Eq.~]{%
  \hyperref[{#2}]{#1(\ref*{#2})}%
}
\def\equationautorefname#1#2\null{%
  Eq.#1(#2\null)%
}

\newcommand{\Figref}[1]{\hyperref[{#1}]{Figure~\ref*{#1}}}% sentence start: "Figure 1"
\newcommand{\figref}[1]{\hyperref[{#1}]{Fig.~(\ref*{#1})}}% within sentence: "Fig. (1)"
\newcommand{\Eqref}[1]{\hyperref[{#1}]{Equation~\ref*{#1}}}% sentence start: "Equation 1"
\begin{document}

\title{\textbf{A Flexible Approach to Augmenting a Bayesian VAR with Nonlinear Factors}\thanks{We gratefully acknowledge helpful comments from  seminar participants at the Central Bank of Colombia.  The views expressed herein are solely those of the authors and do not necessarily reflect the views of the Federal Reserve Bank of Cleveland or the Federal Reserve System.  Please address correspondence to: Florian Huber. Department of Economics, University of Salzburg. \textit{Address}: M\"{o}nchsberg 2a, 5020 Salzburg, Austria. \textit{Email}: \href{mailto:florian.huber@plus.ac.at}{florian.huber@plus.ac.at}. We acknowledge the use of Claude (Anthropic) as an aid for drafting, editing, and brainstorming during the preparation of this manuscript.}}

\author{Todd E.\ Clark\thanks{Economist Emeritus, Federal Reserve Bank of Cleveland, and Fellow, Dept.\ of Economics, Johns Hopkins University} \and
Florian Huber\thanks{Professor of Economics, University of Salzburg} \and
Gary Koop\thanks{Professor of Economics, University of Strathclyde}}

\date{}

\maketitle
\thispagestyle{empty}

\doublespacing
% Tighten the vertical white space around display equations (kept compatible
% with setspace's double spacing by re-applying after each \normalsize).
\makeatletter
\g@addto@macro\normalsize{%
  \setlength\abovedisplayskip{6pt plus 2pt minus 2pt}%
  \setlength\abovedisplayshortskip{2pt plus 2pt}%
  \setlength\belowdisplayskip{6pt plus 2pt minus 2pt}%
  \setlength\belowdisplayshortskip{4pt plus 2pt minus 2pt}%
}
\makeatother
\begin{center}
\begin{minipage}{0.9\textwidth}
\noindent\small
\begin{center}
    \textbf{Abstract}\\
\end{center}
This paper proposes a  vector autoregression augmented with nonlinear factors that are modeled nonparametrically using regression trees. There are four main advantages of our model.  First, the use of factor methods ensures that departures from linearity are modeled parsimoniously. In particular, they exhibit functional pooling where a small number of nonlinear factors are used to model common nonlinearities across variables.  Second, modeling potential nonlinearities nonparametrically lessens the risk of misspecification. Third, Bayesian computation using MCMC is straightforward even in very high-dimensional models, allowing for efficient, equation-by-equation estimation, thus avoiding computational bottlenecks that arise in popular alternatives such as the time-varying parameter VAR. Fourth, existing methods for identifying structural economic shocks in linear factor models can be adapted for the nonlinear case in a straightforward fashion using our model. Exercises involving artificial  and macroeconomic data illustrate the properties of our model and its usefulness for forecasting and structural economic analysis.      
\end{minipage}
\end{center}

\setcounter{page}{0}
\normalsize\newpage\renewcommand{\footnotelayout}{\setstretch{2}}
\section{Introduction}
To illustrate the rationale for the more sophisticated model developed in this paper, consider modeling $M$ time series $\{y_{it}\}_{t=1}^T$ as a nonlinear function of some variable $x_t$:
\begin{equation}
y_{it} = \mu_i(x_{t}) + \varepsilon_{it}, \label{example}
\end{equation}
where $\mu_i$ is a nonlinear function which we aim to learn and that is specific to series $i$ and $\varepsilon_{it}$ is a shock term. If we estimate this model in an unrestricted manner we end up learning $M$ separate functions with $T$ being potentially small. This gives rise to over-fitting concerns and, if viewed as a system (if shocks are correlated), estimating the model in \autoref{example} can be computationally cumbersome. 

\begin{figure}[h!]
    \centering
    \includegraphics[width=1\textwidth]{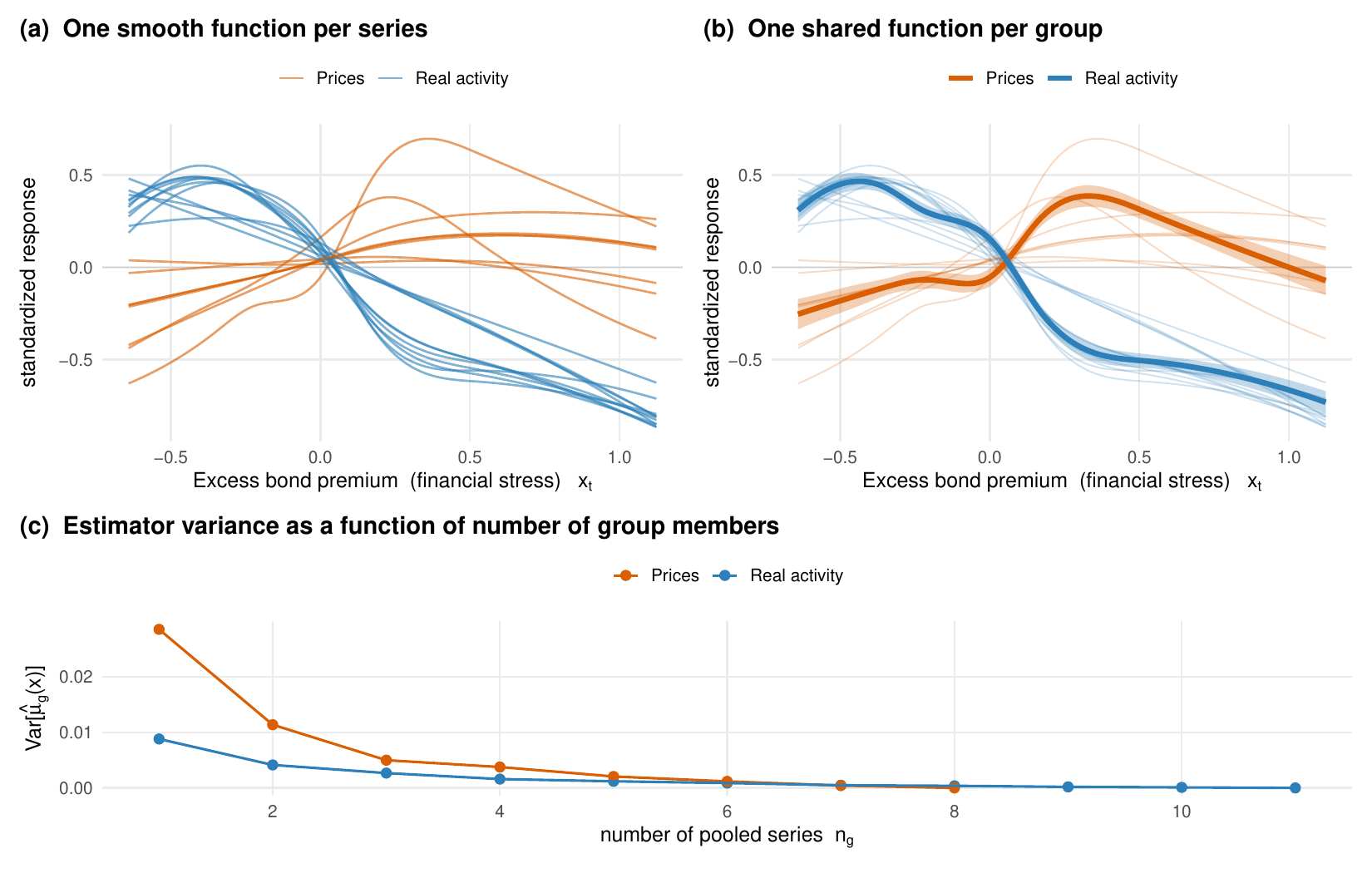}
    \caption{Illustrative example of function pooling applied to US price and output series}\label{fig:pooling}
\end{figure}
In an example, we estimate a simple formulation of the model in \autoref{example} relating different measures of aggregate output and prices in the US to the excess bond premium (EBP) of \cite{GilchristZakrajsek:2012:AER}.\footnote{The online appendix provides additional details on the data and model.} Panel (a) of \autoref{fig:pooling} provides estimates of these nonlinear functions, with function values on the vertical axis and EBP values on the horizontal axis.  The functions for output (in blue) look very similar across measures, declining in a nonlinear way with increases in the EBP.   The functions for prices (in orange) show more heterogeneity but still some commonality, with some tendency to rise and then eventually decline with higher values of the EBP.  This pattern of commonality gives rise to the idea that instead of estimating separate functions for each of these series, estimating one function per group would be more efficient (both computationally and statistically). Assuming a nonlinear factor structure gives rise to function pooling by sharing information across series to estimate a small number of common functions (instead of $M$ separate ones). The result can be seen in panel (b) of  \autoref{fig:pooling}. In this case, two functions are estimated that track most of the variation in the unit-level function estimates and with a substantially increased effective sample size that implies a lower estimator variance (see panel (c), which shows how the estimator variance declines as the number of members in a group increases).

These considerations motivate the high-dimensional nonlinear time series model we propose in this paper.  Nonlinearities are modeled as deviations from the linear VAR component also included in the model. In \cite{clark2023tail}, this is done for every variable in the system using regression trees. Since this leads to an enormous computational burden and risks over-fitting, in this paper we assume that the number of nonlinear functions is much smaller than the number of endogenous series, leading to favorable computational and statistical properties.  For reasons detailed below, we view the model and our estimation approach as effectively balancing tradeoffs associated with over-parameterization, misspecification, and computation.

From an econometric perspective, our approach can be seen as featuring functional pooling, with information pooled across variables to improve the estimation of the nonlinear functions determining the factors.  Our functional pooling approach extends to VARs with nonlinear functions the pooling of information across series that \cite{MullerWatsonJAE} achieve with hierarchical priors in a linear AR model setting.  This pooling of functions also has precedents in other Bayesian literature on clustering and pooling of similar time series \citep[see, e.g.,][]{frohwirth2008model, chan2020reducing}. % , MullerWatsonJAE %Papers such as \cite{chan2020reducing} exploit the fact that time-varying parameters (TVPs) in a large TVP vector autoregression (TVP-VAR) feature a factor structure to increase statistical efficiency. This is similar to what we aim to do in this paper. Instead of assuming linearity we, however, assume nonlinear relations between the covariates and a set of macro outcomes . However, in any macroeconomic or financial dataset there are so many empirically plausible nonlinearities that might occur that the researcher who adopts any particular parametric model (e.g., a TVP-VAR or a Markov switching DFM) runs a risk of misspecification. 

In light of possible misspecification, inclusion of a nonlinear component (that does not increase the computational burden appreciably) can be interpreted as a defensive modeling strategy \citep[see][]{linero2025},  with the nonparametric component serving to control for possible misspecification of the linear model.  As discussed in more detail below, our use of regression trees for the nonlinear part of the model is predicated on the previously established empirical success of Bayesian additive regression trees in common macroeconomic datasets.

For the purpose of using the model for structural analysis, we specify the innovation component of the model to include a (linear) static factor model, which permits structural inference through sign restrictions along the lines of \cite{BH2015} and \cite{KOROBILIS2022}.\footnote{This assumption is consistent with papers such as \cite{bai2007determining} and \cite{forni2025american}, who postulate that the economy is driven by a small number of fundamental shocks.}  This piece of our specification also contributes to computational efficiency; conditional on the linear static factors (as well as the nonlinear dynamic factors), the VAR block of the model can be estimated equation-by-equation.  Together, the factor structures of our model make it scalable to large variable sets.

We consider simulations and a real data analysis. In simulations, our approach produces much better forecasts  and impulse responses than linear models when the DGP is nonlinear, while nearly matching the linear BVAR benchmark when it is not.  In the real data analysis, we apply our semiparametric nonlinear model to forecast key US macroeconomic aggregates and to analyze the impacts of financial conditions and monetary policy shocks. The forecasting application shows noticeable gains in out-of-sample density forecast accuracy over a linear BVAR. The structural application, using sign restrictions for identification, uncovers significant asymmetries in the responses of most indicators of economic activity and inflation: large tightenings of financial conditions and monetary policy have bigger absolute impacts than easings.

The paper proceeds as follows. Section \ref{sec:multiBART} presents our proposed model, discusses identification and provides additional information on the prior setup.  Section \ref{sec:MCsimulations} provides Monte Carlo evidence on the ability of the model to uncover various types of nonlinearity in common factors.  Sections \ref{sec:forecastingapp} and \ref{sec:IRFapp} provide application results on, respectively, out-of-sample forecasting and structural analysis.  Section \ref{sec:conclusion} summarizes and concludes.  An online appendix provides additional details on the dataset, estimation algorithm and properties of it, and additional empirical results. 

\section{Econometric framework}\label{sec:multiBART}

\subsection{The factor-BART VAR} \label{sec: fBART-VAR}
In this section, we develop a nonlinear model that can be applied to large macroeconomic and financial {datasets}. Existing nonlinear models \citep[see][among many others]{primiceri2005time, cogley2005drifts, sims2006were, DGG_2013, huber2020inference, HUBER202352} have the shortcoming that they do not scale well to large dimensions and  are prone to over-fitting.  There are also recent papers that propose nonparametric Bayesian factor models \citep[see, e.g.,][]{velasco2025lettreedecidefabart, chernis2025bayesiangaussianprocessdynamic}. These papers assume nonparametric observation equations while restricting the state equations of the factors to be linear. The model we propose differs since the number of unknown functions to estimate is much smaller than $M$.

Prior regularization in these high-dimensional models is highly effective in lowering the likelihood of over-fitting, but imposing constraints can also be beneficial when they are supported by empirical evidence or theory. This is the route we take in this paper, relying on a nonlinear factor structure to pool information across variables to estimate a small number of common nonlinear functions.
The factor-BART VAR assumes that $\bm y_t =(y_{1t}, \dots, y_{Mt})'$ evolves according to the following semiparametric model:
\begin{equation}
     \bm y_t = \bm A \bm x_t + \bm \Lambda_\mu \bm \mu(\bm x_t) + \bm \varepsilon_t, \label{eq: nonlinVAR}  %  \bm f_t
\end{equation}
where $\bm A$ is an $M \times K (=Mp)$ matrix that linearly links $\bm y_t$ to its $p$ lags, stored in $\bm x_t = (\bm y'_{t-1}, \dots, \bm y'_{t-p})'$.  

Nonlinear dynamics are captured by a sequence of $Q_\mu$ latent factors contained in the vector $\bm \mu(\bm x_t) = (\mu_1(\bm x_t), \dots, \mu_{Q_\mu}(\bm x_t))'$. Each component function $\mu_j: \mathbb{R}^K \to \mathbb{R}$ takes $\bm x_t$ as input and returns a scalar output. The $M \times Q_\mu$ coefficient matrix $\bm \Lambda_\mu$ then maps the factors back into the observed series.  A key point to note here is that the functional form of $\mu_j(\bm x_t)$ and $\mu_i(\bm x_t)$ can differ. The loadings $\bm \Lambda_\mu$ are then used to weight each of the functions such that the dynamics for a particular variable are best explained.  To understand this more clearly, the $s^{th}$ equation of \autoref{eq: nonlinVAR} can be written out as:
\begin{equation}
    y_{st} = \bm a'_s \bm x_t + \lambda_{\mu, s1} \mu_1(\bm x_t) + \dots + \lambda_{\mu, s Q_\mu} \mu_{Q_\mu}(\bm x_t) + \varepsilon_{st},
\end{equation}
with $\bm a_s$ denoting the $s^{th}$ row of $\bm A$ and $\lambda_{\mu, sj}$ denoting the $(s, j)^{th}$ element of $\bm \Lambda_\mu$.  

Our model breaks the conditional mean for each variable into a parametric part (i.e., a linear regression component) and a nonparametric part (i.e., the nonlinear factors) that is modeled using BART. In macroeconomics, the most popular approach to modeling $\bm y_t$ would be a VAR that sets $\bm \Lambda_\mu= \bm 0$. However, this is only optimal if the DGP is linear. If there is evidence that this condition does not hold, we end up with a misspecified model and the estimates of $\bm A$ could lead to misleading inference. As noted above, the presence of the factor can also be interpreted as a defensive modeling strategy \citep[see][]{linero2025},  meaning that the nonparametric component serves to control for possible misspecification of the linear model.  From this perspective, the main advantage of the factor structure is that similar types of variables might also exhibit similar forms of misspecification of the linear model and our approach can pick this up through the factor loadings.

We assume that the shocks in $\bm \varepsilon_t$ are Gaussian with a covariance matrix $\bm \Sigma_\varepsilon$. Specific assumptions about $\bm \Sigma_\varepsilon$ will be discussed below.  Of course, this Gaussian specification constitutes a parametric choice, which (for reasons that will become clear below) is convenient for structural inference with the model.  The specification of the model's innovation term could be made less parametric, such as with an additional BART specification (something considered in some of the models of \cite{clark2023tail}) or infinite mixtures of Gaussians \citep[see, e.g.,][]{huber2024fast}. However, such a specification would make  structural analysis more difficult.  In view of the combination of a nonparametric representation of the nonlinear $\bm \Lambda_\mu \bm \mu(\bm x_t)$ component of the model with the linear VAR and Gaussian innovation components, we characterize our specification as semiparametric.
 
 %common factors  any event, with the considerable flexibility afforded by the nonlinear
%\footnote{Our use of that characterization is consistent with \citeauthor{linero2025}'s \citeyearpar{linero2025}  characterization of a similar model.} 
% \cite{linero2025}'s

The factor-BART VAR specification nests three special cases. First, if $\lambda_{\mu, s1} = \dots = \lambda_{\mu, s Q_\mu}= 0$, we end up with a linear regression model for equation $s$. This observation allows us to design a Bayesian shrinkage prior that allows for a data-based assessment of whether nonlinear features are necessary. Second, if we set $\bm \Lambda_\mu = \bm I_M$ (and thus set $Q_\mu = M$), we end up with the mixBART specification proposed in \cite{clark2023tail} and applied in \cite{BHMoil2024}.  Relative to the latter model, our proposed specification is very parsimonious because instead of having to estimate $M$ different functions, we only need to estimate $Q_\mu (\ll M)$ functions.\footnote{ \cite{BaumeisterFrankHuberKoop2025} consider a yet more restricted model in a different setting. In a dynamic panel model, they assume that a single factor determines the nonlinearities that directly impact a set of microeconomic variables, with loadings associated with this factor driven by observed characteristics.  The moments of the microeconomic indicators in turn impact the dynamics of macroeconomic variables.} If $M = 100$ and $Q_\mu = 5$, we are estimating only five unknown functions instead of $100$, reducing functional complexity by 95\%.  (The online appendix provides evidence of the associated gains in computation time.)  Third, if we maintain $\bm \Lambda_\mu = \bm I_M$ (and $Q_\mu = M$) and impose $\bm A$ = $\bm 0$, we end up with the BART VAR  specification of \cite{huber2020inference} and \cite{lima2025minnesotabart}.\footnote{\cite{lima2025minnesotabart} extend the treatment of BART to allow for time dependence and sparsity in trees, as well as --- in the spirit of conventional Minnesota priors --- more shrinkage on longer lags than shorter lags in trees.} % This assumption is motivated by the fact that if series display similar forms of nonlinearities (such as different measures of aggregate inflation or subcomponents of GDP), it makes sense to introduce structure on the functional relationship and we achieve this through the factor specification that leads to function pooling.  Flo: we had this sentence before third. I think with the new example in the intro it's redundant.

Beyond computation, these nested cases clarify when each placement of the nonlinearity is the natural one, which in turn guides the choice of $Q_\mu$ relative to $M$. Our factor placement is preferable when the nonlinearities are \emph{shared}  across series, which is  a common situation in macroeconomic datasets,  where many different series measure a particular concept (such as output, prices or financial conditions) and hence are likely to share the same form of nonlinearity. Exploiting this by estimating $Q_\mu \ll M$ common functions in $\bm \mu(\bm x_t)$, with $\bm \Lambda_\mu$ distributing them across equations, gives statistical efficiency (fewer functions to learn, hence less over-fitting in the short samples often encountered with quarterly datasets) and scalability. When instead the nonlinearities are genuinely idiosyncratic, a per-variable specification that places BART on the conditional mean of each equation ---the mixBART model of \cite{clark2023tail}--- is the more appropriate choice, albeit at  higher computational and statistical cost. Our framework nests this model, so the practitioner need not commit ex-ante. Indeed, the choice is partly data-driven, since the column-wise shrinkage prior on $\bm \Lambda_\mu$ adapts the effective number of active factors, collapsing redundant columns when the data do not call for them. A conceptually different alternative is to place the nonlinearity on the VAR coefficients rather than on the conditional mean directly. This leads to the model of \cite{hauzenberger2022bayesian}. This model, however, implies linear relations between $\bm x_t$ and $\bm y_t$ for a particular point in time and requires the researcher to choose suitable effect modifiers that determine the time-variation in the parameters.

% old example formatting at the bottom of the page
% tried Claude and ChatGPT for options but didn't like what I got

We illustrate how function pooling works through three simple examples. All three analytical examples assume $M=3$ observed time series and $Q_\mu=2$ latent nonlinear factors so that $\bm \mu(\bm x_t) = (\mu_1(\bm x_t) , \mu_2(\bm x_t))'$.

\begin{example}%[Shared Nonlinearity Between Series 1 and 2]
Series 1 and 2 share the same nonlinear function $\mu_1(\bm x_t)$, while the third series depends on a separate nonlinear function $\mu_2(\bm x_t)$. This setup captures common nonlinear behavior among subsets of series.  Specifically:
\[
\bm \Lambda_\mu =
\begin{bmatrix}
1 & 0 \\
1 & 0 \\
0 & 1 \\
\end{bmatrix}\quad
%\text{then}\quad
\Rightarrow \
 \bm \Lambda_\mu \bm \mu(\bm x_t) =
\begin{bmatrix}
\mu_1(\bm x_t) \\
\mu_1(\bm x_t) \\
\mu_2(\bm x_t)
\end{bmatrix}.
\]
%This example implies that series 1 and 2 share the same nonlinear function $\mu_1(\bm x_t)$, while the third series depends on a separate nonlinear function $\mu_2(\bm x_t)$. This setup captures common nonlinear behavior among subsets of series.
\end{example}

\begin{example}%[Idiosyncratic vs Shared Nonlinearity]
%Now assume:
Series 1 and 2 each have  their own nonlinear functions, while series 3 exhibits only linear dynamics. The model flexibly allows for nonlinearities to be omitted when not supported by the data.  Specifically:
\[
\bm \Lambda_\mu =
\begin{bmatrix}
1 & 0 \\
0 & 1 \\
0 & 0 \\
\end{bmatrix}\quad
\Rightarrow \
%\text{ then }\quad
\bm \Lambda_\mu \bm \mu(\bm x_t) =
\begin{bmatrix}
\mu_1(\bm x_t) \\
\mu_2(\bm x_t) \\
0
\end{bmatrix}.
\]
%Series 1 and 2 each have  their own nonlinear functions, while series 3 exhibits only linear dynamics. The model flexibly allows for nonlinearities to be omitted when not supported by the data.
\end{example}

\begin{example}%[Low-Rank Nonlinear Structure]
All three series depend on the same two latent functions, but through different linear combinations. This introduces a complex nonlinear structure while maintaining parsimony if each $\mu_j$ takes a simple form.  Specifically:
%Finally, suppose:
\[
\bm \Lambda_\mu =
\begin{bmatrix}
0.5 & 0.5 \\
0.5 & -0.5 \\
0 & 1 \\
\end{bmatrix}
\Rightarrow \
%\quad \text{then} \quad
\bm \Lambda_\mu \bm \mu(\bm x_t) =
\begin{bmatrix}
0.5\mu_1(\bm x_t) + 0.5\mu_2(\bm x_t) \\
0.5\mu_1(\bm x_t) - 0.5\mu_2(\bm x_t) \\
\mu_2(\bm x_t)
\end{bmatrix}.
\]
%All three series depend on the same two latent functions, but through different linear combinations. This introduces a complex nonlinear structure while maintaining parsimony if each $\mu_j$ takes a simple form.
\end{example}

\subsection{Modeling the error variance and approximating the unknown functions} \label{sec: factors_approx}
Two components of our model still require discussion. First, the error variance $\bm \Sigma_\varepsilon$. For this, we introduce a factor model so that:
\begin{equation}
    \bm \Sigma_\varepsilon = \bm \Lambda_q \bm \Lambda_q' + \bm \Omega, \label{eq: factormodel}
\end{equation}
with $\bm \Lambda_q $ denoting an $M \times Q_q$ matrix of factor loadings, and $\bm \Omega = \text{diag}(\omega_1^2, \dots, \omega_M^2)$ being a diagonal matrix of measurement error variances. 

Notice that \autoref{eq: factormodel} means the innovation term of the model can be equivalently written as:
\begin{equation}
    \bm \varepsilon_t = \bm \Lambda_q \bm q_t + \bm \eta_t, \quad \bm \eta_t \sim \mathcal{N}(\bm 0, \bm \Omega), \quad \bm q_t \sim \mathcal{N}(\bm 0, \bm I_{{Q_q}}),
    \label{eq: error}
\end{equation}
where $\bm q_t$ are zero mean static factors. This representation shows that the reduced-form errors of the model can be decomposed into a common component $\bm \Lambda_q \bm q_t$ and a sequence of idiosyncratic measurement errors. If we wish to use the model for structural inference, restrictions can be used on $\bm \Lambda_q$ to identify the factors in $\bm q_t$ as fundamental economic shocks as in \cite{KOROBILIS2022}. The assumption that the factors have unit variance fixes their scaling.  In principle, we could allow for stochastic volatility in the measurement errors and factors as well, leading to a factor stochastic volatility model \citep[see, e.g.,][]{aguilar2000bayesian}. In addition, one could also augment the variance specification with an explicit outlier component in the spirit of \cite{carriero2024addressing}. Given the evidence in \cite{clark2023tail}, where homoskedastic BART-based models produce predictive distributions almost as accurate as those of models with stochastic volatility, we opt for the simpler homoskedastic specification. 

Second, we need to discuss how we approximate the unknown functions $\mu_i$. In principle, we can use any nonlinear learning technique such as neural networks \citep[see, e.g.,][]{farrell2021deep, HAUZENBERGER2025112260}, Gaussian processes \citep[see, e.g.,][]{williams2006gaussian, fox2015bayesian, tang2024hierarchical, HauzenbergerJBES2025, chernis2025bayesiangaussianprocessdynamic}, or splines \citep[see, e.g.,][]{shin2020functional}. Because of its empirical success with many different datasets (including datasets commonly employed in macroeconomics and finance) and the existence of a set of hyperparameters that work well in all of these, we  approximate the functions $\mu_i$ through Bayesian additive regression trees \citep{chipman2010bart}. BART is a sum-of-trees model that approximates:
\begin{equation}
    \mu_i (\bm x_t) \approx \sum_{s=1}^S \mathfrak{t}(\bm x_t| \bm m_{i, s}, \mathcal{T}_{i, s}),
\end{equation}
where $\mathfrak{t}$ is a tree function, $\bm m_{i, s}$ a vector of terminal node parameters, and $\mathcal{T}_{i, s}$ a tree structure and $S$ is the total number of trees. We set $S=250$ in all applications.  These tree structures consist of decision rules that take the form $\{x \le c\}$ or $\{ x > c\}$ and hence split the input space defined by $x$ into a sequence of disjoint sets. For each of these disjoint sets, there is a terminal node parameter  that plays the role of a fitted value in a regression model.  Trees are prone to over-fitting if no regularization is introduced. Summing over many (possibly) complex trees further increases the risk of over-fitting. As a solution, BART uses Bayesian regularization priors to force each of the trees to take a particularly simple form and thus act as a weak learner. Summation over many such simple trees produces a model with a great deal of representation flexibility, while regularization to keep the trees simple limits the risk of over-fitting.

The factor-BART VAR is capable of capturing a wide range of possible nonlinear relations through the factor structure.  For instance (as employed in our Monte Carlo simulations presented in the next section), the model might feature (i) nonlinearities arising with threshold effects, so that above and below a certain threshold, the dynamics of the common factors are governed by different sets of coefficients, or (ii) factors that are governed by highly nonlinear functions, such as trigonometric functions.  As this discussion suggests, an advantage of using BART or other nonparametric or semiparametric methods is that the range of nonlinear forms they are capable of modeling is enormous. In contrast, \cite{nonlinearDFM} is an example of a nonlinear dynamic factor model where the nonlinearities involve a particular quadratic form. They show how their model is closely related to a nonlinear DSGE model. Quadratic forms such as this are easily captured using BART. %Todd: word change idea here is that BART is more capturing than directly modeling a quadratic (?) % 
\subsection{Identification}\label{sec: identification}
Our model has several latent components that give rise to identification issues. In this section, we discuss how all of them can be solved. The first identification issue arises from the fact that a nonlinear specification can soak up linear dynamics. To see this, consider a simplified version of our model that sets $\bm \Lambda_q = \bm 0$. Stacking the rows of \autoref{eq: nonlinVAR} yields:
\begin{equation*}
    \bm Y = \bm X \bm A' + \bm M \bm \Lambda_\mu' + \bm E,
\end{equation*}
where $\bm Y = (\bm y'_1, \dots, \bm y'_T)', \bm X = (\bm x'_1, \dots, \bm x'_T)', \bm M = (\mu(\bm x_1)', \dots, \mu(\bm x_T)')'$ and $\bm E = (\bm \varepsilon_1, \dots, \bm \varepsilon_T)'$. This model is observationally equivalent to:
\begin{equation*}
    \bm Y = \bm X (\bm A' - \bm W \bm \Lambda_\mu') + (\bm M + \bm X \bm W) \bm \Lambda_\mu' + \bm E,
\end{equation*}
with $\bm W$ being any $K \times Q_\mu$ matrix. Hence, the nonlinear component of the model can soak up linear dynamics, leading to an identification problem. This problem is related to the common identification problem that arises in generalized additive and partially linear models, where the nonlinear component can soak up linear effects \citep{hooker2007generalized, lengerich2020purifying}. 

A common assumption in this literature is to impose orthogonality between $\bm X$ and $\bm M$ so that $\bm M' \bm X = \bm 0$, ensuring that all linear effects are captured by $\bm A$ and all nonlinear effects by $\bm \Lambda_\mu \bm \mu(\bm x_t)$. In our model, this can be achieved in two ways. The first restricts the posterior of $\bm M$ to the space orthogonal to $\bm X$ during sampling. While conceptually straightforward, this requires a substantial modification of the standard BART algorithm since the terminal node parameters of the trees are no longer independent from each other. 

The second way is simpler and relies on ex-post identification of the nonlinear factors. We first estimate the model without any identification restrictions and then post-process the MCMC draws of $\bm M$ to make them orthogonal to $\bm X$ (which can then be combined with additional identifying restrictions). This is done by projecting $\bm M$ onto the space orthogonal to $\bm X$ and adjusting $\bm A$ accordingly so that the product $\bm X \bm A' + \bm M \bm \Lambda_\mu'$ remains unchanged. Letting $\bm M^*$ denote the orthogonalized factors and $\bm A^*$ the corresponding linear coefficients, these can be obtained as
\begin{align}
    \bm M^* &= \bm M - \bm X (\bm X' \bm X)^{-1} \bm X' \bm M = (\bm I - \bm P_X) \bm M, \label{eq: Mstar} \\
    \bm A^* &= \bm A + \bm \Lambda_\mu \bm M' \bm X (\bm X' \bm X)^{-1}, \label{eq: Astar}
\end{align}
where $\bm P_X = \bm X (\bm X' \bm X)^{-1} \bm X'$ is the orthogonal projection onto  the column space of $\bm X$ and we assume $\bm X$ has full column rank. Within the equivalence class parameterized by $\bm W$, this choice corresponds to the unique selection $\bm W^\star = -(\bm X' \bm X)^{-1} \bm X' \bm M$, namely the only $\bm W$ that satisfies $\bm X' (\bm M + \bm X \bm W) = \bm 0$.\footnote{Uniqueness of $\bm W^\star$ follows immediately from standard OLS theory: The orthogonality condition $\bm X' (\bm M + \bm X \bm W) = \bm 0$ is equivalent to the system of normal equations $\bm X' \bm X \, \bm W = -\bm X' \bm M$, which admits the unique solution $\bm W^\star = -(\bm X' \bm X)^{-1} \bm X' \bm M$ whenever $\bm X' \bm X$ is invertible, i.e., whenever $\bm X$ has full column rank $K$. Equivalently, $\bm W^\star$ is, column-by-column, the negative of the OLS coefficient vector obtained by regressing the corresponding column of $\bm M$ on $\bm X$.} Direct substitution into the observationally equivalent model verifies that the conditional mean is unchanged:
\begin{equation*}
    \bm X (\bm A^*)' + \bm M^* \bm \Lambda_\mu' = \bm X \bm A' + \bm M \bm \Lambda_\mu',
\end{equation*}
so the likelihood, predictive distributions, and impulse response functions are invariant under the normalization; only the attribution of explanatory power between the linear block $\bm A$ and the nonlinear block $\bm \Lambda_\mu \bm \mu(\bm x_t)$ changes. This approach is closely related to the post-processing reparameterization studied in \citet[Lemma~3.3]{chakraborty2026orthogonal}.  Hence, transforming the MCMC draws of $\bm M$ and $\bm A$ according to \autoref{eq: Mstar} and \autoref{eq: Astar} yields a posterior distribution for the nonlinear factors that is orthogonal to the linear regressors. 

Another identification issue arises because the factor loadings  $\bm \Lambda_\mu$ and factors $\bm \mu(\bm x_t)$ are not separately identified. This is the standard identification issue that also occurs with linear factor models and arises since $\bm \Lambda_\mu \bm \mu(\bm x_t) =\bm \Lambda_\mu \bm R \bm R' \bm \mu(\bm x_t)$ for any $\bm R$ that satisfies $\bm R \bm R' = \bm I$. This lack of identification gives rise to non-identification of the scale and sign of factors and factor loadings. It also implies that one can permute the columns of $\bm \Lambda_\mu$ and $\bm \mu(\bm x_t)$ without changing the likelihood.

These identification issues can be, again, fixed in various ways.  For instance, similarly to \cite{BBE}, we could restrict the upper $Q_\mu \times Q_\mu$ part of $\bm \Lambda_\mu$ to an identity matrix. This restriction implies that $\mu_1(\bm x_t)$ is the nonlinear component of series $1$, $\mu_2(\bm x_t)$ the nonlinear part of series $2$, and so on. For series $j > Q_\mu$ we then again allow for linear combinations of all the functions. This identification scheme has the shortcoming that it leads to order-dependence in terms of how the elements in $\bm y_t$ are ordered.  To fix the ordering, one can use economic intuition. For instance, in the case that $Q_\mu = 1$, if one plans to model GDP and its subcomponents, a natural ordering puts GDP on top so that the corresponding value of $\mu_1(\bm x_t)$ would imply a nonlinear relationship between GDP and $\bm x_t$, whereas the subcomponents of GDP have nonlinear relations that are proportional to $\mu_1(\bm x_t)$.  However, these identification assumptions impact the likelihood and the ordering of the series becomes another modeling choice that could affect predictive and structural inference in our model. There are also more general solutions using generalized lower triangular matrices, as proposed in \cite{fruhwirth2023counts, fruhwirth2025sparse}.

Another, and in our view preferable, approach relies on \emph{ex-post} identification and leaves the likelihood unrestricted. Conditional on the orthogonalized draws $\bm M^*$ from \autoref{eq: Mstar}, we resolve the remaining scale, sign, ordering, and rotational indeterminacies in two steps.

We first compute the standard deviation $\sigma_{\mu, j}$ of each column $\bm \mu_j^* = (\mu_j^*(\bm x_1) , \dots, \mu_j^*(\bm x_T))'$ of $\bm M^*$ and rescale (for each MCMC draw):
\begin{equation*}
    \widetilde{\bm \Lambda}_\mu = \bm \Lambda_\mu ~ \text{diag}(\sigma_{\mu, 1}, \dots, \sigma_{\mu, Q_\mu}) \quad \text{ and } \quad \widetilde{\bm M} = \bm M^* ~ \text{diag}(\sigma^{-1}_{\mu, 1}, \dots, \sigma^{-1}_{\mu, Q_\mu}).
\end{equation*}
The columns of $\widetilde{\bm M}$ then have unit variance: the scale of the nonlinear factors is absorbed into the loadings, the product $\bm \Lambda_\mu \bm \mu(\bm x_t)$ is preserved, and the likelihood is left unchanged.

Once the scale is fixed, the remaining indeterminacy is with respect to orthogonal transformations through $\bm R$ (with $\bm R \bm R' = \bm I$) of the columns of $\widetilde{\bm \Lambda}_\mu$ and $\widetilde{\bm M}$. This involves sign changes, column permutations, and, when $Q_\mu > 1$, continuous rotations. Our shrinkage prior on the columns of $\bm \Lambda_\mu$ solves column permutations by imposing a soft ordering on the columns, but it does not fix the sign or the continuous rotation. However, to solve all of these a posteriori we can use the \texttt{MatchAlign} procedure of \cite{poworoznek2021matching}. For each draw $s$, the procedure applies a \emph{varimax} rotation $\bm V^{(s)}$ to $\widetilde{\bm \Lambda}_\mu^{(s)}$, orienting the loadings towards a simpler, sparse representation, and matches the rotated columns against a common reference draw to obtain a signed permutation $\bm P^{(s)}$. The reference is the draw whose loadings attain the median condition number $\kappa(\widetilde{\bm \Lambda}_\mu) = \sigma_{\max}(\widetilde{\bm \Lambda}_\mu) / \sigma_{\min}(\widetilde{\bm \Lambda}_\mu)$ across the posterior sample; since $\kappa$ summarizes how well separated the columns are, its median yields a representative, outlier-robust pivot that avoids both near-degenerate and atypically orthogonal draws. Crucially, this pivot resolves only the \emph{relative} sign and column labelling \emph{across} draws, not the final reported ordering. Our column-wise prior on $\bm \Lambda_\mu$ imposes shrinkage that softly increases with the column index $j$ (through $\varpi_j \propto j^{-2}$), in the spirit of the increasing-shrinkage priors of \cite{bhattacharya2011sparse} and \cite{legramanti2020bayesian}, so the columns carry a natural \emph{a priori} ordering by magnitude. Because individual draws need not honor this soft ordering, and because the reference draw fixes only relative labels, we apply a single global permutation $\bm \Pi$ that sorts the aligned columns by the $L_2$ norm of the posterior mean (in decreasing order). This anchors the final column labels by data-driven magnitude rather than by the reference draw itself. Writing $\bm R^{(s)} = \bm V^{(s)} \bm P^{(s)} \bm \Pi$, the aligned draws are:
\begin{equation*}
    \widehat{\bm \Lambda}_\mu^{(s)} = \widetilde{\bm \Lambda}_\mu^{(s)} \, \bm R^{(s)} \quad \text{ and } \quad \widehat{\bm M}^{(s)} = \widetilde{\bm M}^{(s)} \, \bm R^{(s)}.
\end{equation*}
Because $\bm V^{(s)}$, $\bm P^{(s)}$, and $\bm \Pi$ are each orthogonal, so is their product $\bm R^{(s)}$, and applying the same $\bm R^{(s)}$ to loadings and factors preserves the product, $\widehat{\bm M}^{(s)} \widehat{\bm \Lambda}_\mu^{(s)\prime} = \widetilde{\bm M}^{(s)} \widetilde{\bm \Lambda}_\mu^{(s)\prime}$, and retains the orthogonality to $\bm X$ established above, $\bm X' \widehat{\bm M}^{(s)} = \bm 0$.
 This single step subsumes the sign normalization that the factor-modeling literature usually performs separately \citep[e.g.,][]{Kastner_2017}.

As with the orthogonalization step, the alignment is applied \emph{ex-post} to the stored posterior draws and leaves the product $\bm \Lambda_\mu \bm \mu(\bm x_t)$ --- and with it the likelihood, the predictive densities, and the generalized impulse response functions --- numerically unchanged. Resolving the rotation therefore concerns only the interpretation of the individual factors and loadings. As a diagnostic, inefficiency factors computed on $\bm \Lambda_\mu \bm \mu(\bm x_t)$ remain low even though those for the unaligned $\bm \mu(\bm x_t)$ are inflated by rotational drift (see \autoref{app:mixing} of the online appendix).

A related observation concerns the role of the prior in selecting among observationally equivalent factorizations. While the product $\bm \Lambda_\mu \bm \mu(\bm x_t)$ is identified, its factorization into loadings and latent functions is not, and different factorizations endow the functions $\mu_j$ with different degrees of smoothness. To illustrate, suppose the nonlinear contribution to three series is $\bm \Lambda_\mu \bm \mu(\bm x_t) = (x_{1t}^2, \, x_{1t}^2, \, x_{1t}^2 + x_{2t}^2)'$ with $Q_\mu = 2$. Both $(\mu_1, \mu_2) = (x_{1t}^2, \, x_{2t}^2)$ and $(\mu_1, \mu_2) = (x_{1t}^2, \, x_{1t}^2 + x_{2t}^2)$ reproduce this product exactly, yet the second carries a function of higher curvature. Any criterion for choosing among the equivalent factorizations therefore matters substantively for the recovered functions. In the unidentified posterior this choice is made implicitly through the BART prior, which places most of its mass on simple trees. The alignment above instead makes the choice explicit, through the simple-structure criterion that varimax optimizes. Either way, the selection concerns only the latent functions and loadings, whereas quantities such as predictive distributions, impulse responses  and the product $\bm \Lambda_\mu \bm \mu(\bm x_t)$ are invariant in that respect.

The final identification issues are related to $\bm \Lambda_q$ and $\bm q_t$ and its interaction with $\bm \Lambda_\mu \bm \mu(\bm x_t)$. These are sorted out by fixing the variance of $\bm q_t$ to $\bm I_{Q_q}$, using sign restrictions on the loadings (see Sub-section \ref{sec:signs}) and noting that $\bm \mu(\bm x_t)$ is a deterministic function of lagged observables while $\bm q_t$ is a sequence of white noise shocks. Hence, nonlinear factors cannot absorb the shock factors (and vice versa).

\subsection{Priors} \label{sec: priors} 

Our model involves different blocks of parameters or latent states (e.g., the VAR block, the factor loadings, and the BART factors). For each of these blocks, earlier work in the literature has proposed a range of different priors and any combination of these can be used with our factor-BART VAR. We make particular choices from this set of standard priors. On the VAR parameters we use a horseshoe prior \citep{carvalho2009handling} with row-specific global shrinkage terms; on the BART factors we use precisely the same prior setup proposed in \cite{chipman2010bart}; on the $\omega_j^2$ we use inverse Gamma priors; and on $\bm \Lambda_q$ we use either truncated Gaussian priors (as discussed in Section \ref{sec:IRFapp}) or weakly informative Gaussian priors.

The only departure from standard choices is the prior we use on $\bm \Lambda_\mu$. In this case, we consider a row- and column-wise shrinkage prior \citep[see, e.g.,][]{huber2019adaptive, kastner2019sparse}. Particularly, we consider a shrinkage prior  that has row- and column-specific shrinkage terms:
\begin{align*}
    \lambda_{\mu, ij} \sim \mathcal{N}(0,  \psi_{\mu, ij}^2 \tau^2_{\mu, i} \varpi_j), \quad &\psi_{\mu, ij} \sim \mathcal{C}^+(0,1), \quad \tau_{\mu, i} \sim \mathcal{C}^+(0, 1), \quad &\varpi_j = \frac{\varpi}{j^2}, \quad \varpi \sim \mathcal{G}^{-1}(a_\varpi, b_\varpi),
\end{align*}
for $i=1, \dots, M$ and $j=1, \dots, Q_\mu$ where $\mathcal{C}^+(0, 1)$ denotes the half-Cauchy distribution and $\mathcal{G}^{-1}(a_\varpi, b_\varpi)$ the inverse Gamma distribution with hyperparameters $a_\varpi, b_\varpi$. We set these equal to $a_\varpi = 3$ and $b_\varpi = 0.03$. Conditional on $\varpi_j$, this is a standard horseshoe prior with row-specific global shrinkage terms. The row-wise scaling parameters $\tau_{\mu, i}$ shrink a particular row toward zero. Setting $\tau_{\mu, i} \approx 0$ would therefore imply that the resulting  $i^{th}$ equation is linear. The presence of the local shrinkage parameters $\psi_{\mu, ij}^2$ effectively allows for selecting a particular nonlinear form, if warranted by the data.  

This prior also serves as a device for selecting the effective number of nonlinear factors by increasingly forcing the columns in $\bm \Lambda_\mu$ to zero. The presence of the local scaling terms then allows for deviations from this shrinkage pattern, if necessary. This approach automatically removes unnecessary factors from the model by shrinking factor loadings associated with irrelevant factors to zero.   

Our model can be estimated using a sophisticated Markov chain Monte Carlo algorithm that involves sampling from (mostly) well-known full conditional posterior distributions. The detailed algorithm and all full conditional distributions are provided in the online appendix.

\section{Artificial data exercises}\label{sec:MCsimulations}

In this section, to assess the ability of our model to uncover various types of nonlinearity in common factors, we carry out an empirical exercise involving artificial data.  Across a range of data-generating processes, we use simulated data to compare the efficacy of our proposed model and a restricted linear specification in forecasting and impulse response analysis --- the two purposes for which we use the model in the empirical applications of Sections \ref{sec:forecastingapp} and \ref{sec:IRFapp}.  The online appendix provides some additional results that confirm that our model successfully discriminates linear from nonlinear equations.

\subsection{Design of the Monte Carlo study}

We generate artificial datasets of $M=16$ time series of length $T \in \{250, 500\}$. The choice of $M$ is moderately large and typical of what is used in the VAR literature. The choices for $T$ are the sort that often occur with monthly and quarterly macroeconomic datasets.%\footnote{In  results omitted in the interest of brevity, we also conducted experiments with $T = 10,000$, a sample large enough to effectively make priors irrelevant and ensure that the data determine the posteriors.  Qualitatively, results on the MSEs of coefficients for the very large sample are very similar to those reported for $T = 500$, with our model yielding coefficient accuracy gains with nonlinear DGPs and largely matching the accuracy of the linear VAR.}   

Our first four DGPs use the following general specification:
\begin{align*}
\bm{y}_t = \bm{A}_1 \bm{y}_{t-1} + \bm A_2 \bm y_{t-2}+ \bm{\Lambda}_1 \bm{\mu}_t + \bm{\Lambda}_2 \bm{q}_t + \bm{\eta}_t, 
\end{align*}
with $
\bm{\eta}_t \sim \mathcal{N}(\bm{0}, 0.01^2 \times \bm I_{16}),
\bm{q}_t \sim \mathcal{N}(\bm{0}, \bm{I}_{2}),
\bm y_0 = \bm y_{-1} = \bm 0,$
and the elements of $\bm \Lambda_1$ being drawn from $\mathcal{N}(0.5, 0.25^2)$ and then sparsified by randomly setting to zero $50\%$ of its rows. Note that this specification implies that half of the equations in our DGPs are linear and do not contain the nonlinear factor.  The coefficient matrices $\bm A_1$ and $\bm A_2$ are sparse. Their non-zero elements are drawn from $\mathcal N(0,0.1^2)$ and $\mathcal N(0,0.05^2)$, respectively, with $90\%$ of the entries set to zero at random. The diagonal elements are fixed at $\text{diag}(\bm A_1)=0.25$ and $\text{diag}(\bm A_2)=0.10$ to ensure stability. 
The elements of $\bm \Lambda_2$ are drawn from standard Gaussian distributions, with two diagonal elements set to 1 for scale normalization.  We assume $Q_q=2$.  Our fifth DGP departs from the general specification by making $\bm{\eta}_t$ $t$-distributed with 4 degrees of freedom, rather than Gaussian. 

With this general setup, we consider five specific DGPs. \texttt{DGP 1} and \texttt{DGP 5} set $\bm{\Lambda}_1=0$ and, thus, are linear. Accordingly, there is no need to specify the factors. For the remaining DGPs, we assume $Q_\mu = 3$ factors and introduce different laws of motion for their conditional means. They all assume that the factor evolves according to the nonlinear functions of lagged observables. 

In particular, the nonlinear factor depends on  $\tilde{\bm y}_{t-1} = \tfrac12(\bm y_{t-1} + \bm y_{t-2})$ in three different ways. \texttt{DGP 2} sets $\mu(\bm x_t) = \bm{B}_1 \cos\!\left( \tilde{\bm y}_{t-1} \cdot \tilde y_{1,t-1} \cdot \tilde y_{3,t-1} \cdot \pi \right) 
+ \bm{B}_2 \left| \tilde{\bm y}_{t-1} - 0.5 \right| 
+ \bm{B}_3 \sin(\tilde{\bm y}_{t-1}), $ with  operations inside trigonometric functions and the absolute value being element-wise. This DGP is inspired by the statistics literature \citep[see, e.g.,][]{chipman2010bart}. In terms of the properties of the simulated time series this DGP implies cyclical movements of the factors (consistent with business cycle fluctuations), with the length and intensity of the waves being driven by lagged endogenous series.

\texttt{DGP 3} is an endogenous regime-switching specification, with the switching process depending on the sign of the first component of the smoothed lag vector. Specifically, this DGP sets $\mu (\bm x_t) = [\bm{B}_1 \tilde{\bm y}_{t-1} ] \times \mathbb{I}(\tilde y_{{1,t-1}} < 0) + [\bm{B}_2 \tilde{\bm y}_{t-1}] \times \mathbb{I}(\tilde y_{{1,t-1}} \ge 0)$.
 Similar processes have been considered in, e.g., \cite{kolesar2025dynamic}.

\texttt{DGP 4}  resembles the one studied in, e.g., \cite{gonccalves2024nonparametric} and \cite{HAUZENBERGER2025112260} and implies size and sign asymmetries in how $\bm y_t$ reacts to lagged values. This DGP sets $\mu(\bm x_t) = \tfrac14 \bm{B}_1 \tilde{\bm y}_{t-1}^{\,3} + \bm{B}_2 \bm r_t,$ where $\tilde{\bm y}_{t-1}^{\,3} = \tilde{\bm y}_{t-1} \odot \tilde{\bm y}_{t-1} \odot \tilde{\bm y}_{t-1}$ (multiplication is applied element-wise) and the $j^{th}$ element of $\bm r_t$ is given by $r_{jt} = \max(0, \tilde y_{j,t-1})$ for $j=1,\dots,16$.

In all of the nonlinear DGPs, the elements of the matrices $\bm B_j$ in the factor equations are sampled from Gaussian distributions with mean zero and variances for $\bm B_1, \bm B_2, \bm B_3$ equal to $0.15^2, 0.075^2$ and $0.1^2$, respectively.

In what follows, with each artificial dataset we estimate two models: the factor-BART VAR and a restricted linear specification that serves as a benchmark. For the factor-BART VAR we set $Q_\mu = 8$ nonlinear factors, which is a much larger value than that used in the DGPs. Thus our estimating model is over-parameterized. We do this so as to investigate the ability of our prior to shrink and choose the correct more parsimonious specification. In both of the models we set $Q_q=3$, which is larger than the value used in the DGPs. The benchmark linear specification is identical to the factor-BART VAR except that $Q_\mu=0$.  In lieu of adding terminology to reflect the linear factor structure that we retain in the model's innovation term, we simply refer to this benchmark as a linear BVAR. Hence, accuracy differences solely reflect performance gains from adding nonlinear factors.

\subsection{Monte Carlo findings}

We begin the {discussion of the} results of the simulation exercise with out-of-sample forecast accuracy, specifically, the joint log predictive likelihoods (LPLs) reported in \autoref{tab: Sims}. The table shows the different DGP configurations in the columns. The rows include the joint LPLs across forecast horizons and for different values of $T$. All numbers are relative to the linear BVAR benchmark and are averaged across $500$ replications from the DGP.

\begin{table}[h!]
\centering
\begin{tabular}{rr ccccc}
  \hline
 &  & \multicolumn{5}{c}{DGP} \\
 \cmidrule(lr){3-7}
 &  & 1 & 2 & 3 & 4 & 5 \\ 
  \hline
$T=250$ & $h=1$ & 0.16 & 4.94 & 5.19 & 6.13 & 0.02 \\ 
         & $h=4$ & -0.01 & 3.75 & 2.43 & 3.08 & 0.13 \\ 
         & $h=8$ & 0.07 & 3.40 & 2.39 & 4.46 & -0.03 \\ \midrule
$T=500$ & $h=1$ & -0.03 & 4.46 & 4.27 & 6.68 & -0.24 \\ 
         & $h=4$ & 0.08 & 3.89 & 2.89 & 2.81 & -0.14 \\ 
         & $h=8$ & -0.18 & 3.87 & 2.35 & 3.94 & -0.33 \\ 
   \hline
\end{tabular}
\caption*{\tiny \textbf{Notes}: The table shows  joint log score differences across $500$ replications from each of the DGPs to the linear BVAR (which sets $Q_\mu=0$). Numbers greater than zero imply a better performance in terms of LPLs.}
\caption{Forecast performance based on synthetic data for $500$ replications from the DGP.} \label{tab: Sims}
\end{table}
Our main finding is that, for the nonlinear DGPs, our factor-BART VAR forecasts better than the linear benchmark. This is unsurprising, but also illustrates how BART methods can uncover a diverse range of nonlinear forms; reassuringly, BART performs roughly as well across our three very different nonlinear DGPs, including those with abrupt breaks (\texttt{DGP 3}) and smooth nonlinearities (\texttt{DGP 2}). It is also reassuring that even for the linear DGP, \texttt{DGP 1}, our factor-BART VAR performs roughly the same (albeit not quite as well, as expected) as the true linear model, so that the semiparametric model approximates linearity without over-fitting. For the fat-tailed DGP \texttt{DGP 5}, both estimating models are misspecified and there is very little difference in forecast performance, although with $T=500$ the linear BVAR forecasts slightly better.

As noted above, in addition to using our model for forecasting, we also use it for structural analysis.  Accordingly, we next consider in the Monte Carlo analysis the estimation accuracy and coverage properties of the generalized impulse responses \citep{KoopPesaranPotter:1996:girf} to the first structural shock in the model. We use sign restrictions to be consistent with the signs of $\bm \Lambda_q$ in the DGP  (see \autoref{sec:signs}) and obtain the GIRFs through the Monte Carlo procedure described in \autoref{app: girf_appendix}.  The true GIRFs are means of responses obtained from the same forward simulation with $2{,}000$ paths from the known DGP.

  \begin{table}[h!]
  \centering
  \begin{tabular}{l r >{\columncolor{gray!15}} r r}
  \toprule
   & \multicolumn{2}{c}{68\% coverage} & \\
  \cmidrule(lr){2-3}
  DGP & factor-BART & linear BVAR & MAE \\
  \midrule
  \multicolumn{4}{l}{\textit{Panel A: $T=250$}} \\
  DGP 1 & 0.56 & 0.57 & 1.00 \\
  DGP 2 & 0.73 & 0.56 & 0.76 \\
  DGP 3 & 0.70 & 0.61 & 0.71 \\ 
  DGP 4 & 0.77 & 0.65 & 0.45 \\
  DGP 5 & 0.64 & 0.59 & 0.82 \\ \midrule
  \multicolumn{4}{l}{\textit{Panel B: $T=500$}} \\
  DGP 1 & 0.56 & 0.55 & 0.96 \\
  DGP 2 & 0.78 & 0.48 & 0.83 \\
  DGP 3 & 0.72 & 0.55 & 0.72 \\
  DGP 4 & 0.78 & 0.60 & 0.61 \\ 
  DGP 5 & 0.67 & 0.59 & 0.81 \\
  \bottomrule
  \end{tabular}
    \caption*{\tiny \textbf{Notes}: Coverage is the empirical frequency with which the model's 68\% (16/84) posterior band contains the true GIRF, averaged across horizons (0--12), variables, and Monte Carlo replications. Column 4 reports the median across replications of (factor-BART per-rep MAE) / (linear BVAR per-rep MAE), where the per-rep MAE is taken across response variables and impulse-response horizons.  Values below 1 favor factor-BART.  Replications: 500 per cell.}
  \caption{Generalized impulse-response coverage and relative MAEs: factor-BART vs.\ linear BVAR.}
  \label{tab:girf-mc}
  \end{table}

\begin{figure}[h!]
    \centering
    \includegraphics[trim={0 0 0 1cm},clip, width=1\linewidth]{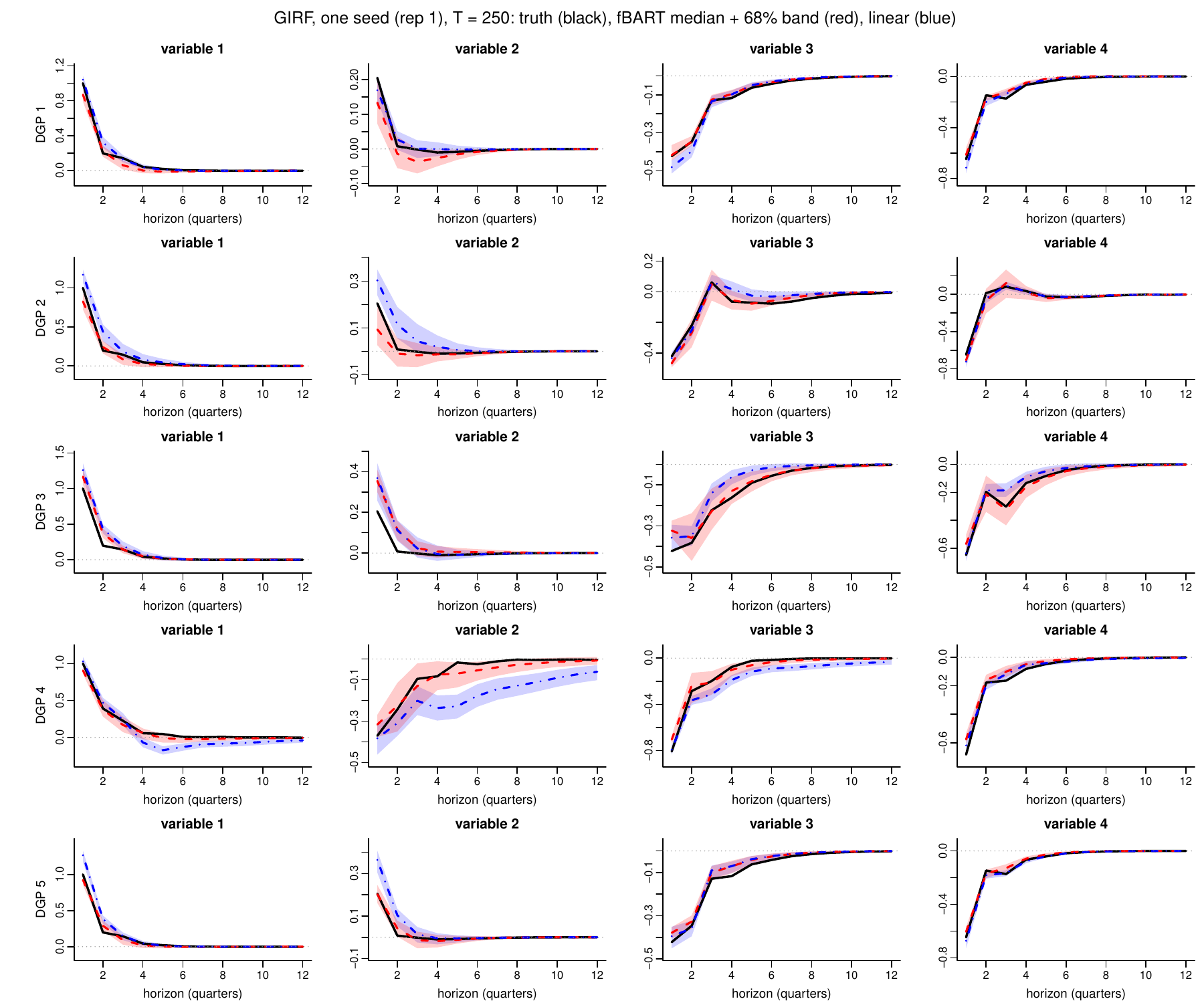}
     \caption*{\tiny \textbf{Notes}: The red (blue) area shows the 16th and 84th credible intervals, the dashed (dot-dashed) red (blue) lines show the posterior median of factor-BART (linear BVAR). The black line is the true GIRF based on the DGP.}
    \caption{GIRF comparison of factor-BART to the linear BVAR model based on a single repetition from the DGP with $T=250$.}
    \label{fig:girf-T250}
\end{figure}

For each replication, we store the posterior median and the $68\%$ (16/84) posterior band of the GIRFs. Table \ref{tab:girf-mc} reports two frequentist diagnostics across replications: the empirical coverage of the posterior band, averaged over horizons ($0$--$12$) and responding variables, and the relative mean absolute error (MAE) of the posterior median of the GIRF of factor-BART relative to the linear BVAR. Figure~\ref{fig:girf-T250} complements the aggregate statistics by plotting GIRFs from a single representative replication at $T=250$ against the truth, so that within-sample posterior bands and point estimates can be inspected jointly. 

With DGP 1, factor-BART and the linear BVAR baseline are statistically indistinguishable, indicating that the additional flexibility of factor-BART carries little efficiency cost when the data-generating process is linear.  When we consider nonlinear DGPs, factor-BART delivers uniformly higher (and more accurate relative to nominal) coverage rates. In terms of point estimation, we also find that factor-BART produces much lower MAEs than the linear BVAR, with sizable improvements that increase with the sample size. 

The single-replication GIRFs confirm the aggregate picture: factor-BART's posterior median tracks the truth far more closely than the linear BVAR, whose bands are often too narrow to cover the true response. Together, the results of the Monte Carlo analysis indicate our model and estimation approach to be successful in forecast accuracy and impulse response estimation, capturing nonlinearities when they exist and largely matching a linear specification when they do not.

\section{Application: Forecasting the US economy}\label{sec:forecastingapp} 

\subsection{Data, design of the forecasting exercise and model specification}
In our applications, we use $M=22$ major quarterly macroeconomic and financial time series.  With two exceptions, we take the  data from the FRED-QD database developed in \cite{McCrackenNg2021} and maintained at the Federal Reserve Bank of St.\ Louis.  As detailed in the variable list provided in  \autoref{tab:data} of the online appendix, the selected series include growth of real GDP and many of its main components, various measures of aggregate inflation, other indicators of the labor market and economic activity such as payroll employment and industrial production, the federal funds rate and various other interest rates covering longer maturities.  One exception to the data source is that we also include in the model the excess bond premium (EBP) of \cite{GilchristZakrajsek:2012:AER} \citep[using the current updated series available from][]{FavaraEtAl2016}.  In the second exception, we include in the model the utilization-adjusted measure of growth in total factor productivity developed in \cite{fernald2014tfp} (and downloaded from the author's website).  Our set of selected variables is very similar to that used in the large Bayesian VAR of \cite{Crumpetal:LargeBVAR:ijcb:2025}.  We transform all variables to be stationary following recommendations from   \cite{McCrackenNg2021}; see  \autoref{tab:data} of the online appendix for details.  With transformations implemented, our data sample runs from 1976:Q3 to 2023:Q4. 

The forecast design is as follows. We start with an initial estimation sample that ends in 2001:Q4. After estimating the model we iteratively compute predictive distributions from $h=1$ up to ${h=12}$ quarters ahead. We then add one more observation to the model estimation sample (i.e., use a recursive scheme) and repeat this procedure until we reach the end of the sample. Finally, after obtaining the predictive distributions for all observations in the hold-out, we compute energy scores to evaluate density accuracy for the full set of variables and continuous ranked probability scores (CRPSs) to evaluate density accuracy on a variable-by-variable basis.\footnote{The energy score is a generalization of the CRPS, to which it collapses in the univariate case:
$\text{ES}_t(\bm{y}_t) = E_\mathfrak{f}||\hat{\bm{y}}_{t} - \bm{y}_{t}|| - 0.5 E_\mathfrak{f}||\hat{\bm{y}}_{t} - \hat{\bm{y}}_{t}'||$, where $\hat{\bm{y}}_{t}$ and $\hat{\bm{y}}_{t}'$ are independent random vectors with distribution $\mathfrak{f}$. The energy score and the CRPS are less sensitive to outliers (like the extremes experienced during the COVID pandemic) than the LPL.}

We consider models that set $p=2$ and $Q_q=3$ and focus on multiple values of $Q_\mu \in \{1, 3, 6, 8\} $.  The case $Q_\mu=8$ implies an over-fitting model for which our shrinkage prior detects the effective number of nonlinear factors. As additional competing models, we include the mixBART specification  of \cite{clark2023tail}, a Gaussian process (GP) VAR \citep{HauzenbergerJBES2025}, as well as a time-varying parameter VAR with a horseshoe shrinkage prior on the time-invariant parameters and the state innovation variances.\footnote{Our GP-VAR model differs from the original implementation of \cite{HauzenbergerJBES2025} by featuring a factor structure in the shocks.} As a benchmark, we consider a restricted linear specification that sets $Q_\mu=0$ while keeping $Q_q=3$, i.e., keeping a linear factor structure in the model's innovation term.  Again, for notational brevity, we refer to this benchmark as a linear BVAR.

\subsection{Predictive evidence}
\autoref{tab: ES} provides average energy scores for our factor-BART VAR specification and the alternatives described above, all relative to average scores for the linear BVAR.  Over the full sample (top panel of the table), the factor-BART VAR model consistently improves on the accuracy of the linear model, by small margins (1 to 2\%) at the one-step horizon but more sizable margins (10 to 13\%) at multi-step horizons.  Results are largely the same with 1 or 3 nonlinear factors as with 6 or 8 nonlinear factors.  The mixBART specification yields accuracy similar to that of the factor-BART VAR, and both models belong to the $75\%$ Model Confidence Set \citep{HansenLundeNason2011}.  But it is worth stressing that estimating the model with $Q_\mu=8$ takes around 9 minutes on a state-of-the-art MacBook Pro laptop, whereas estimating the mixBART specification takes around 30 minutes on the same machine, so that the factor structure in our model provides substantial computational and appreciable forecast accuracy gains over the linear benchmark.  From the perspective of overall density forecast accuracy, it follows that our introduction of a factor structure to capture nonlinearities rather than allowing a nonlinear component for each variable provides computational gains facilitating large models without reducing forecast accuracy.  Of the other specifications considered, the alternative of the GP-VAR slightly improves on the linear BVAR's accuracy for multi-step horizons but falls short of the accuracy of the BART-based specifications.  Finally, the TVP-VAR falls short of the accuracy of all the other models.  This may stem from the rich parameterization that comes with using TVPs in a large model.

The lower panels of \autoref{tab: ES} break forecast accuracy down across three sub-samples.\footnote{The evolution of the relative energy score over the full hold-out is reported in \autoref{fig:energyscoresovertime} of \autoref{app:addforecast} in the online appendix.} The improvements are not the artifact of a single episode. The gains are largest over the 2008--2019 window---roughly $13$ to $14\%$ at the one-year horizon---and remain clear in the pre-crisis 2002--2007 window. The short post-COVID window (2020--2023) is the one period in which the picture is mixed: At the one-step horizon factor-BART is no more accurate than the linear BVAR, but by the three-years horizon its relative energy score falls to around $0.86$--$0.90$, although with only sixteen quarters in this window the tests are correspondingly less decisive. In terms of statistical significance, in the full sample and in the two pre-COVID windows the factor-BART specifications belong to the $75\%$ Model Confidence Set (shown in orange) at most horizons, and the Giacomini--White tests reject equal predictive accuracy against the linear BVAR (asterisks) for the large majority of the multi-step comparisons.  
\begin{table}[h!]
  \centering
  \begin{tabular}{lllllllll}
    \hline
    & & \multicolumn{4}{c}{\texttt{fBART}} & & & \\
  $h \downarrow$ &  & $Q_\mu{=}1$ & $Q_\mu{=}3$ & $Q_\mu{=}6$ & $Q_\mu{=}8$ & \texttt{mixBART} & \texttt{GP-VAR} & \texttt{TVP-VAR} \\
    \hline
  \multicolumn{9}{l}{\textit{Panel: Full sample (2002--2023) ($n=88$ quarters)}} \\
  1  & & \textcolor{mcsorange}{0.99}         & \textcolor{mcsorange}{0.99}         & \textcolor{mcsorange}{0.99}         & \textcolor{mcsorange}{0.99}         &
  \textcolor{mcsorange}{1.01}         & \textcolor{mcsorange}{1.02}         & 1.12$^{**}$ \\
  4  & & 0.91$^{***}$                        & 0.90$^{***}$                        & \textcolor{mcsorange}{0.89$^{***}$} & \textcolor{mcsorange}{0.89$^{***}$} &
  \textcolor{mcsorange}{0.90$^{***}$} & 0.94$^{***}$                        & 1.15$^{*}$  \\
  8  & & \textcolor{mcsorange}{0.90$^{***}$} & 0.90$^{***}$                        & \textcolor{mcsorange}{0.89$^{***}$} & \textcolor{mcsorange}{0.89$^{***}$} &
  \textcolor{mcsorange}{0.91$^{***}$} & 0.95$^{***}$                        & 1.38$^{**}$ \\
  12 & & 0.88$^{***}$                        & \textcolor{mcsorange}{0.88$^{***}$} & \textcolor{mcsorange}{0.87$^{***}$} & \textcolor{mcsorange}{0.88$^{***}$} &
  \textcolor{mcsorange}{0.89$^{***}$} & 0.94$^{**}$                         & 0.99        \\
  \hline
  \multicolumn{9}{l}{\textit{Panel: 2002--2007 ($n=24$ quarters)}} \\
  1  & & 0.95$^{*}$                          & 0.93$^{***}$                        & 0.89$^{***}$                        & \textcolor{mcsorange}{0.87$^{***}$} &
  \textcolor{mcsorange}{0.88$^{***}$} & 1.06$^{*}$                          & 1.22$^{***}$ \\
  4  & & 0.97$^{***}$                        & 0.96$^{***}$                        & \textcolor{mcsorange}{0.93$^{***}$} & \textcolor{mcsorange}{0.93$^{***}$} &
  \textcolor{mcsorange}{0.94$^{**}$}  & 1.03                                & 1.13$^{***}$ \\
  8  & & 0.99                                & 0.98$^{*}$                          & \textcolor{mcsorange}{0.96$^{***}$} & \textcolor{mcsorange}{0.95$^{*}$}   &
  \textcolor{mcsorange}{0.98}         & \textcolor{mcsorange}{1.01}         & \textcolor{mcsorange}{1.06}  \\
  12 & & 0.97$^{***}$                        & 0.97$^{***}$                        & \textcolor{mcsorange}{0.95$^{***}$} & \textcolor{mcsorange}{0.94$^{***}$} &
  \textcolor{mcsorange}{0.95$^{***}$} & \textcolor{mcsorange}{1.01}         & \textcolor{mcsorange}{0.97$^{**}$} \\
  \hline
  \multicolumn{9}{l}{\textit{Panel: 2008--2019 ($n=48$ quarters)}} \\
  1  & & 0.97                                & \textcolor{mcsorange}{0.95}         & \textcolor{mcsorange}{0.94}         & \textcolor{mcsorange}{0.96}         &
  \textcolor{mcsorange}{0.98}         & 1.04                                & 1.20$^{***}$ \\
  4  & & 0.87$^{***}$                        & \textcolor{mcsorange}{0.86$^{***}$} & \textcolor{mcsorange}{0.86$^{***}$} & \textcolor{mcsorange}{0.87$^{***}$} &
  0.89$^{***}$                        & 0.93$^{***}$                        & 1.03         \\
  8  & & 0.85$^{***}$                        & \textcolor{mcsorange}{0.84$^{***}$} & \textcolor{mcsorange}{0.84$^{***}$} & \textcolor{mcsorange}{0.85$^{***}$} &
  0.88$^{***}$                        & 0.93$^{***}$                        & 1.38$^{*}$   \\
  12 & & 0.85$^{***}$                        & \textcolor{mcsorange}{0.84$^{***}$} & \textcolor{mcsorange}{0.85$^{***}$} & 0.85$^{***}$                        &
  0.88$^{***}$                        & 0.92$^{***}$                        & 0.96         \\
  \hline
  \multicolumn{9}{l}{\textit{Panel: 2020--2023 ($n=16$ quarters)}} \\
  1  & & 1.03$^{*}$                          & \textcolor{mcsorange}{1.04}         & \textcolor{mcsorange}{1.05$^{*}$}   & \textcolor{mcsorange}{1.06$^{*}$}   &
  \textcolor{mcsorange}{1.08}         & \textcolor{mcsorange}{0.98}         & \textcolor{mcsorange}{1.03}  \\
  4  & & 0.97                                & 0.97                                & 0.95                                & 0.96                                &
  \textcolor{mcsorange}{0.90}         & \textcolor{mcsorange}{0.89$^{***}$} & 1.56$^{*}$   \\
  8  & & 0.96                                & 1.03                                & 1.01                                & 1.00                                &
  0.92$^{*}$                          & 0.94$^{***}$                        & 1.98$^{***}$ \\
  12 & & 0.89$^{***}$                        & 0.89$^{**}$                         & 0.86$^{**}$                         & 0.90$^{*}$                          &
  0.82$^{**}$                         & 0.86$^{***}$                        & 1.42$^{***}$ \\
  \hline
  \end{tabular}
  \caption*{\tiny \textbf{Notes}: Each cell reports the multivariate Energy Score of the column model divided by the Energy Score of the linear BVAR ($Q_\mu =
  0$), restricted to the sub-sample shown in the panel header. The exercise is a recursive out-of-sample evaluation. Asterisks summarize Giacomini--White \citep{giacomini2006tests} tests
  of equal predictive ability against the linear-BVAR baseline, with HAC variance estimated by Newey--West using $h-1$ lags: $^{***}\,p<0.01$, $^{**}\,p<0.05$,
  $^{*}\,p<0.10$. ES values shown in \textcolor{mcsorange}{orange} indicate that the column model belongs to the $75\%$ Model Confidence Set
  \citep{HansenLundeNason2011} at that horizon (computed within each sub-sample over the eight headline models using the $T_R$ statistic and a circular block
  bootstrap).}
  \caption{Energy scores relative to the linear BVAR, overall and by sub-sample.}
  \label{tab: ES}
  \end{table}

\begin{figure}[h!]
    \centering
    \includegraphics[width=1\textwidth, trim={0cm 0cm 0cm 0cm}, clip]{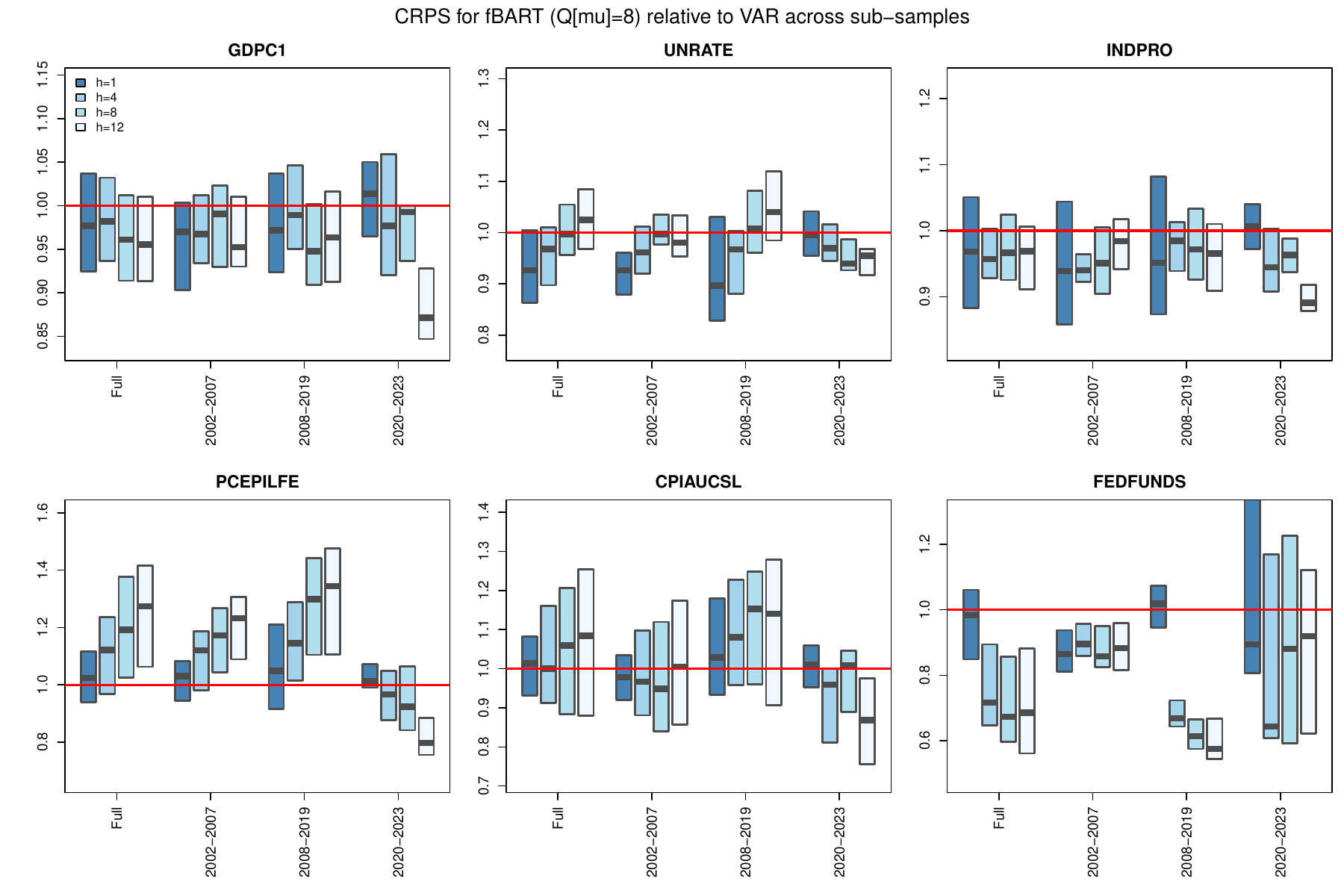}
    \caption{Boxplots of relative CRPSs to the linear BVAR for the factor-BART VAR with $Q_\mu = 8$, across variables, forecast horizons and sub-samples.}
    \label{fig:relativeCRPS_subsample}
\end{figure}

To shed further light on overall predictive performance, \autoref{fig:relativeCRPS_subsample} reports boxplots of the CRPS for our factor-BART VAR with $Q_\mu = 8$ relative to the linear BVAR, for a subset (in the interest of brevity) of key variables  and horizons $h \in \{1, 4, 8, 12\}$, over the full hold-out and the three sub-samples (2002--2007, 2008--2019, 2020--2023); values below one indicate that factor-BART improves upon the benchmark.

CRPS ratios for the real-activity variables (real GDP growth, unemployment, industrial production) are mostly below 1 across samples, with gains that vary across horizons and sometimes larger in the COVID era than in other samples. The federal funds rate shows the largest and most consistent improvements (exceeding 30\% during 2008--2019), while the two inflation measures are broadly competitive with the linear BVAR before 2020 and improve clearly post-COVID (median score ratios around 0.85--0.95). Comparisons against the remaining nonlinear specifications, and the evolution of the energy score over time, are provided in the online appendix.

\section{Structural analysis using a large nonlinear model of the US economy}\label{sec:IRFapp} 

Having established some benefits of our proposed model for out-of-sample forecasting, this section's application shows its ability to capture nonlinearities in responses to structural shocks.  In this application, we use precisely the same set of $M = 22$ variables and factor-BART VAR specification as in the forecasting application (with $Q_\mu = 8)$.  We use sign restrictions (detailed below) as in studies such as \cite{KOROBILIS2022} to identify shocks to demand, supply, monetary policy, and financial conditions.  Accordingly, we set the number of common shocks to $Q_q=4$. 

%The first subsection below starts with some reduced-form results on the importance of the factors and how the observed variables vary nonlinearly with the covariates. The second subsection provides further information on how we identify the structural shocks and presents the impulse response functions for shocks to financial conditions and monetary policy.

\subsection{Reduced-form results}
\begin{figure}[h!]
    \centering
    \includegraphics[width=0.9\linewidth]{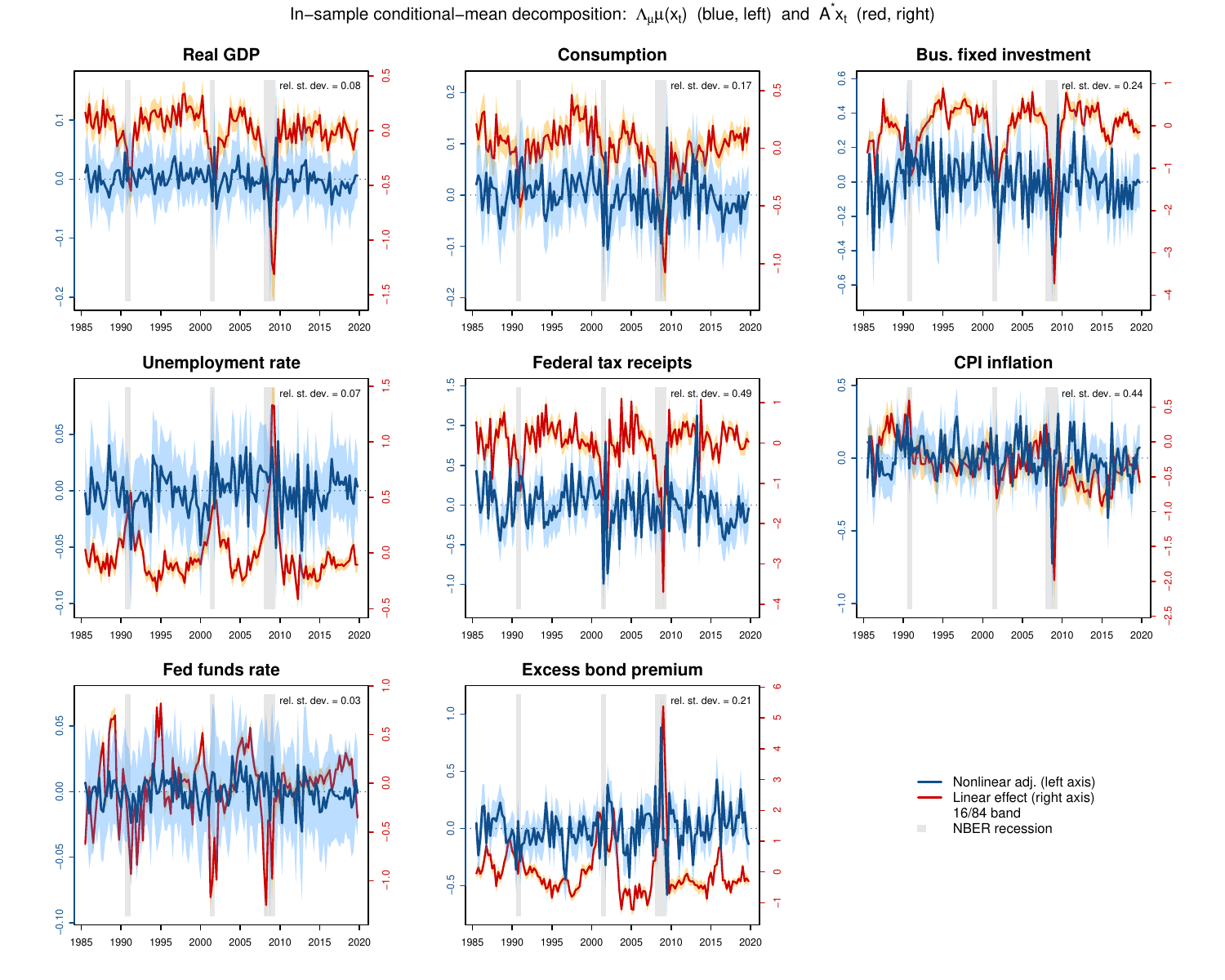}
    \caption*{\tiny \textbf{Notes}: Blue (left axis): posterior median and $16/84$ band of the identified nonlinear adjustment $\bm \Lambda_\mu \bm \mu(\bm x_t)$.  Red (right axis): posterior median and $16/84$ band of the identified linear effect $\bm X \bm A^\ast$.  The two components are on different vertical scales (the nonlinear adjustment is the smaller of the two); grey shaded periods are NBER recessions.}
    \caption{In-sample conditional-mean decomposition for the eight focus variables}
    \label{fig:factorvarianceshares}
\end{figure}

To provide a sense of the magnitudes of nonlinearities, \autoref{fig:factorvarianceshares} reports the time series evolution of the orthogonalized linear and nonlinear components of the model. Notice that without the ex-post orthogonalization, the resulting fit coming from $\bm \Lambda_\mu \mu(\bm x_t)$ would also capture linear common trends or other linear, reduced-rank effects.  In each panel, an inset box provides the ratio of (i) the standard deviation of the nonlinear component to (ii) the standard deviation of the linear component.

\autoref{fig:factorvarianceshares}  gives rise to  two major observations. First, for every focus variable the nonlinear adjustment $\bm \Lambda_\mu \bm \mu(\bm x_t)$ (in blue and on the left axis) is an order of magnitude smaller than the linear component $\bm X \bm A^\ast$ (in red, right axis), so the linear BVAR continues to account for the bulk of the conditional mean while the nonlinear block supplies a parsimonious correction.  Across variables, the standard deviation of the nonlinear component relative to that of the linear component ranges from less than 0.1 for real GDP, unemployment, and the federal funds rate to about 0.2 for consumption and business investment and more than 0.4 for CPI inflation and tax receipts.  Second, that correction is far from constant, being muted in normal times and becoming relevant around NBER recessions (grey bands). This is most visible for the real-activity variables (real GDP, consumption, business fixed investment, and the unemployment rate) and for the excess bond premium, where the nonlinear adjustment spikes in 2008--09 and again in 2020. These are precisely the episodes in which a purely linear model tends to mis-predict, and the nonlinear block pulls the linear extrapolation back toward the realized path. For smoother and more persistent series such as the federal funds rate and inflation, the adjustment is comparatively small and flat. %, mirroring the forecasting result that these variables gain little from the added flexibility (at least at the one-quarter horizon). %A complementary, shock-specific view of the same nonlinearity---how the focus variables bend with the level of financial stress---is provided by the partial-dependence plots in \autoref{fig:PDPs_EBP_agg}.}

\begin{figure}[t!]
    \centering
    \includegraphics[width=\linewidth, clip, trim=0cm 0cm 2cm 1.5cm]{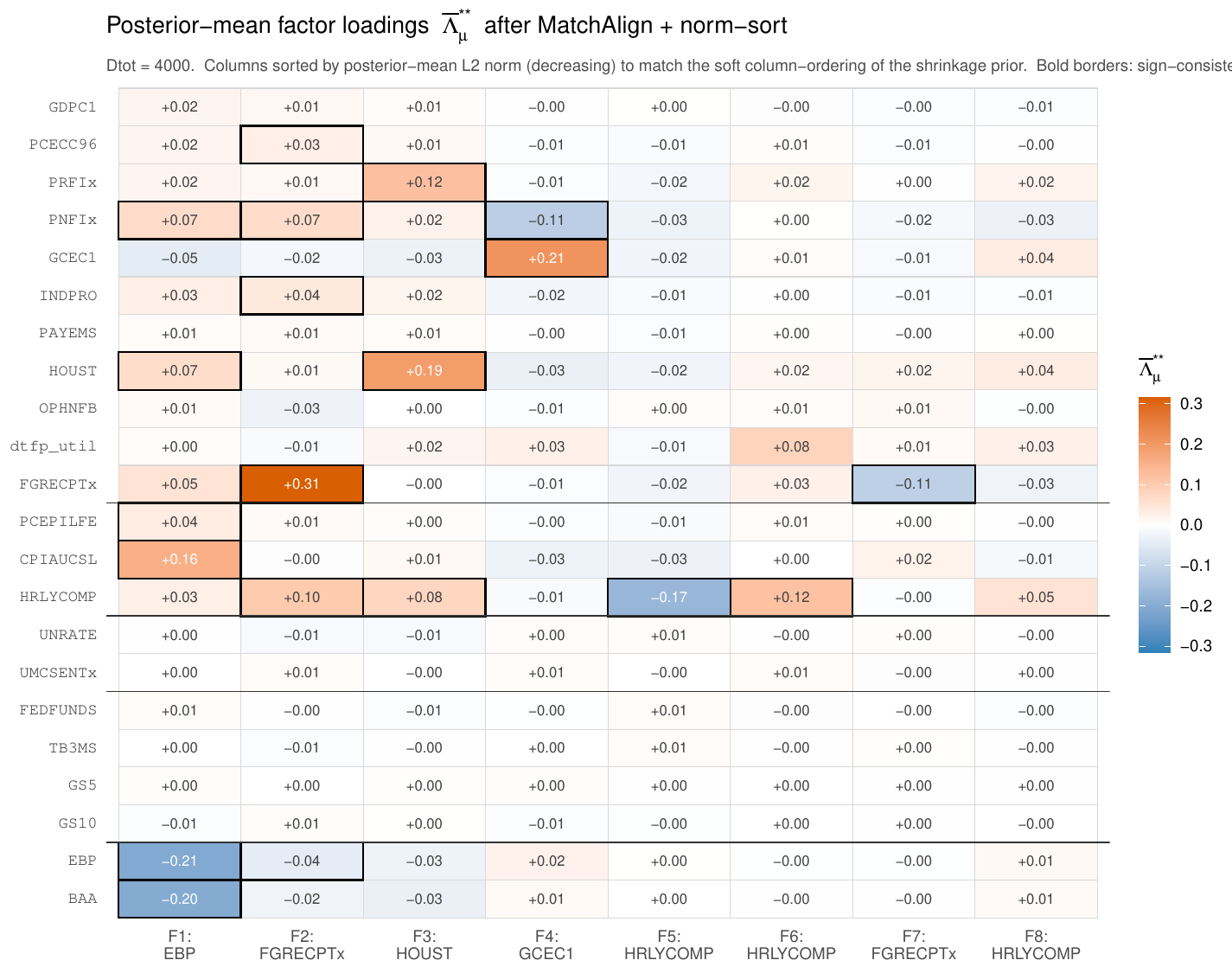}
        \caption*{\tiny\textbf{Notes:} Cells show posterior-mean loadings $\bar{\Lambda}_{\mu,ij}$ after the alignment procedure described in \autoref{sec: identification}. Columns are sorted by decreasing posterior-mean $L_2$ norm; each column header reports the dominant series, i.e.\ the variable with the largest $|\hat{\Lambda}_{\mu,ij}|$ in that column. Cells with a bold border are sign-consistent across draws, $\max\!\bigl(\Pr(\Lambda_{\mu,ij} > 0), \Pr(\Lambda_{\mu,ij} < 0)\bigr) \ge 0.84$. Rows are grouped by economic block.}
    \caption{Posterior-mean factor loadings $\widehat{\bm \Lambda}_\mu$ after MatchAlign with posterior-mean $L_2$ norm-sort.}
    \label{fig:loadings_aligned}
\end{figure}

\autoref{fig:loadings_aligned} reports the posterior mean of the  identified  loadings, with the columns ordered by decreasing $L_2$ norm. Two features stand out. First, the loadings decay sharply from left to right: Only the first four to five factors carry economically and statistically significant loadings, while the trailing columns are essentially shrunk to zero. This is direct evidence that, although we estimate $Q_\mu = 8$ factors, the column-wise shrinkage prior selects a much smaller effective number of nonlinear factors --- an empirical confirmation of the low-rank, shared view of the nonlinearity that motivates the factor structure in \autoref{sec: fBART-VAR}. Second, the factors are economically interpretable. The leading factor loads negatively on the excess bond premium and the BAA spread and positively on prices, housing, and investment; the subsequent factors pick up federal tax receipts, housing starts, government spending, and labor compensation. Notably, the interest-rate block (the federal funds rate, the three-month bill, and the five- and ten-year yields) loads close to zero on every factor, indicating that the estimated nonlinearity is concentrated in real activity, credit and financial conditions, and prices rather than in the rates block---consistent with the largely symmetric interest-rate responses documented in \autoref{sec:signs}.

\subsection{Nonlinear responses to  {financial and }monetary policy shocks}\label{sec:signs} 

Turning to impulse response estimates, as noted above we use sign restrictions to identify shocks to demand, supply, and monetary policy, and a financial shock. Sign restrictions are implemented in the tradition of \cite{BH2015} and \cite{KOROBILIS2022} through priors truncated to the respective set implied by the sign restrictions. In our framework, this implies changing the Gaussian prior on $\bm \Lambda_q$ to a truncated Gaussian prior to attach an economic meaning to the elements in $\bm q_t$ (and thus avoid identification issues related to column switching). We achieve this by setting:
{
\footnotesize
\begin{equation*}
    \lambda_{q, ij} \sim \begin{cases}
        \mathcal{N}_{0, \infty}(0, 10^2), \quad &\text{ if shock $j$ is restricted to have a positive impact effect on variable $i$},\\
        \mathcal{N}_{-\infty, 0}(0, 10^2), \quad &\text{ if shock $j$ is restricted to have a negative impact effect on variable $i$},\\
        d_c(\lambda_{q, ij}) , \quad &\text{ if shock $j$ is restricted to have a fixed impact effect $c$ on variable $i$},\\
        \mathcal{N}(0, 10^2), \quad &\text{ if shock $j$ exerts an unrestricted impact effect on variable $i$}.
    \end{cases} 
\end{equation*}
} 
\noindent Here, we let $d_c(\lambda_{ij})$ denote the Dirac delta function, which puts point mass on $c$. If we use a truncated prior, the posterior simulation step for $\bm \Lambda_q$ changes slightly. Instead of simulating the rows of $\bm \Lambda_q$ from unrestricted Gaussian distributions, the truncated prior implies that the posterior is truncated to the set implied by the prior.

\autoref{tab: signs} lists the sign restrictions.  For example, following typical practice, a one standard deviation monetary policy shock (tightening) is identified as a shock that raises the federal funds rate  and reduces GDP growth, increases the unemployment rate, and reduces core PCE inflation.  A shock to financial conditions is identified as a shock that raises the  EBP  and reduces GDP growth, consumption, investment, consumer sentiment, and core PCE inflation, and raises the unemployment rate, leading to a reduction of the federal funds rate.  These identifying restrictions are similar to those used in studies such as \cite{Chanetal2025large}.
\begin{table}[t!]
\centering
\scalebox{0.85}{
\begin{tabular}{lcccccccccccccccccccccc}
\toprule
\textbf{Shock} 
& \rotatebox{90}{GDPC1}
& \rotatebox{90}{PCECC96}
& \rotatebox{90}{EBP}
& \rotatebox{90}{PRFIx}
& \rotatebox{90}{PNFIx}
& \rotatebox{90}{UNRATE}
& \rotatebox{90}{GCEC1}
& \rotatebox{90}{FGRECPTx}
& \rotatebox{90}{PCEPILFE}
& \rotatebox{90}{CPIAUCSL}
& \rotatebox{90}{HRLYCOMP}
& \rotatebox{90}{OPHNFB}
& \rotatebox{90}{dtfp\_util}
& \rotatebox{90}{PAYEMS}
& \rotatebox{90}{UMCSENTx}
& \rotatebox{90}{INDPRO}
& \rotatebox{90}{HOUST}
& \rotatebox{90}{FEDFUNDS}
& \rotatebox{90}{TB3MS}
& \rotatebox{90}{GS5}
& \rotatebox{90}{GS10}
& \rotatebox{90}{BAA} \\
\midrule

Demand 
& + &  &  &  & - & - &  & + & + & + &  &  & 0 &  & + & + &  & + & + &  &  &  \\

Monetary 
& - &  &  &  &  & + &  &  & - & - &  &  & 0 & - &  & - &  & + & + & + & + & + \\

Supply 
& - & - &  &  &  & + &  &  & + & + &  &  &  &  &  &  &  &  &  &  &  &  \\

Financial 
& - & - & + &  & - & + &  &  & - &  &  &  &  &  & - &  &  & - &  &  &  &  \\

\bottomrule
\end{tabular}
}
\caption*{\tiny \textbf{Notes}: `+' refers to a positive impact reaction, `-' refers to a negative one, and `0' implies that we fix the impact effect, whereas no sign means that we have introduced no restriction.}
\caption{Sign restrictions used to identify the structural factors in the model}
\label{tab: signs}
\end{table}
In the interest of brevity, we focus on financial and monetary policy shocks.  To assess possible nonlinearities, we consider the impacts of shocks differing in sign and size. Specifically, we estimate impulse response functions (IRFs) to positive and negative shocks, with the responses to the negative shock multiplied by $-1$ to facilitate comparing responses to positive and negative shocks.  We consider both small and large shocks.  The shocks are scaled so that the monetary policy shock (in the tightening case) triggers a median increase in FEDFUNDS of about 20 basis points in the small shock case and  about 55 basis points in the large shock case.  The financial shock causes an on-impact increase in the EBP of about 0.35 in the small case and 1.0 in the large case.  In all cases, we compute the impulse response functions (IRFs) using the generalized IRF approach of \cite{KoopPesaranPotter:1996:girf}; see \autoref{app: girf_appendix} for a full description of the Monte Carlo implementation. 

%\textcolor{purple}{Check whether we can delete this sentence if we already link to the Appendix.}

\begin{figure}[p]
    \centering
    \begin{subfigure}{\linewidth}
        \centering
        \includegraphics[width=\linewidth]{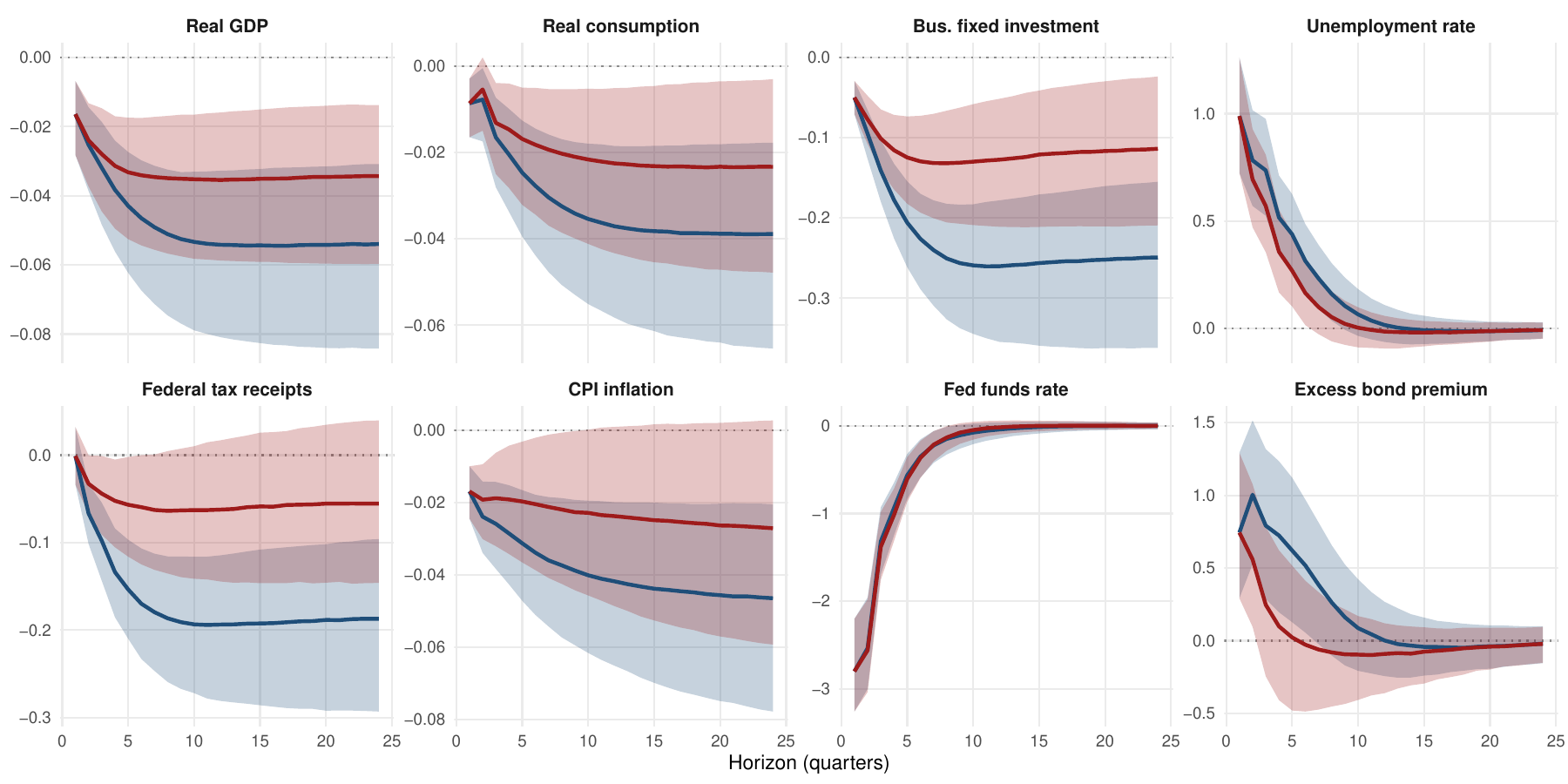}
        \caption{Large financial-conditions shock}
        \label{fig:irfs_financial_large}
    \end{subfigure}

    \begin{subfigure}{\linewidth}
        \centering
        \includegraphics[width=\linewidth]{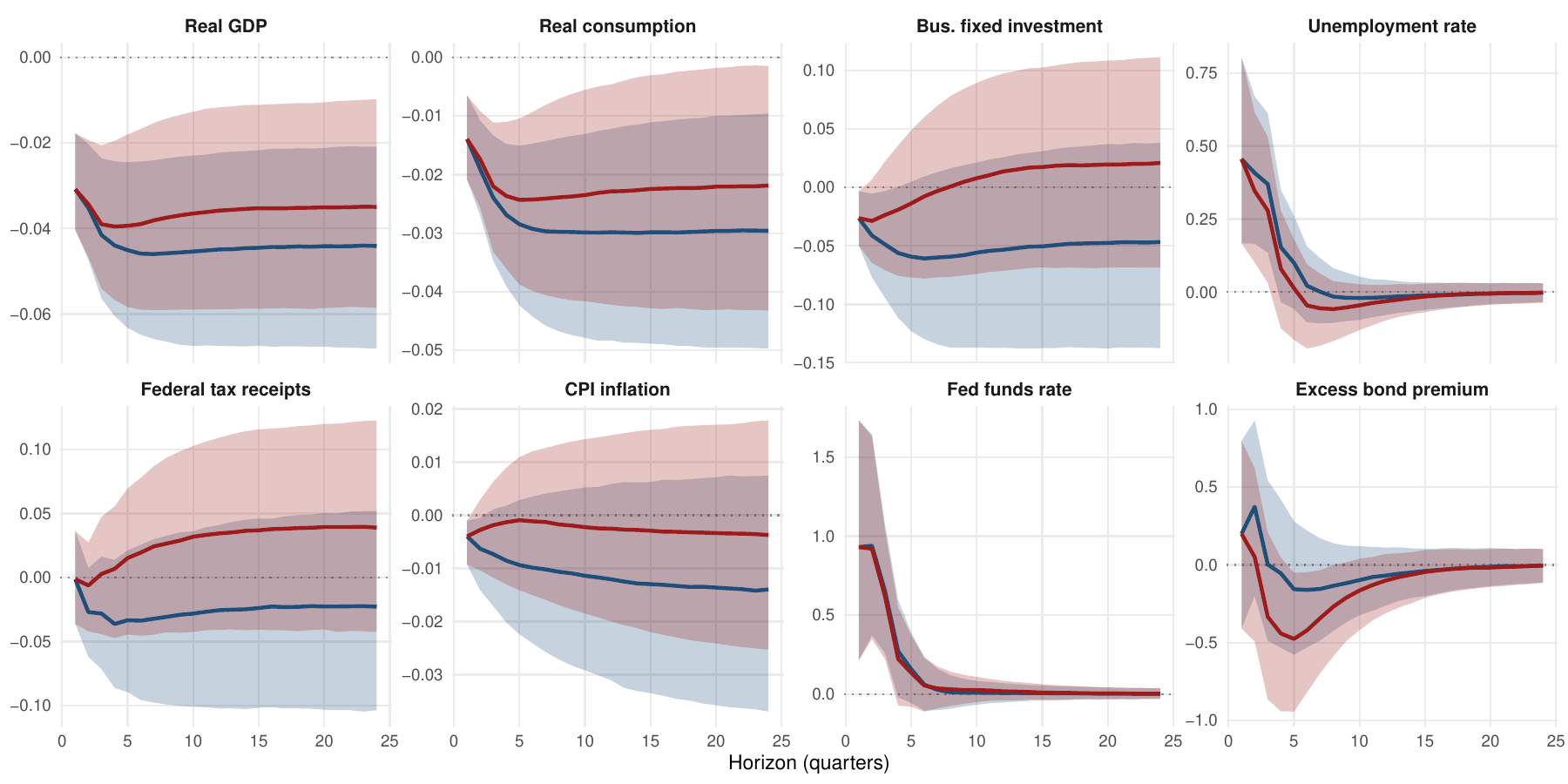}
        \caption{Large monetary policy shock }
        \label{fig:irfs_monetary_large}
    \end{subfigure}
        \caption*{\tiny\textbf{Notes:} Blue solid line and shaded blue band: posterior median and $16/84$ band of the response to the contractionary ($+$) shock.  Red solid line and shaded red band: posterior median and $16/84$ band of the response to the expansionary ($-$) shock, multiplied by $-1$.  Grey dotted line marks zero.  Eight focus response variables shown; the remaining $14$ responses and the small-shock variants are reported in \autoref{app:fullirf}.}
    \caption{GIRFs of the eight focus variables to large financial-conditions and monetary policy shocks.}
    \label{fig:irfs_large}
\end{figure}

In the main text we concentrate on the \emph{large} shocks, for which the model's nonlinearities are most apparent, and report the eight focus variables in \autoref{fig:irfs_financial_large} and \autoref{fig:irfs_monetary_large}. The responses to the corresponding small shocks, together with the remaining fourteen variables, are collected in \autoref{app:fullirf} (Figures~\ref{fig:focus_financial_small} and~\ref{fig:focus_monetary_small} for the focus variables). The  responses to small shocks are qualitatively  similar in shape and sign but exhibit only mild asymmetries between tightenings and easings.

Starting with financial shocks, \autoref{fig:irfs_financial_large} reports the responses of the focus variables to large tightening (positive) and easing (negative) shocks, the latter multiplied by $-1$. A tightening of financial conditions slows economic activity by lowering GDP, consumption, investment, industrial production, housing starts, and employment while raising unemployment. Inflation also declines. At this large shock size the responses are clearly asymmetric. A tightening has bigger impacts on activity and inflation than the corresponding easing, with the gap most pronounced six to eight quarters out for GDP, consumption, investment, industrial production, payroll employment, and federal tax receipts. Consumer sentiment and interest rates remain largely symmetric. These findings are consistent with \cite{barnichon2022effects}, who show that contractionary financial shocks trigger larger effects than benign ones.

\autoref{fig:irfs_monetary_large} reports the responses to large monetary policy shocks. A policy tightening slows economic activity  (similarly to the financial shock) but it also lowers headline CPI and core PCE inflation. As with financial shocks, the asymmetry is evident at the large shock size: A tightening has bigger impacts on activity and inflation than a corresponding easing, with the largest gaps six to eight quarters out for GDP, investment, industrial production, and payroll employment, while consumer sentiment and interest rates respond largely symmetrically.

\section{Concluding remarks}\label{sec:conclusion}
We have proposed an extension to the standard VAR where departures from linearity are modeled using a nonlinear factor structure. The nonlinear factor is modeled using Bayesian regression tree methods.  The use of a factor structure ensures parsimony and is motivated by the empirical findings that macroeconomic time series are typically found to be driven by a small number of factors. The fact that these factors are nonlinear and estimated nonparametrically reduces the risk of misspecification. These features are of particular use for the researcher who expects their dataset to exhibit nonlinearities but has little guidance as to the precise form these nonlinearities will take.  In an artificial data exercise and an empirical study involving US macroeconomic data, we demonstrate the effectiveness of our methods and their ability to pick out patterns in the data that cannot be found using linear methods.  

%It would be straightforward to extend our model to allow for other features of the VAR to be modeled more flexibly. For instance, the assumption of Gaussian errors could be easily relaxed through a mixture of Gaussians. Or the error variances could be allowed to be time varying using BART methods as in \cite{pratola2020heteroscedastic}. But our factor-BART VAR already exhibits a high degree of flexibility. 

\small{\setstretch{0.85}
\addcontentsline{toc}{section}{References}
\bibliographystyle{frbcle.bst}
\bibliography{lit_FRBCWP}}\normalsize\clearpage

\begin{center}\LARGE\textbf{Online Appendix}\end{center}
\begin{appendices}
\setcounter{equation}{0}
\setcounter{table}{0}
\renewcommand\thetable{A.\arabic{table}}
\renewcommand\theequation{A.\arabic{equation}}
\section{Data Appendix}\label{app: data}
\begin{table}[h!]
\centering
\caption{Overview of the dataset}
\label{tab:data}
\begin{tabular}{lll}
\toprule
\textbf{Variable} & \textbf{Description} & \textbf{Transformation} \\
\midrule
GDPC1         & Real Gross Domestic Product                     & 5 \\
PCECC96       & Real Personal Consumption Expenditures          & 5 \\
EBP           & Excess Bond Premium                             & 1 \\
PRFIx         & Residential fixed investment                    & 5 \\
PNFIx         & Non-residential fixed investment                & 5 \\
UNRATE        & Unemployment rate                               & 2 \\
GCEC1         & Government consumption expenditures             & 5 \\
FGRECPTx      & Federal tax receipts                            & 5 \\
PCEPILFE      & Core PCE price index                            & 6 \\
CPIAUCSL      & Consumer Price Index                            & 6 \\
HRLYCOMP      & Hourly compensation                             & 6 \\
OPHNFB        & Labor productivity (nonfarm business)           & 5 \\
dtfp\_util    & Utilization-adjusted total factor productivity  & 1 \\
PAYEMS        & Payroll employment                              & 5 \\
UMCSENTx      & Consumer sentiment index                        & 1 \\
INDPRO        & Industrial production                           & 5 \\
HOUST         & Housing starts                                  & 5 \\
FEDFUNDS      & Federal funds rate                              & 2 \\
TB3MS         & 3-month Treasury bill rate                      & 2 \\
GS5           & 5-year Treasury yield                           & 2 \\
GS10          & 10-year Treasury yield                          & 2 \\
BAA           & BAA corporate bond yield                        & 2 \\
\bottomrule
\end{tabular}
\caption*{\footnotesize \textbf{Notes}: The time series are sourced from the McCracken \& Ng database (except EBP and dtfp\_util).
The transformation codes mean: 1 = raw series, 2 = first differences, 3 = second differences, 4 = logarithm, 5 = log first differences, 6 = second differences of logs.}
\end{table}

\subsection*{More details on the application of Section 1}

To illustrate the relationship between the excess bond premium and real output and prices, we plot the estimated functional form $\hat{f}_i(\text{EBP})$ for each outcome variable $i$. The horizontal axis displays values of the EBP, while  the vertical axis displays the corresponding function estimate $\hat{f}_i(\text{EBP})$.  Each curve represents the posterior mean of a smoothing spline fit, chosen here for simplicity of illustration.

%visual clarity; alternative nonparametric estimators such as BART yield  qualitatively similar results but produce less regular curves that are harder to  interpret in a stylized setting.

The underlying data are drawn from the FRED-MD dataset \citep{mccracken2016fred}. 
All series enter the model as year-over-year percentage growth rates. We consider 
two blocks of variables: a \emph{real-activity block} comprising $n = 11$ 
industrial production sub-aggregates, and a \emph{price block} comprising $n = 8$ 
CPI and PCE inflation series. The full list of series is provided in 
Table~\ref{tab:series}.

\begin{table}[ht]
\centering
\caption{FRED-MD series used in the analysis} %A  (year-over-year \% growth rates)
\label{tab:series}
\begin{tabular}{lll}
\toprule
\textbf{Mnemonic} & \textbf{Description} & \textbf{Block} \\
\midrule
\multicolumn{3}{l}{\textit{Panel A: Real-Activity Block ($n = 11$) --- 
    Industrial Production sub-aggregates}} \\[4pt]
\texttt{INDPRO}    & Industrial Production Index (headline)      & Real \\
\texttt{IPFPNSS}   & IP: Final Products and Nonindustrial Supplies & Real \\
\texttt{IPFINAL}   & IP: Final Products                          & Real \\
\texttt{IPCONGD}   & IP: Consumer Goods                          & Real \\
\texttt{IPDCONGD}  & IP: Durable Consumer Goods                  & Real \\
\texttt{IPNCONGD}  & IP: Nondurable Consumer Goods               & Real \\
\texttt{IPBUSEQ}   & IP: Business Equipment                      & Real \\
\texttt{IPMAT}     & IP: Materials                               & Real \\
\texttt{IPDMAT}    & IP: Durable Materials                       & Real \\
\texttt{IPNMAT}    & IP: Nondurable Materials                    & Real \\
\texttt{IPMANSICS} & IP: Manufacturing (SIC)                     & Real \\
\midrule
\multicolumn{3}{l}{\textit{Panel B: Price Block ($n = 8$) --- 
    CPI / PCE inflation series}} \\[4pt]
\texttt{CPIAUCSL}     & CPI: All Items                  & Price \\
\texttt{CPIAPPSL}     & CPI: Apparel                    & Price \\
\texttt{CPIMEDSL}     & CPI: Medical Care               & Price \\
\texttt{CUSR0000SAC}  & CPI: Commodities                & Price \\
\texttt{CUSR0000SAD}  & CPI: Durables                   & Price \\
\texttt{CUSR0000SAS}  & CPI: Services                   & Price \\
\texttt{CPIULFSL}     & CPI: All Items Less Food        & Price \\
\texttt{PCEPI}        & PCE: Chain-Type Price Index     & Price \\
\bottomrule
\end{tabular}
\end{table}

\section{Technical Appendix}\label{app:B}
\setcounter{equation}{0}
\setcounter{table}{0}
\renewcommand\thetable{B.\arabic{table}}
\renewcommand\theequation{B.\arabic{equation}}

\subsection{Specific prior choices}

We use a horseshoe prior on the VAR coefficients:\footnote{The prior on the VAR's coefficients could alternatively take a Minnesota form.  For VAR forecast accuracy, global-local priors such as the horseshoe often perform comparably to, or slightly better than, the Minnesota prior \citep[e.g.,][]{huber2019adaptive}.}
\begin{equation*}
    a_{ij} \sim \mathcal{N}(\underline{a}_{ij}, \psi_{a, ij}^2 \tau^2_{a}), \quad \psi_{a, ij} \sim \mathcal{C}^+(0, 1), \quad \tau_a \sim \mathcal{C}^+(0, 1).
\end{equation*}
Here, $\underline{a}_{ij}$ is a prior mean that can be set such that the elements in $\bm y_t$ are forced toward a random walk (if they are non-stationary) or toward a white noise process (if they are stationary). Note that this prior hyperparameter choice is centered over coefficient values that imply that each element of $\bm y_t$ is a persistent AR process. Hence the nonlinear factor part of the model has a strong tendency to soak up deviations from the persistent part of $\bm y_t$.

In contrast to the prior on the factor loadings $\bm \Lambda_\mu$, the prior on the linear VAR part of the model features a single global shrinkage parameter and hence small values of $\tau_a$ force all elements in $\bm A$ to zero. The idiosyncratic scaling parameters then allow for non-zero effects by leading to a heavy-tailed marginal (of $\psi_{a, ij}$) prior on $a_{ij}$.

On the elements of $\bm \Lambda_q$, $\lambda_{q, ij}$, we use Gaussian priors with zero mean and variance $10^2$. This noninformative choice implies little shrinkage on the elements of $\bm \Lambda_q$. In our structural analysis in Section \ref{sec:IRFapp} where we use sign restrictions, we modify this prior to be a truncated Gaussian distribution.  For the diagonal elements in $\bm \Omega$ we consider conjugate inverse Gamma priors with hyperparameters $a_\omega = b_\omega = 0.01$ to render the prior weakly informative.

For the regression trees, we adopt the standard prior specification proposed in \cite{chipman2010bart} with $S = 250$ trees. We use these standard settings because they have been shown to perform well across a wide variety of applications, particularly when modeling US macroeconomic data. In addition, \citet{HUBER202352} conduct robustness checks and find that deviations from this baseline specification have only a minor impact on predictive accuracy.

This prior involves three aspects. First, the probability that a given node is non-terminal decreases in the depth $d$ of the node:\ $\mathfrak{a}(1+d)^{-\mathfrak{b}}$ with $\mathfrak{a}=0.95$ and $\mathfrak{b}=2$. Second, the splitting variable assignments are uniformly distributed so that, a priori, each splitting variable is equally likely. Third, conditional on the chosen splitting variable, we use a uniform prior on the discrete set of available splitting values.

On the terminal node parameters $\mu_{ij,s}$ we use a Gaussian prior. We follow \cite{chipman2010bart} and normalize each factor to lie between $-0.5$ and $0.5$. Then:
\begin{equation*}
    \mu_{ij, s} \sim \mathcal{N}\!\left(0,\, \frac{0.5}{\sqrt{S}~{\mathfrak{m}}}\right),
\end{equation*}
where $\mathfrak{m} = 2$ is a variance hyperparameter. Larger values of $S$ reduce the prior variance, so that each tree acts as a weak learner and the risk of over-fitting is controlled.

\subsection{Details on the full conditional posterior distributions}

The following section provides the full conditional posterior distributions required for posterior simulation of the factor-BART VAR. The joint posterior is intractable but, given that the priors are conditionally conjugate, we can implement a Gibbs sampler that cycles through the following steps.

\begin{itemize}
\item \textbf{Step I --- VAR coefficients $\bm A$:}
Let $\bm a_s'$ denote the $s^{th}$ row of $\bm A$. Conditional on all other parameters, define the partially adjusted dependent variable $\bm Y^{(s)} = \bm Y_s - \bm M \bm \lambda_{\mu,s} - \bm Q \bm \lambda_{q,s}$, where $\bm Y_s$ is the $T\times 1$ vector stacking $y_{st}$ and $\bm Q =(\bm q_1, \dots, \bm q_T)'$ is the $T\times Q_q$ matrix of structural factors. Then:
\begin{equation*}
\bm a_s \mid \bullet \sim \mathcal{N}(\bar{\bm a}_s,\, \bar{\bm V}_{a,s}),
\end{equation*}
\begin{equation*}
\bar{\bm V}_{a,s} = \!\left(\frac{\bm X'\bm X}{\omega_s^2} + \bm D_{a,s}^{-1}\right)^{-1}\!, \qquad
\bar{\bm a}_s = \bar{\bm V}_{a,s}\!\left(\frac{\bm X' \bm Y^{(s)}}{\omega_s^2} + \bm D_{a,s}^{-1}\underline{\bm a}_s\right),
\end{equation*}
where $\bm X$ is the $T\times K$ regressor matrix stacking $\bm x_t$, $\bm D_{a,s}$ is diagonal with entries $\psi_{a,sj}^2\tau_a^2$, and $\underline{\bm a}_s$ is the horseshoe prior mean.

\item \textbf{Step II --- Factor loadings $\bm\Lambda_\mu$:}
Each row $\bm\lambda_{\mu,s}$ (for $s = 1,\dots,M$) is drawn equation-by-equation from
\begin{equation*}
\bm\lambda_{\mu,s}\mid\bullet \sim \mathcal{N}(\bar{\bm\lambda}_{\mu,s},\, \bar{\bm V}_{\mu,s}),
\end{equation*}
\begin{equation*}
\bar{\bm V}_{\mu,s} = \!\left(\frac{\bm M'\bm M}{\omega_s^2} + \bm D_{\mu,s}^{-1}\right)^{-1}\!, \qquad
\bar{\bm\lambda}_{\mu,s} = \bar{\bm V}_{\mu,s}\,\frac{\bm M'}{\omega_s^2}\!\left(\bm Y_s - \bm X\bm a_s - \bm Q\bm\lambda_{q,s}\right),
\end{equation*}
where $\bm D_{\mu,s}$ is diagonal with entries $\tilde{\psi}_{\mu,sj}^2 = \psi_{\mu,sj}^2\tau_{\mu,s}^2\varpi_j$. % When using the unidentified scheme (recommended), rows $1,\dots,i_{\text{ident}}-1$ of $\bm\Lambda_\mu$ are either fixed (identity scheme) or left-triangularly constrained (lower triangular scheme) as described in Section~\ref{sec:multiBART}.

\item \textbf{Step III --- Nonlinear factors $\bm\mu(\bm x_t)$ via weighted Bayesian backfitting:}
Each factor function $\mu_j$ is updated in turn using BART \citep{chipman2010bart}. For factor $j$, define the residual excluding the contribution of factor $j$:
\begin{equation*}
    \bm R_j = \bm Y - \bm X\bm A' - \bm Q \bm \Lambda'_q - \sum_{k\neq j}\bm\mu_k\bm\lambda_{\mu,k}'  ,
\end{equation*}
where $\bm\mu_k$ is the $T\times 1$ vector $(\mu_k(\bm x_1),\dots,\mu_k(\bm x_T))'$ and $\bm\lambda_{\mu,k}$ is the $k^{th}$ column of $\bm\Lambda_\mu$. The factor $\mu_j$ enters each equation through the scalar loading $\lambda_{\mu,ij}$. To reduce this $M$-dimensional regression to a univariate target, we project $\bm R_j$ onto $\bm\lambda_{\mu,j}$:
\begin{equation*}
    \tilde r_{jt} = \bm\lambda_{\mu,j}'\bm R_j(t)\Big/\|\bm\lambda_{\mu,j}\|^2,
\end{equation*}
where $\bm R_j(t)$ denotes the $t^{th}$ row of $\bm R_j$. This scalar response satisfies
\begin{equation*}
    \tilde r_{jt} = \mu_j(\bm x_t) + \tilde\varepsilon_{jt},
\end{equation*}
with time-varying variances $\varsigma_{jt} = \left(\bm\lambda_{\mu,j}' \bm H_t \bm\lambda_{\mu,j}\right)/\|\bm\lambda_{\mu,j}\|^4$, where $\bm H_t = \text{diag}(h_{1t},\dots,h_{Mt})$. The BART sampler for $\mu_j$ is then run as a standard weighted BART regression of $\tilde r_{jt}$ on $\bm x_t$ with weights $w_{jt}=1/\varsigma_{jt}$. Within this step, the sum-of-trees representation of $\mu_j$ is updated via Bayesian backfitting, cycling through each tree and sampling the tree structure $\mathcal{T}_{j,s}$ and its terminal node parameters $\bm m_{j,s}$ from their full conditionals using the BART MCMC steps of \cite{chipman2010bart}.

\item \textbf{Step IV --- Factor loadings $\bm\Lambda_q$:}
Each row $\bm\lambda_{q,s}$ is drawn from
\begin{equation*}
\bm\lambda_{q,s}\mid\bullet \sim \mathcal{N}(\bar{\bm\lambda}_{q,s},\, \bar{\bm V}_{q,s}),
\end{equation*}
\begin{equation*}
\bar{\bm V}_{q,s} = \!\left(\frac{\bm Q'\bm Q}{\omega_s^2} + \frac{1}{100}\bm I_{Q_q}\right)^{-1}\!, \qquad
\bar{\bm\lambda}_{q,s} = \bar{\bm V}_{q,s}\,\frac{\bm Q'}{\omega_s^2}\!\left(\bm Y_s - \bm X\bm a_s - \bm M \bm\lambda_{\mu,s}\right).
\end{equation*}
When sign restrictions are used (Section~\ref{sec:signs}), we replace the Gaussian prior with a truncated Gaussian, so that the posterior is also truncated to the set implied by the sign restrictions.

\item \textbf{Step V --- Static factors $\bm q_t$:}
The innovation factors are drawn independently for each $t$:
\begin{equation*}
\bm q_t\mid\bullet \sim \mathcal{N}(\bar{\bm q}_t,\, \bar{\bm V}_q),
\end{equation*}
\begin{equation*}
\bar{\bm V}_q = \!\left(\bm\Lambda_q'\bm\Omega^{-1}\bm\Lambda_q + \bm I_{Q_q}\right)^{-1}\!, \qquad
\bar{\bm q}_t = \bar{\bm V}_q\,\bm\Lambda_q'\bm\Omega^{-1}\!\left(\bm y_t - \bm A\bm x_t - \bm\Lambda_\mu\bm\mu(\bm x_t)\right).
\end{equation*}

\item \textbf{Step VI --- Error variances $\bm\Omega$:}
For each $s=1,\dots,M$, let $\hat y_{st} = \bm a_s'\bm x_t + \bm\lambda_{\mu,s}'\bm\mu(\bm x_t) + \bm\lambda_{q,s}'\bm q_t$. Then:
\begin{equation*}
\omega_s^2\mid\bullet \sim \mathcal{G}^{-1}\!\left(a_\omega + \tfrac{T}{2},\; b_\omega + \tfrac{1}{2}\sum_{t=1}^T(y_{st} - \hat y_{st})^2\right).
\end{equation*}

\item \textbf{Step VII --- Horseshoe hyperparameters for $\bm A$:}
The global and local shrinkage parameters $\tau_a$, $\nu_a$, $\{\psi_{a,ij}\}$, $\{\nu_{a,ij}\}$ are sampled using the efficient auxiliary variable sampler of \cite{makalic2015simple}.% applied to the vector $\text{vec}(\bm A) - \text{vec}(\underline{\bm A})$.

\item \textbf{Step VIII --- Horseshoe hyperparameters for $\bm\Lambda_\mu$:}
The shrinkage parameters for $\bm\Lambda_\mu$ are updated row-by-row. For row $i$ (looping over $i =1,\dots,M$), let $\bm\lambda_{\mu,i\cdot}$ denote the relevant elements of the $i^{th}$ row. We apply the \cite{makalic2015simple} sampler to the scaled loadings $\lambda_{\mu,ij}/\sqrt{\varpi_j}$, obtaining updated local parameters $\psi_{\mu,ij}$ ($j = 1,\dots,Q_\mu$) and the row-specific global parameter $\tau_{\mu,i}$. The combined prior variance is then set to $\tilde\psi_{\mu,ij}^2 = \psi_{\mu,ij}^2\tau_{\mu,i}^2\varpi_j$.

\item \textbf{Step IX --- Column-shrinkage hyperparameter $\varpi$:}
The global column-shrinkage level $\varpi$ controls the overall strength of shrinkage on later factors. Its full conditional is derived by noting that $\lambda_{\mu,ij} \mid \psi_{\mu,ij}, \tau_{\mu,i}, \varpi \sim \mathcal{N}(0, \psi_{\mu,ij}^2\tau_{\mu,i}^2\varpi/j^\kappa)$. Collecting all elements of $\bm\Lambda_\mu$ across rows and columns and combining with the $\mathcal{G}^{-1}(a_\varpi, b_\varpi)$ prior gives
\begin{equation*}
    \varpi \mid \bullet \sim \mathcal{G}^{-1}\!\left(a_\varpi + \frac{M Q_\mu}{2},\; b_\varpi + \frac{1}{2}\sum_{i=1}^M\sum_{j=1}^{Q_\mu}\frac{j^\kappa\,\lambda_{\mu,ij}^2}{\psi_{\mu,ij}^2\,\tau_{\mu,i}^2}\right),
\end{equation*}
with $a_\varpi = 3$, $b_\varpi = 0.03$, and $\kappa = 2$. After drawing $\varpi$, the column-specific penalties are updated as $\varpi_j = \varpi/j^\kappa$, and the combined prior variances $\tilde\psi_{\mu,ij}^2$ are recomputed accordingly.

\end{itemize}

\subsection{Mixing properties of MCMC algorithm}\label{app:mixing}
We find that, despite its complexity, this algorithm mixes well.  We back our claim by showing two MCMC convergence diagnostics: inefficiency factors and the Raftery \& Lewis diagnostic  of the total number of runs required to achieve a certain level of precision for the large US macro dataset we consider in our empirical work.\footnote{We specify the parameters of this statistic along the lines suggested in \cite{primiceri2005time}: quantile=0.025, desired accuracy=0.025 and the required probability of obtaining the required accuracy p=0.95.} The quantiles are computed across equations and covariates/factors (for $\bm A$ and $\bm \Lambda_q$, respectively) and across factors and time (for $\bm \mu(\bm x_t)$). These results are based on the structural application and we run eight chains (with 10,000 draws, out of which we drop the first 5,000 as burn-in, each) in parallel. For each of these parallel runs we take every 10th draw from the 5,000 retained draws. This gives, per chain, 500 draws.  Combining these draws yields 4,000 draws from which we compute our functions of interest. In all cases, inefficiency factors are well below 30 in terms of the median. Only for $\bm \mu(\bm x_t)$ we find a 90\% quantile of the IF of around 53. This is consistent with the sampler navigating a non-identified subspace. Hence, we also compute inefficiency factors on the rotation invariant quantity $\bm \Lambda_\mu \bm \mu(\bm x_t)$. In this case, inefficiency factors are very low.  In terms of the Raftery \& Lewis diagnostic we find that the required number of runs is often below the total number of iterations. 
\begin{table}[h!]
\centering
\begin{tabular}{lrrrrr}
  \toprule
 Quantiles & 10\% & 25\% & 50\% & 75\% & 90\% \\ 
  \midrule
    \multicolumn{6}{c}{Inefficiency factors} \\
$\bm A$ & 1.00 & 1.02 & 1.13 & 1.40 & 1.98 \\ 
  $\bm \Lambda_q$ & 3.04 & 4.22 & 6.99 & 10.90 & 14.83 \\ 
  $\bm \mu(\bm x_t) $ & 6.09 & 7.96 & 12.97 & 30.02 & 53.14 \\ 
  $\bm \Lambda_\mu \bm \mu(\bm x_t)$ & 1.08 & 1.12 & 1.26 & 1.60 & 1.97 \\ \midrule
  \multicolumn{6}{c}{Raftery \& Lewis diagnostic} \\
  $\bm A$ & 148.00 & 151.00 & 157.00 & 165.00 & 182.00 \\ 
   $\bm \Lambda_q$ & 198.00 & 243.75 & 527.00 & 807.50 & 1070.00 \\ 
  $\bm \mu(\bm x_t)$ & 349.80 & 460.00 & 660.00 & 1158.00 & 1711.20 \\ 
  $\bm \Lambda_\mu \bm \mu(\bm x_t)$ & 151.20 & 154.00 & 157.00 & 165.00 & 175.00 \\
   \bottomrule
\end{tabular}
\end{table}

\begin{figure}[!h]
    \centering
    \includegraphics[width=.76\linewidth]{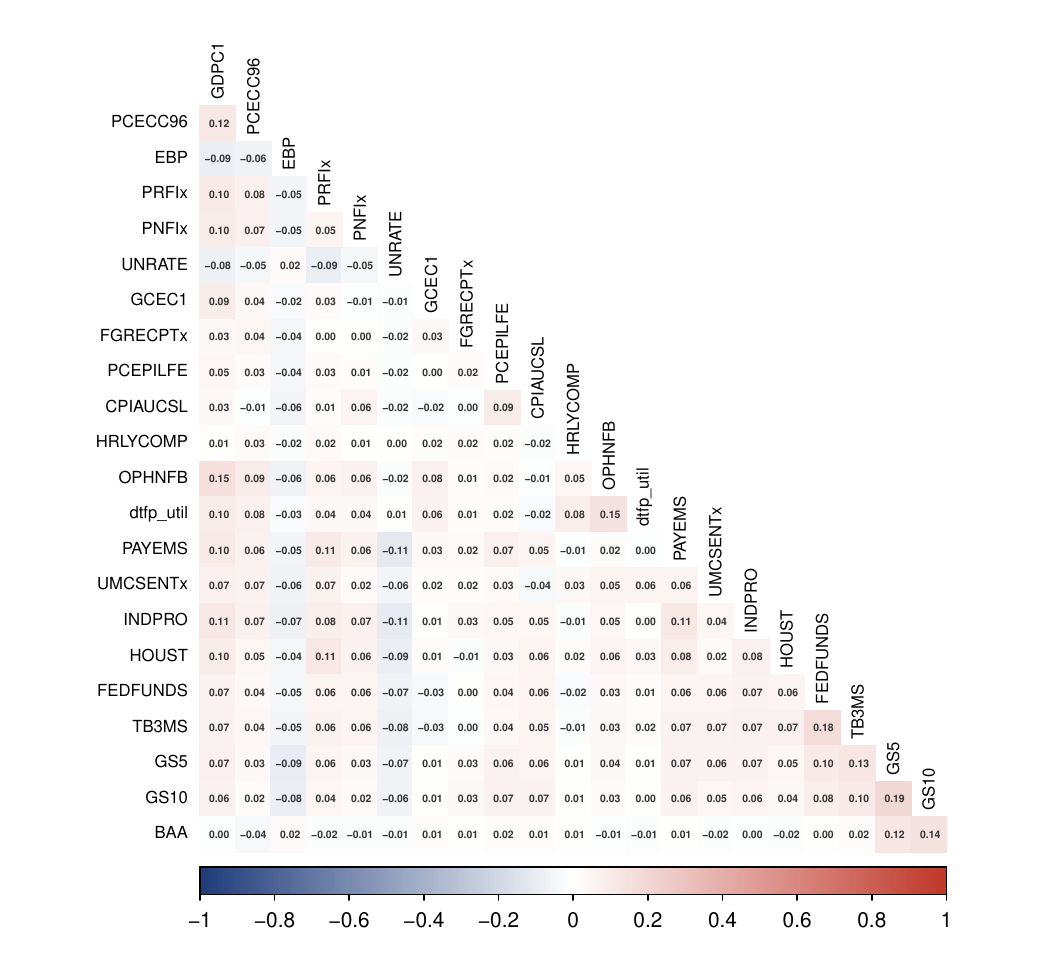}
    \caption{Shrinkage correlation matrix of the posterior-mean reduced-form residual $\bar{\bm \varepsilon}_t$, with shrinkage intensity $\lambda = 0.8$ toward the identity.}
    \label{fig:residcorr}
\end{figure}

Finally, another sanity check for our MCMC algorithm is whether the reduced-form errors based on a particular draw from the conditional mean function are actually independent from each other. If this is not the case, our model, which treats the equations as independent (given the latent factors), would be misspecified. We can check the cross-equation correlation of the reduced-form residual $\bar{\bm \varepsilon}_t = \bm y_t - \bar{\bm A} \bm x_t - \bar{\bm \Lambda}_\mu \bar{\bm \mu}(\bm x_t)$, evaluated at the posterior mean of the linear and nonlinear blocks.  Its correlation matrix is plotted in \autoref{fig:residcorr}. The median absolute off-diagonal correlation is $0.04$ and the maximum is $0.19$.  The residual correlations are concentrated near zero so that we can safely treat every equation as independent from each other.

\subsection{Computational efficiency}

\begin{figure}[h!]
    \centering
    \includegraphics[width=.7\linewidth, trim={0cm 0cm 0cm 2cm}, clip]{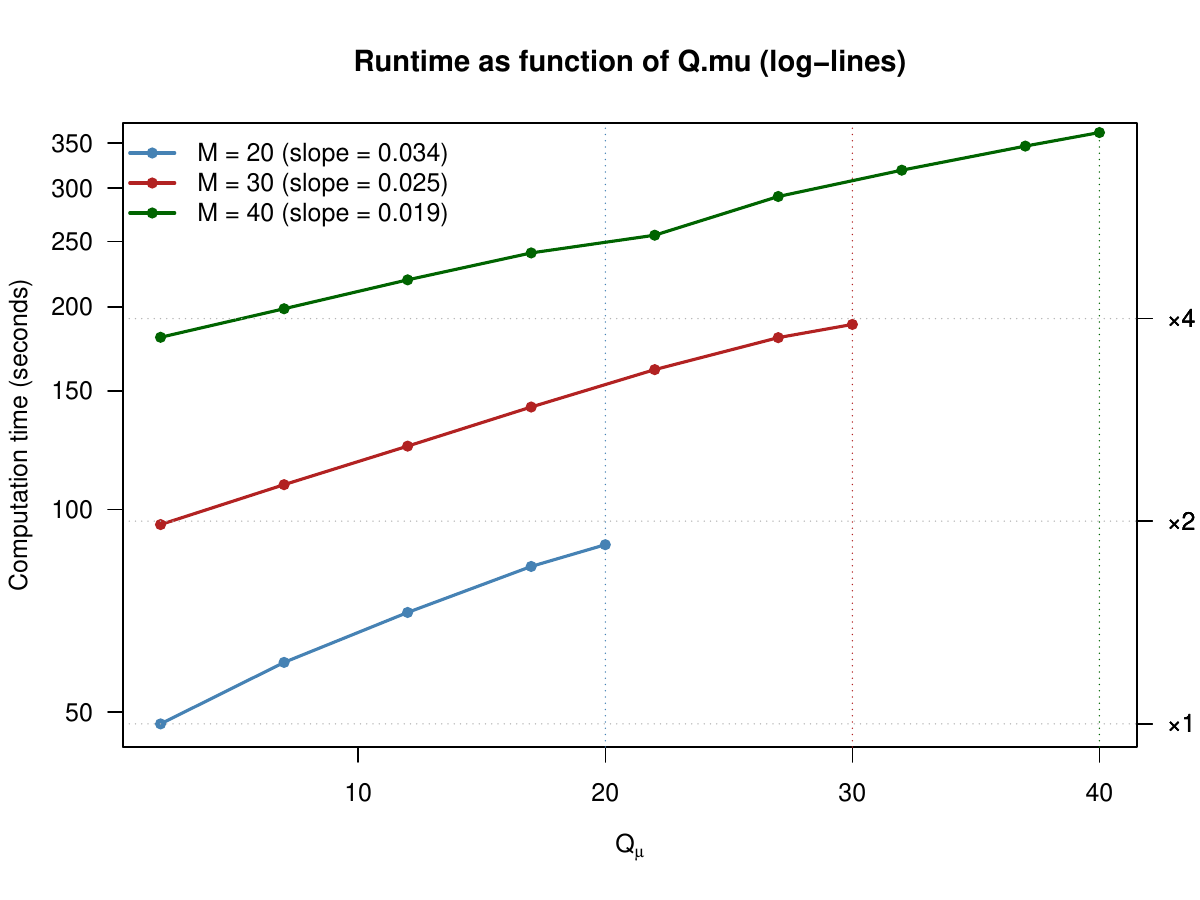}
    \caption{Runtimes to produce 1,000 draws across factor-BART VAR sizes}
    \label{fig:CPUtimes}
\end{figure}

%discussion above of the 
To flesh out the computational efficiency that comes with the factor specification of the factor-BART model, we have run some simulations to quantify computational runtimes for models differing in the number of variables $M$ and of common nonlinear factors $Q_\mu$.  \autoref{fig:CPUtimes} reports the number of seconds necessary to generate $1, 000$ draws from our MCMC algorithm for a dataset (generated randomly) with $T=500$ observations and for different values of $M \in \{20, 30 , 40\}$.  The lines give the log times, with the left side of the vertical axis providing times and the right side of the axis indicating how computation times multiply relative to the baseline, the smallest model having $M = 20$ and $Q_\mu = 2$.  The slope in the legend refers to the percentage increases in runtimes from including additional factors.

For a given $M$, the CPU time to estimate the model rises sharply with the number of BART terms.  For example, with $M=30$, the CPU time roughly doubles as the number of factors increases from $Q_\mu = 2$ to $Q_\mu = 30$ (i.e., as the model moves from our proposed specification to the mixBART model).  However, as the number of model variables ($M$) rises, for a given set of variables the rate of increase in CPU time with additional factors diminishes somewhat (i.e., the reported slopes of the lines decline with $M$).  This pattern is driven by the fixed costs of the estimation algorithm.  Returning to our statement above about scaling advantages that come with the factor-BART specification, the results in \autoref{fig:CPUtimes} indicate that a model with 30 variables and 2 factors can be estimated in little more than the time required to estimate a model with 20 variables and 20 factors.

\subsection{Computation of generalized impulse responses}\label{app: girf_appendix}

Because the conditional mean of the model is nonlinear in $\bm x_t$ through the BART factor block $\bm \Lambda_\mu \bm \mu(\bm x_t)$,  the response of ${\bm y_{T+h}}$ to a shock at ${T}$ depends both on the magnitude and sign of the shock and on the history of the system.  We therefore work with the structural generalized impulse-response function (GIRF) of \citet{KoopPesaranPotter:1996:girf}.

Let ${\bm \omega_{T-1} = (\bm y_{T-1}, \bm y_{T-2}, \ldots, \bm y_{T-p})}$ denote the history at the end of period $T-1$, the same conditioning set that enters $\bm x_T$ in our model.  For a fixed magnitude $\delta$ of the $s^{th}$ identified structural shock at $T$, the conditional GIRF of $\bm y_t$ at horizons $h = 0, 1, \ldots, H$ is the $(H+1)$-vector of differences between two conditional expectations:
\begin{equation}\label{eq: GIRF}
  \bm I_y(H, \delta, \bm \omega_{T-1})
  \;=\;
  \Big(\,
      \mathbb{E}\!\left[\bm y_{T+h} \,\big|\, q_{T, s} = \delta,\; \bm \omega_{T-1}\right]
      \;-\;
      \mathbb{E}\!\left[\bm y_{T+h} \,\big|\, \bm \omega_{T-1}\right]
   \,\Big)_{h = 0, 1, \ldots, H}.
\end{equation}
The first expectation conditions both on $\bm \omega_{T-1}$ and on the realization $\delta$ of the $s$-th structural shock at $T$; the second expectation conditions only on $\bm \omega_{T-1}$.  Both expectations are taken over the joint distribution of \emph{all other} future innovations, i.e.\ the remaining elements of $\bm q_T$, the contemporaneous noise $\bm \eta_T$, and the entire future-shock sequence $\{\bm q_{T+\tau}, \bm \eta_{T+\tau}\}_{\tau \ge 1}$.  By construction, no past structural shock $\{\bm q_\tau, \bm \eta_\tau\}_{\tau \le T-1}$ enters either expectation in~\eqref{eq: GIRF}: conditioning on $\bm \omega_{T-1}$ encodes the full information set up to $T-1$ through the observed lags $\bm y_{T-1}, \ldots, \bm y_{T-p}$ that form $\bm x_T$, and past innovations enter the GIRF only via their effect on these observed lagged values.  

At $h = 0$, equation~\eqref{eq: GIRF} reduces to the on-impact effect $\bm \Lambda_q\, \delta\, \bm e_s$ (where $\bm e_s$ is a $Q_q\times1$ vector with a $1$ in the $s^{th}$ position) so the GIRF inherits the structural interpretation of the identified shock; at $h \ge 1$ the nonlinear factor map $\bm \mu(\cdot)$ is re-evaluated at the iterated state and the response is a convolution of $\bm \Lambda_q\, \delta\, \bm e_s$ with the dynamic mechanism of the model.  As a result, any asymmetries between large/small and positive/negative shocks at $h \ge 1$ reflect both the shape of $\bm \mu$ and the averaging over future shocks built into~\eqref{eq: GIRF}.

It is worth being explicit about why this GIRF retains its structural label beyond impact.  The sign restrictions on $\bm \Lambda_q$ identify $q_{T, s}$ as a structural innovation at $T$; the nonlinearity of $\bm \mu(\cdot)$ does not interfere with this identification because $\bm \mu(\cdot)$ enters only through the lagged state $\bm x_T$, which is held fixed in \autoref{eq: GIRF}.  At $h \ge 1$ the impact effect $\bm \Lambda_q\, \delta\, \bm e_s$ is then propagated through the model's own structural law of motion --- exactly as in a linear SVAR, where horizon-wise responses inherit their structural interpretation from impact identification combined with the dynamic mechanism \citep[see Chapter 18 of][]{KilianLutkepohl2017}.  Beyond the future shock averaging discussed above, the unconditional GIRF in~\eqref{eq: GIRF_unc} below additionally integrates over the empirical distribution of histories, so any size or sign asymmetry reported in \autoref{sec:signs} reflects the average shape of $\bm \mu(\cdot)$ along the sample-relevant region of the state space; asymmetries arising from sample-period composition rather than from $\bm \mu(\cdot)$ itself can in principle be exposed by conditioning on particular regimes, an option discussed below.

Because $\bm I_y(H, \delta, \bm \omega_{T-1})$ in general depends on the specific history $\bm \omega_{T-1}$, an applied researcher interested in the typical effect of a shock of magnitude $\delta$ over the full sample reports the {unconditional} GIRF, obtained by integrating~\eqref{eq: GIRF} over the empirical distribution of histories,
\begin{equation}\label{eq: GIRF_unc}
  \bm I_y(H, \delta) \;=\; \int \bm I_y(H, \delta, \bm \omega^r)\, d\bm \omega^r.
\end{equation}
In practice, $\bm I_y(H, \delta)$ is estimated by averaging $\bm I_y(H, \delta, \bm \omega^r)$ over many histories $\bm \omega^r$ drawn (with replacement) from the in-sample observations \citep[see Section~18.2.2 of][]{KilianLutkepohl2017}.  Conditioning on a particular state of the economy (e.g.\ averaging only over recession histories) is also possible by restricting the resampled histories to the desired subset.  All GIRF figures reported in the paper correspond to the unconditional version~\eqref{eq: GIRF_unc} computed over the full in-sample range of $\bm \omega^r$.

We report GIRFs for both a small and a large shock.  Reporting both magnitudes makes it possible to gauge the degree of size asymmetry in the model -- a feature absent from any linear specification -- while still keeping $\delta$ fixed at the comparison stage so that the corresponding GIRFs are directly comparable across histories and across positive versus negative shocks.

 \eqref{eq: GIRF_unc} is approximated by Monte Carlo simulation for each iteration of our MCMC algorithm.  At each post-burn-in MCMC draw $d = 1, \dots, D$ we proceed as follows.
\begin{steps}
  \item \emph{History.}  Draw a conditioning history $\bm \omega_{T-1}^{(d)}$ uniformly from the set of in-sample dates.
  \item \emph{Future-shock paths.}  For each path $r = 1, \dots, R$ draw a base shock vector $\bm u^{(d, r)} \sim \mathcal{N}(\bm 0, \bm I_{Q_q})$ that will be used to construct the contemporaneous shock at $T$, an independent scalar $\tilde u_s^{(d, r)} \sim \mathcal{N}(0, 1)$ that will serve as the baseline value of the $s^{th}$ element at $T$, a noise vector $\bm \eta_T^{(d, r)} \sim \mathcal{N}(\bm 0, \bm \Omega^{(d)})$ at $T$, and an independent sequence of future shocks $\{\bm q_{T+h}^{(d, r)}, \bm \eta_{T+h}^{(d, r)}\}_{h = 1}^{H}$ with $\bm q_{T+h}^{(d, r)} \sim \mathcal{N}(\bm 0, \bm I_{Q_q})$ and $\bm \eta_{T+h}^{(d, r)} \sim \mathcal{N}(\bm 0, \bm \Omega^{(d)})$.  We set $R = 50$ paths in our US application.
  \item \emph{Coupled forward iteration of the two paths.}  Construct the baseline ($j = 0$) and shocked ($j = 1$) structural-shock vectors at $T$ from $\bm u^{(d, r)}$ by overwriting its $s^{th}$ element,
\[
  \bm q_T^{(0, d, r)} = \bm u^{(d, r)} - u_s^{(d, r)} \bm e_s + \tilde u_s^{(d, r)} \bm e_s, \qquad
  \bm q_T^{(1, d, r)} = \bm u^{(d, r)} - u_s^{(d, r)} \bm e_s + \delta\, \bm e_s,
\]
so that the two paths agree on the elements other than $s$ and differ only in element $s$: on the baseline it equals the independent draw $\tilde u_s^{(d, r)} \sim \mathcal{N}(0, 1)$ and on the shocked trajectory it equals the prescribed magnitude $\delta$.  For $h \ge 1$ both paths share the same future structural shock, $\bm q_{T+h}^{(0, d, r)} = \bm q_{T+h}^{(1, d, r)} = \bm q_{T+h}^{(d, r)}$, and at every $h \ge 0$ they share the same measurement-error draw $\bm \eta_{T+h}^{(d, r)}$.  Iterate the model forward for $h = 0, 1, \dots, H$,
\begin{equation*}
  \bm y_{T+h}^{(j, d, r)} = \bm A^{(d)} \bm x_{T+h}^{(j, d, r)} + \bm \Lambda_\mu^{(d)} \bm \mu^{(d)}\!\left(\bm x_{T+h}^{(j, d, r)}\right) + \bm \Lambda_q^{(d)} \bm q_{T+h}^{(j, d, r)} + \bm \eta_{T+h}^{(d, r)}, \qquad j \in \{0, 1\},
\end{equation*}
where the path-specific state vectors $\bm x_{T+h}^{(j, d, r)}$ are constructed recursively from the iterated $\bm y$'s on the corresponding path and from the history $\bm \omega_{T-1}$; at $h = 0$ both states coincide and equal the history-implied $\bm x_T$.  The nonlinear factor map $\bm \mu^{(d)}(\cdot)$ is the BART ensemble at draw $d$, evaluated at the iterated state via the predict method of the BART sampler.
  \item \emph{Path averaging.}  Set the per-draw GIRF estimate to the average across paths,
\begin{equation*}
  \widehat{\mathrm{GIRF}}_h^{(d)}(\delta, \bm \omega_{T-1}^{(d)})
  = \frac{1}{R} \sum_{r=1}^{R} \!\left(\bm y_{T+h}^{(1, d, r)} - \bm y_{T+h}^{(0, d, r)}\right), \qquad h = 0, 1, \dots, H.
\end{equation*}
  \item \emph{Posterior summary.}  The reported GIRF figures show the posterior median and the 16th and 84th posterior percentiles of $\widehat{\mathrm{GIRF}}_h^{(d)}$ across $d = 1, \dots, D$.
\end{steps}

Steps (1) to (5) implement the  algorithm of Section~18.2.2 of \citet{KilianLutkepohl2017}. The reported credible bands therefore reflect three sources of uncertainty: (i) posterior uncertainty about $(\bm A, \bm \Lambda_\mu, \bm \mu, \bm \Lambda_q, \bm \Omega)$, (ii) sampling uncertainty about the conditioning history ${\bm \omega_{T-1}}$ via the per-draw history draw in step (1), and (iii) sampling uncertainty about the future-shock paths via steps (2)--(4) with $R = 50$.

As a sanity check on the procedure, the Monte Carlo evaluation in \autoref{sec:MCsimulations} (Table~\ref{tab:girf-mc}) shows that this construction recovers the true GIRFs of the simulated data-generating processes with high frequentist coverage of the posterior bands and low MAE relative to the truth, both on the linear baseline (DGP~1), on which it reduces to the standard linear IRF, and on the nonlinear DGPs that build in known asymmetries.

% -------------------------------------------------------------------

\clearpage
\setcounter{table}{0}
\setcounter{figure}{0}
\renewcommand\thetable{C.\arabic{table}}
\renewcommand\thefigure{C.\arabic{figure}}
\section{Empirical Appendix}\label{app:C}
\subsection{Additional Monte Carlo simulation results}
%To get a sense of the uncertainty surrounding the values in \autoref{tab: MSE_A} that average across all replications, we provide boxplots of the MSEs in \autoref{fig:MSE_simulations}.  In these plots, the black lines and boxes display the 25th, 50th, and 75th percentiles of the factor-BART VAR (blue boxes) and linear BVAR (yellow boxes) squared coefficient errors across the 500 simulations. It can be seen that our finding that the factor-BART VAR does a better job of estimating the linear part of the model when the true DGP is nonlinear and does nearly as well when the true DGP is linear holds up robustly across draws from the DGPs.  Having demonstrated that our approach performs well in simulations, we now apply it to real-world data.

%\begin{figure}
%    \centering
%    \begin{minipage}[t]{1\textwidth}
%       \centering  (a) $T=250$
%    \end{minipage}\\
%    \includegraphics[width=1\linewidth]%{Simulation_output/MSEA_short.pdf}
%    \\
%        \begin{minipage}[t]{1\textwidth}
%       \centering  (b) $T=500$
%    \end{minipage}\\
%    \includegraphics[width=1\linewidth]%{Simulation_output/MSEA_long.pdf}
%    \caption{Posterior percentiles of the mean squared errors %between the true VAR coefficients and the posterior median of $\bm A$. }
%    \label{fig:MSE_simulations}
%\end{figure}

We consider how well our specification shrinks various equations toward linearity. To this end, we compute the posterior distribution of the area under the receiver operating characteristic curve (AUC) based on the shrinkage parameters of the prior on $\bm \Lambda_\mu$.  To this end, we define a shrinkage score $S_i$ that is given by:
\begin{equation*}
    S_i = \frac{\sum_{j=1}^{Q_\mu} \log \phi^{-1}_{ij}}{Q_\mu},
\end{equation*}
where $\phi_{ij} = \varpi_{j} \psi^2_{\mu, ij} \tau_{\mu, i}^2$ denotes the actual prior variance associated with the $(i,j)^{th}$ element in $\bm \Lambda_{\mu}$. Notice that larger values of $S_i$ imply stronger shrinkage toward linearity of the $i^{th}$ equation.  Define $\mathcal{Z}$ to be the set of indices for zero rows in $\bm \Lambda_\mu$ and $\mathcal{S}$ the set of non-zero rows in $\bm \Lambda_\mu$.
Given $S_i$, the AUC is defined as:
\begin{equation*}
    \text{AUC} = \frac{1}{|\mathcal{Z}| |\mathcal{S}|} \sum_{i \in \mathcal{Z}} \sum_{s \in \mathcal{S}} \mathbb{I}(S_i > S_s),
\end{equation*}
which approximates the probability that a randomly selected zero row is more strongly shrunk than a randomly selected non-zero row. The AUC is defined as the posterior probability that an equation that is truly linear ($\bm \lambda_i = \bm 0)$ receives stronger shrinkage than an equation that is truly nonlinear ($\bm \lambda_i \neq \bm 0)$, and therefore measures the ordering (separation) ability of the shrinkage mechanism without requiring a threshold.

\begin{figure}[htbp]
    \centering
    \includegraphics[width=1\linewidth]{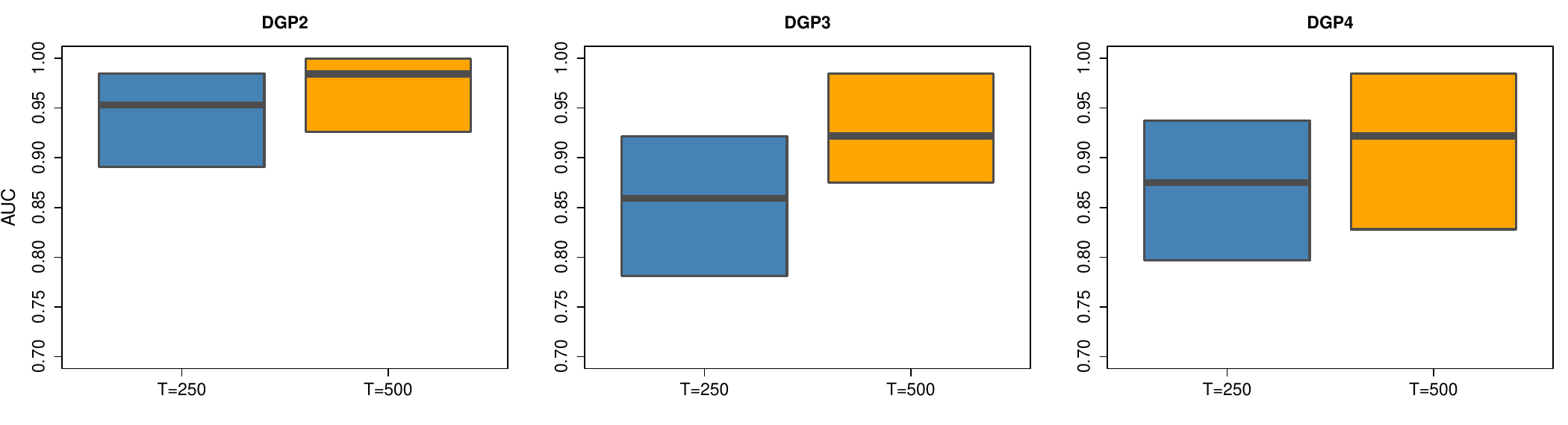}
    \caption{Posterior percentiles of the AUC score}
    \label{fig:AUC_boxplots}
\end{figure}

We compute the AUC value for every MCMC draw, compute the posterior median of the AUC, and then report boxplots that show the 25th, 50th, and 75th percentiles of posterior medians for the 500 replications from the DGP in \autoref{fig:AUC_boxplots}. Notice that the linear DGPs, \texttt{DGP 1} and \texttt{DGP 5}, are absent from the figure since the set $\mathcal{S}$ is empty and the AUC is not  defined in this case. This figure clearly shows that the factor-BART VAR classifies the linear equations well and shrinks strongly toward linearity. This holds for all the different types of nonlinearity in our DGPs.  For example, with $T = 250$, the AUC is about 0.95 for DGP 2 and 0.85 for DGPs 3 and 4; these are very high posterior probabilities that estimates of linear equations are shrunk more strongly than estimates of truly nonlinear equations.  Unsurprisingly, larger sample sizes lead to modestly higher AUCs and implied probabilities. 

\FloatBarrier

\subsection{Additional forecasting application results}\label{app:addforecast}

\begin{figure}[htbp]
    \centering
    \begin{minipage}[t]{0.32\textwidth}
        \centering (a) GDPC1
    \end{minipage}
        \begin{minipage}[t]{0.32\textwidth}
        \centering (b) UNRATE
    \end{minipage}
          \begin{minipage}[t]{0.32\textwidth}
        \centering (c) INDPRO
    \end{minipage}\\
    \begin{minipage}[t]{0.32\textwidth}
        \includegraphics[width=.9\textwidth,  trim={0cm 0cm 0cm 0cm},clip]{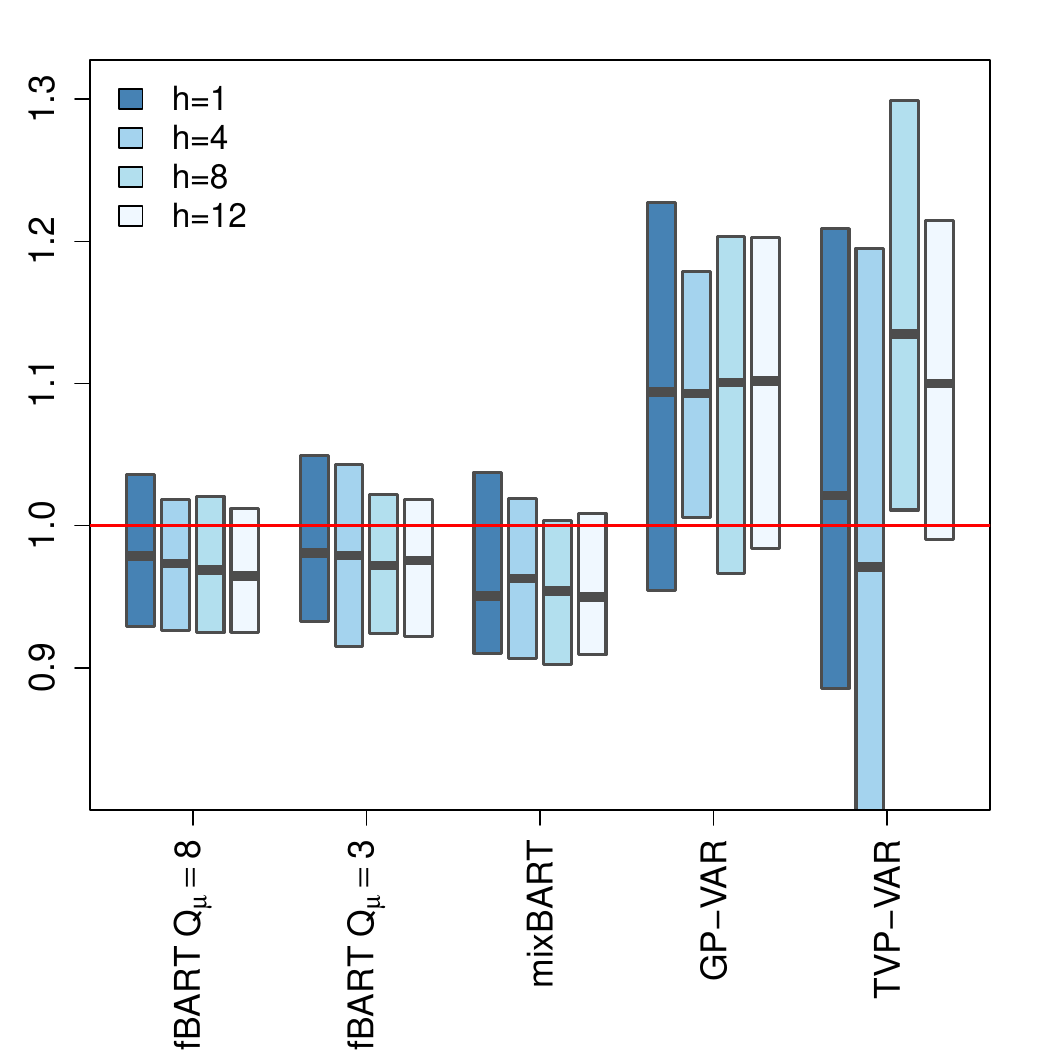}
    \end{minipage}
        \begin{minipage}[t]{0.32\textwidth}
        \includegraphics[width=.9\textwidth,  trim={0cm 0cm 0cm 0cm},clip]{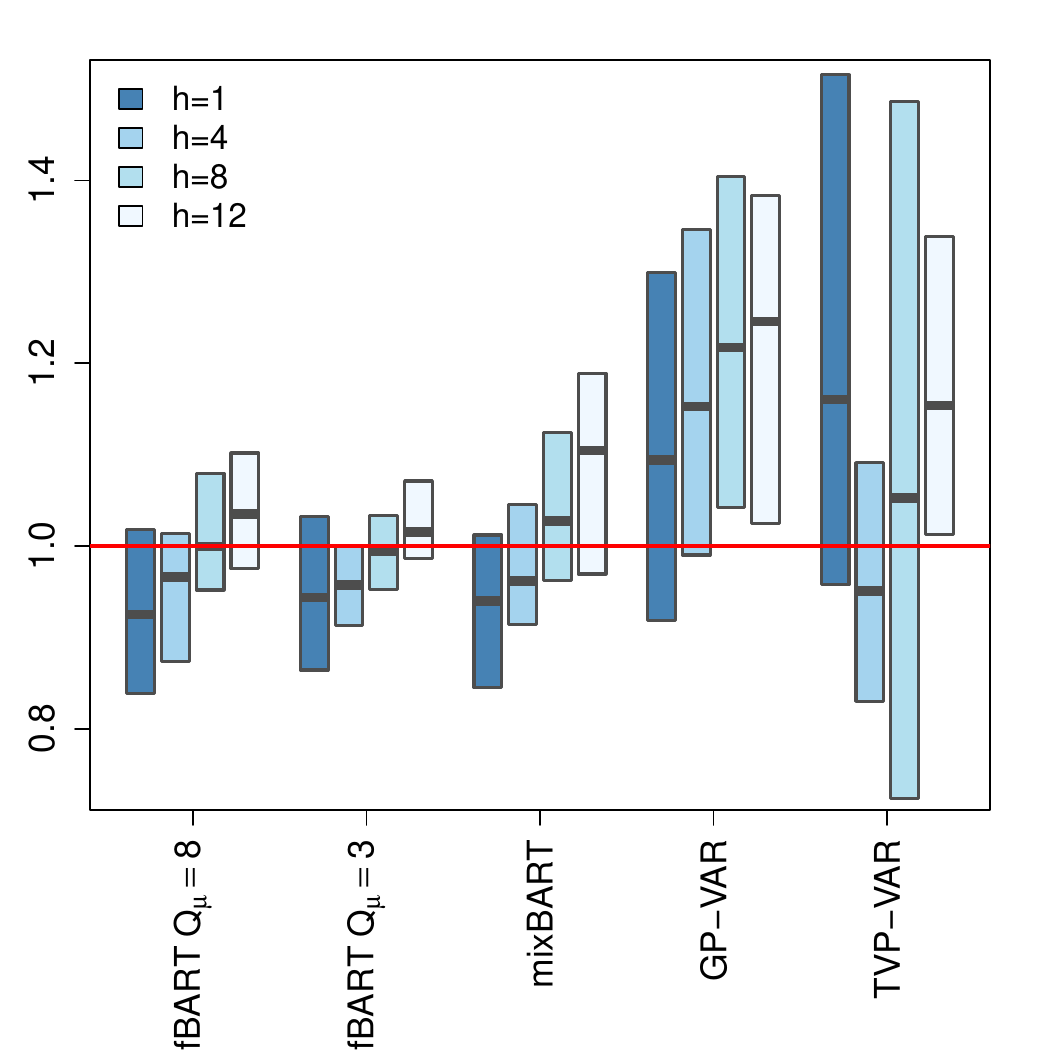}
    \end{minipage}
     \begin{minipage}[t]{0.32\textwidth}
        \includegraphics[width=.9\textwidth,  trim={0cm 0cm 0cm 0cm},clip]{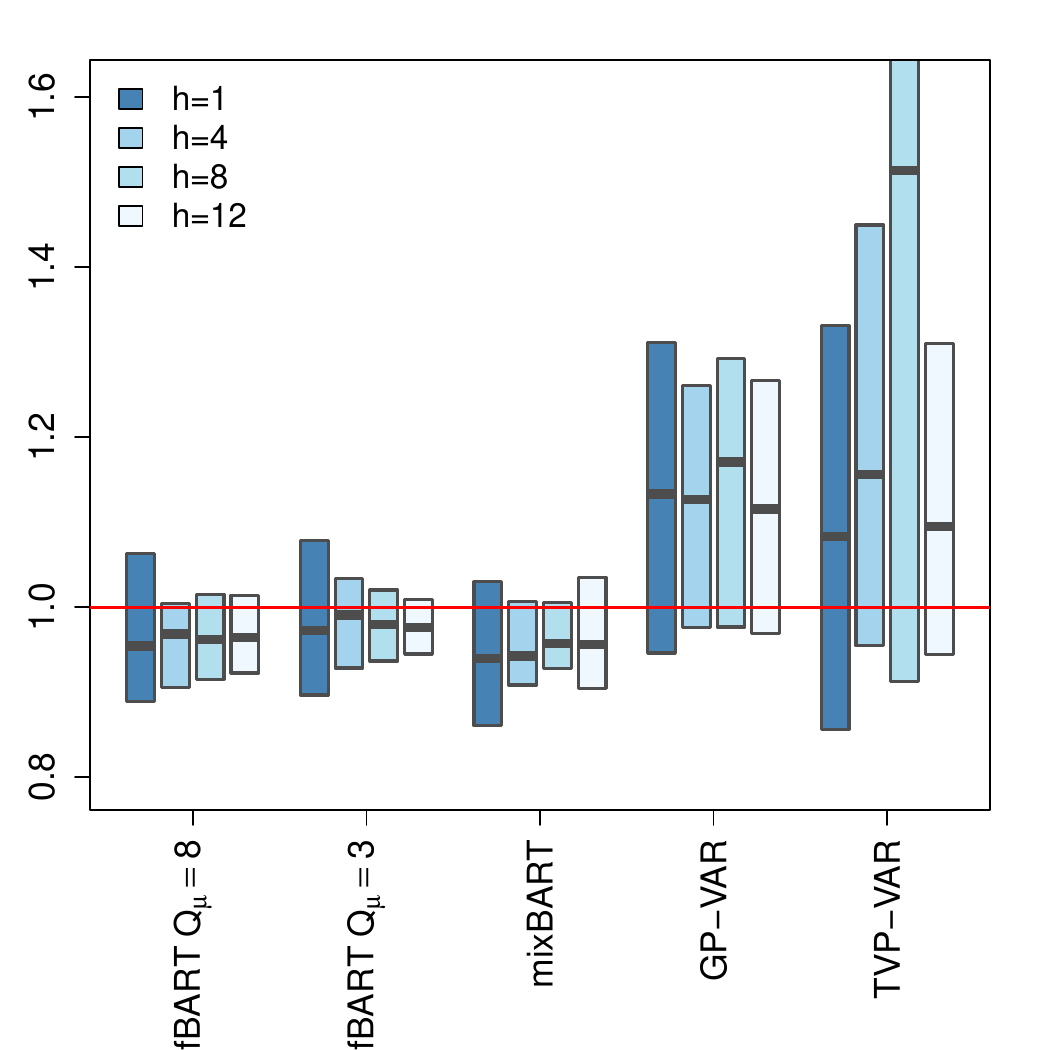}
    \end{minipage}\\
        \begin{minipage}[t]{0.32\textwidth}
        \centering (d) FEDFUNDS
    \end{minipage}
        \begin{minipage}[t]{0.32\textwidth}
        \centering (e) CPIAUCSL
    \end{minipage}
        \begin{minipage}[t]{0.32\textwidth}
        \centering (f) PCEPILFE
    \end{minipage}\\
     \begin{minipage}[t]{0.32\textwidth}
        \includegraphics[width=.9\textwidth,  trim={0cm 0cm 0cm 0cm},clip]{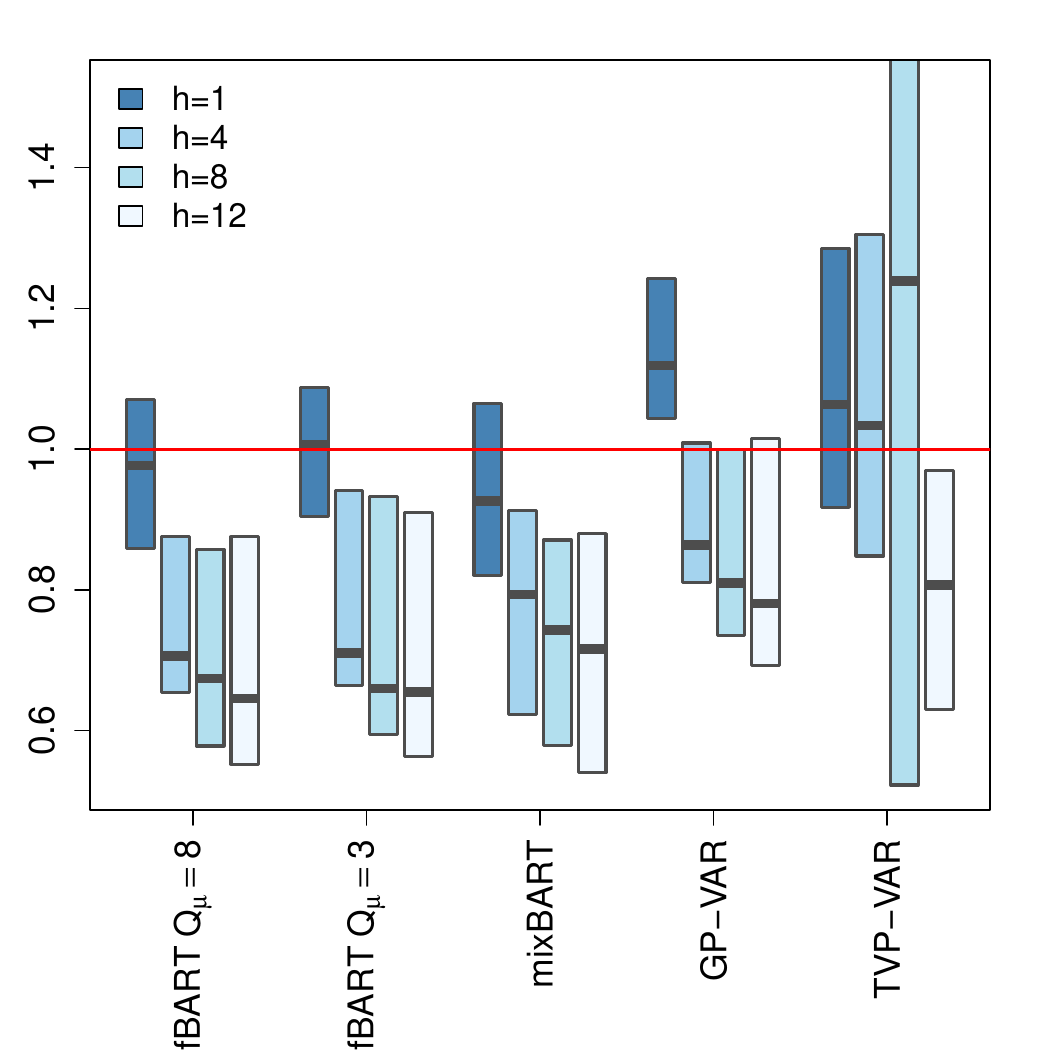}
    \end{minipage}
   \begin{minipage}[t]{0.32\textwidth}
        \includegraphics[width=.9\textwidth,  trim={0cm 0cm 0cm 0cm},clip]{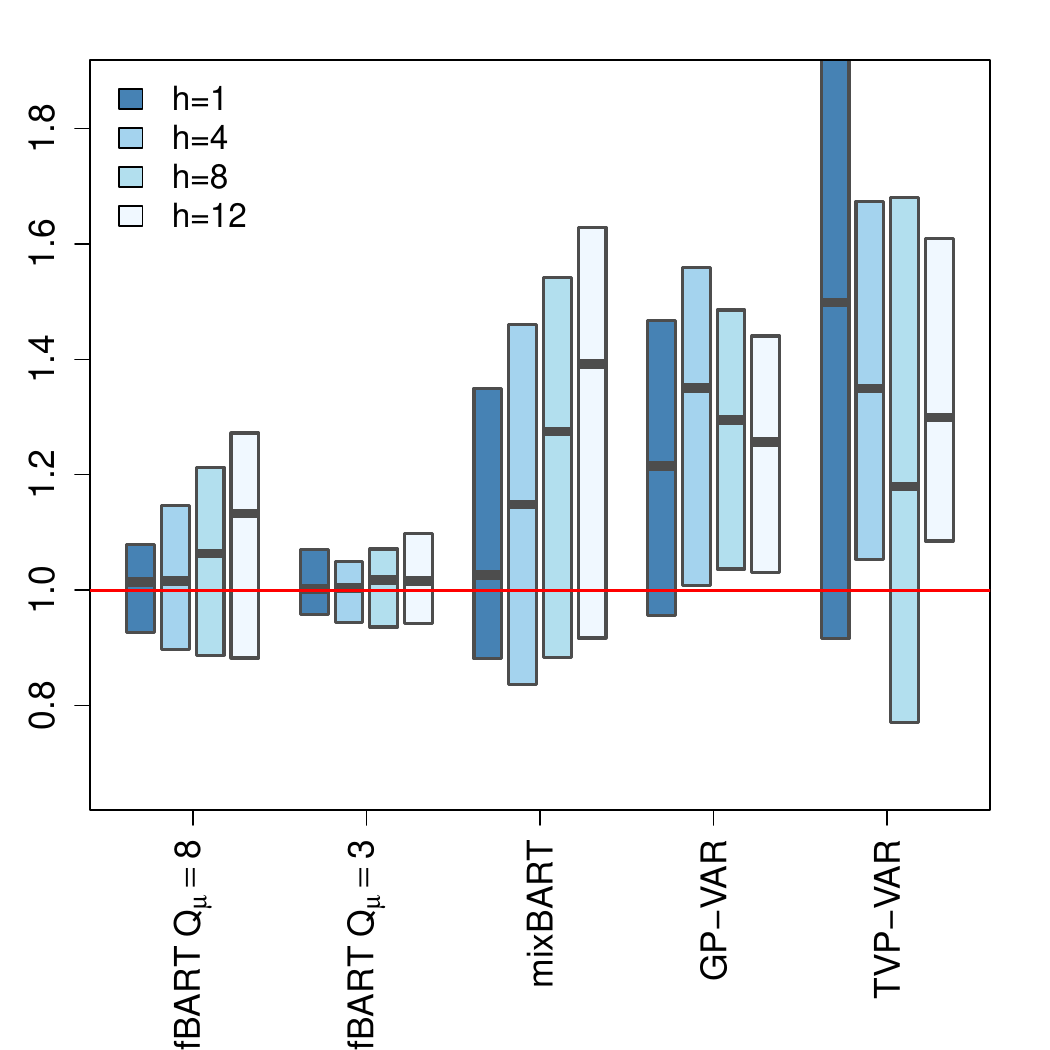}
    \end{minipage}
        \begin{minipage}[t]{0.32\textwidth}
        \includegraphics[width=.9\textwidth,  trim={0cm 0cm 0cm 0cm},clip]{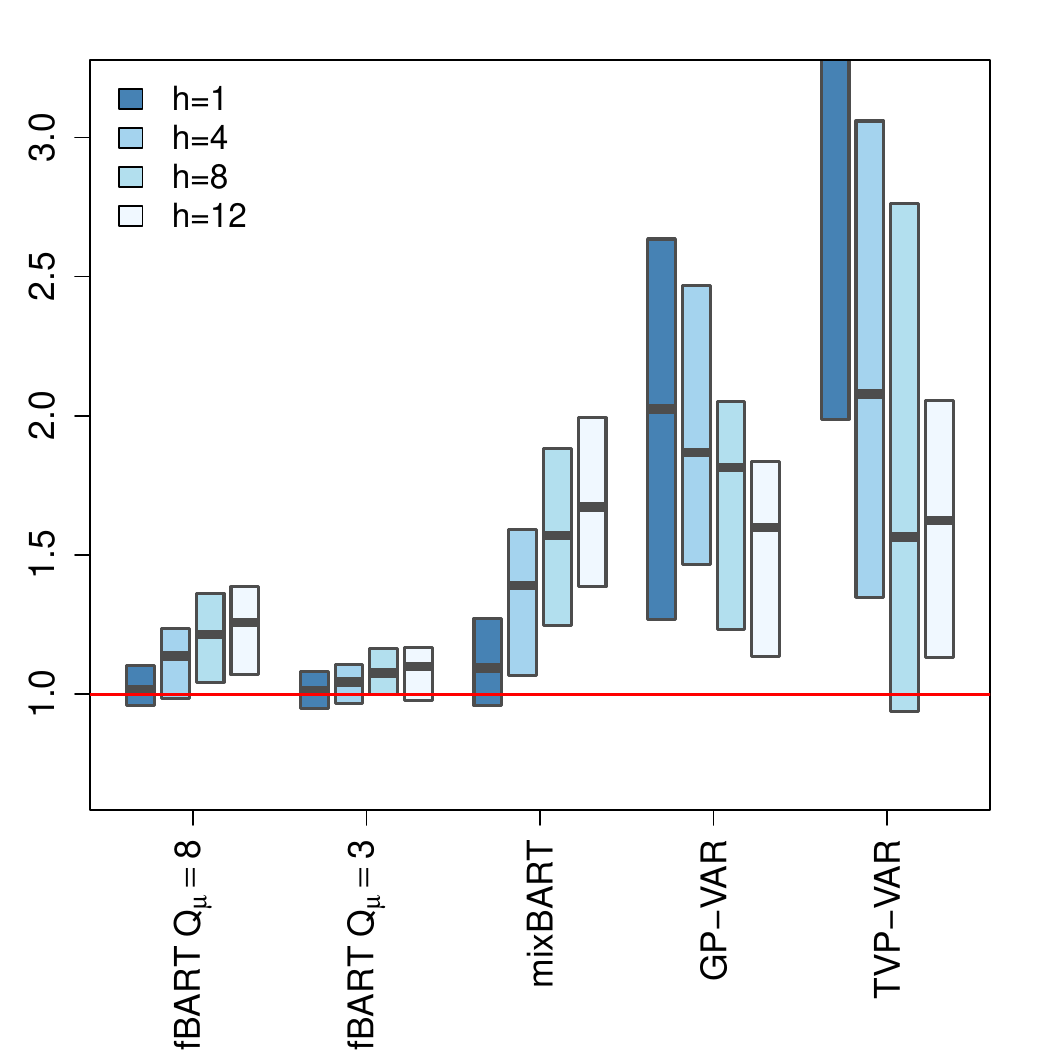}
    \end{minipage}
    \caption{Boxplots of relative CRPSs to the {BVAR} across variables, forecast horizons and competing model specifications.}
    \label{fig:relativeCRPS}
\end{figure}

To compare the factor-BART VAR against the full set of competing specifications considered in the forecasting application, \autoref{fig:relativeCRPS} reports relative CRPS results for the same six headline variables and forecast horizons as in the main text.  For each variable-horizon combination, the figure provides the median (black bar) and interquartile range (box) of the time series of the CRPS for each nonlinear model relative to the CRPS of the linear BVAR baseline.  Broadly, these results confirm the patterns seen in the energy scores for the full set of model variables.  Except for the inflation variables, the specifications including a nonlinear BART component improve on the accuracy of the linear BVAR (and in most cases do so across all horizons).  Our factor-BART VAR once again achieves accuracy gains similar to those achieved with the mixBART model, but with reduced computation requirements.  For these variables, forecasts from the GP-VAR and TVP-VAR models are modestly to substantially less accurate than forecasts from the BART models and the linear BVAR.  In the case of the inflation measures, the factor-BART VARs yield forecasts more accurate than those from all the other nonlinear specifications, including mixBART.  However, for these variables, the factor-BART VAR is not quite as accurate as a linear BVAR.

We do not view this inflation result as a flaw to be hidden but as informative about where nonlinear flexibility pays off. Over our post-2001 hold-out, both CPI and core PCE inflation behave like smooth, highly persistent and near-linearly forecastable processes, so the additional nonlinear factor block has little genuine asymmetry or regime variation to exploit; its main effect is to add estimation variance, which translates into the slight CRPS deterioration relative to the parsimonious linear BVAR seen in \autoref{fig:relativeCRPS}. By contrast, the clear gains concentrate exactly where state-dependence and asymmetry are economically plausible -- real-activity and {labor}-market indicators, the federal funds rate, and financial conditions -- consistent with the sub-sample evidence that even the inflation forecasts improve over the linear baseline in the volatile COVID-era window, when nonlinearity is more pronounced. The pattern therefore tempers, rather than overturns, the case for added flexibility: factor-BART buys robustness and accuracy when the data-generating process is genuinely nonlinear, at a modest and well-{localized} cost for series that remain close to linear over the evaluation period.

\begin{figure}[htbp]
\begin{minipage}[t]{0.48\textwidth}
    \centering (a) $h=1$
\end{minipage}
\begin{minipage}[t]{0.48\textwidth}
    \centering (b) $h=4$
\end{minipage}\\ 
\begin{minipage}[t]{0.48\textwidth}
\includegraphics[width=1\textwidth, trim={0cm 0cm 0cm 2cm},clip]{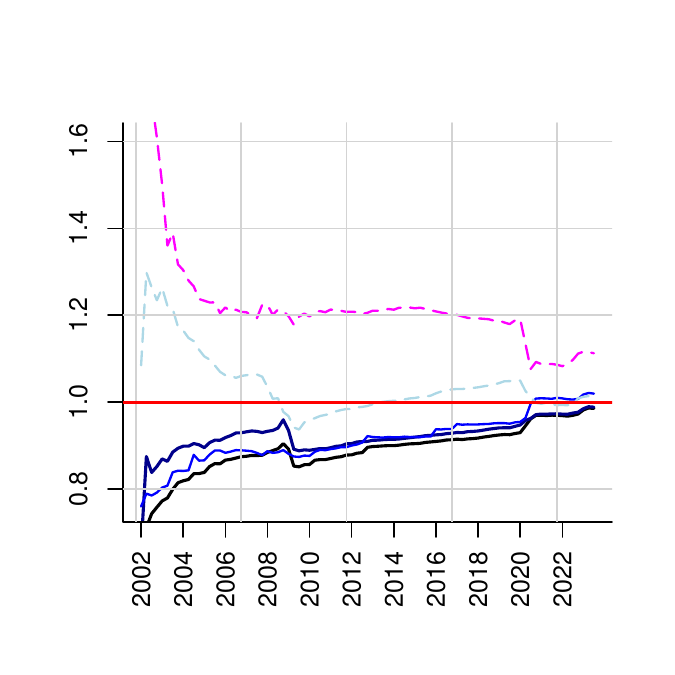}
\end{minipage}
\begin{minipage}[t]{0.48\textwidth}
\includegraphics[width=1\textwidth,  trim={0cm 0cm 0cm 2cm},clip]{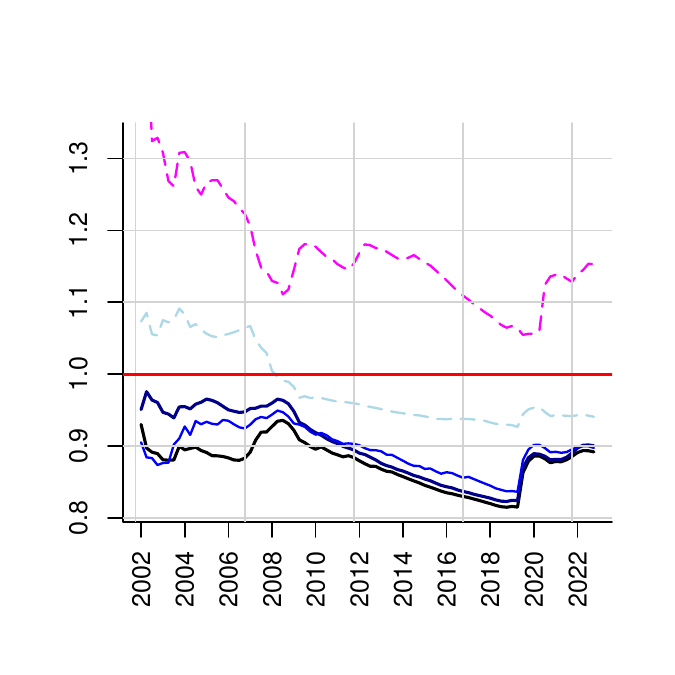}
\end{minipage}\\
\begin{minipage}[t]{0.48\textwidth}
    \centering (c) $h=8$
\end{minipage}
\begin{minipage}[t]{0.48\textwidth}
    \centering (d) $h=12$
\end{minipage}\\ 
\begin{minipage}[t]{0.48\textwidth}
\includegraphics[width=1\textwidth,  trim={0cm 0cm 0cm 2cm},clip]{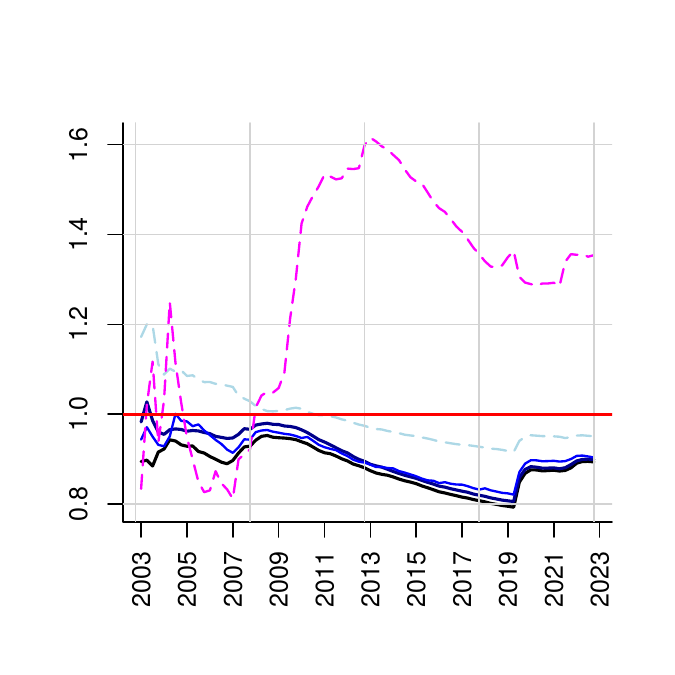}
\end{minipage}
\begin{minipage}[t]{0.48\textwidth}
\includegraphics[width=1\textwidth,  trim={0cm 0cm 0cm 2cm},clip]{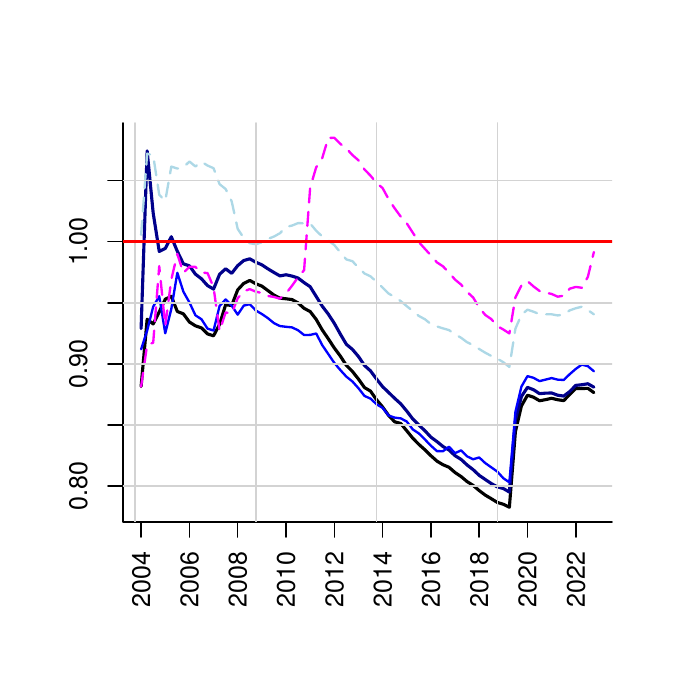}
\end{minipage}\\
\begin{minipage}[t]{1\textwidth}
    \centering \includegraphics[scale=0.6, trim={0cm 2cm 0cm 1.8cm},clip]{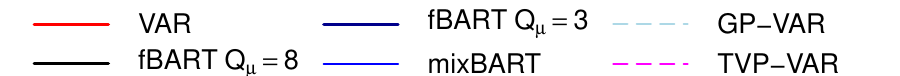}
\end{minipage}
\caption{Evolution of the average energy score over the hold-out period.}
\label{fig:energyscoresovertime}
\end{figure}

To assess how accuracy changes over the hold-out sample, \autoref{fig:energyscoresovertime} reports recursive average score ratios (for each model relative to the linear BVAR) over time.\footnote{At a given date $t$ in the chart, a given line shows the ratio of the energy score for that model from the beginning of the sample through $t$ relative to the score for the linear BVAR over the same period.}  These results show that the specifications that include a nonlinear component capturing nonlinearities achieve forecast accuracy gains over the linear BVAR over time, and not just on average.  At the one-step horizon, the (recursive average) gains are more than 10\% from 2002 through about 2015, and then decline in the remainder of the sample.  At longer horizons, the gains to including BART in the model tend to increase from the Great Recession up until the pandemic.  Among the BART specifications, over good parts of the sample, our factor-BART VAR with $Q_{\mu} = 8$ factors is slightly more accurate than the same model with 3 factors or mixBART, but as indicated above, on average performance is similar.  The alternative GP-VAR and TVP-VAR specifications are clearly less accurate for most of the sample.
\FloatBarrier

\subsection{Additional in-sample results}
\begin{figure}[htbp]
    \centering
    \includegraphics[width=1\linewidth]{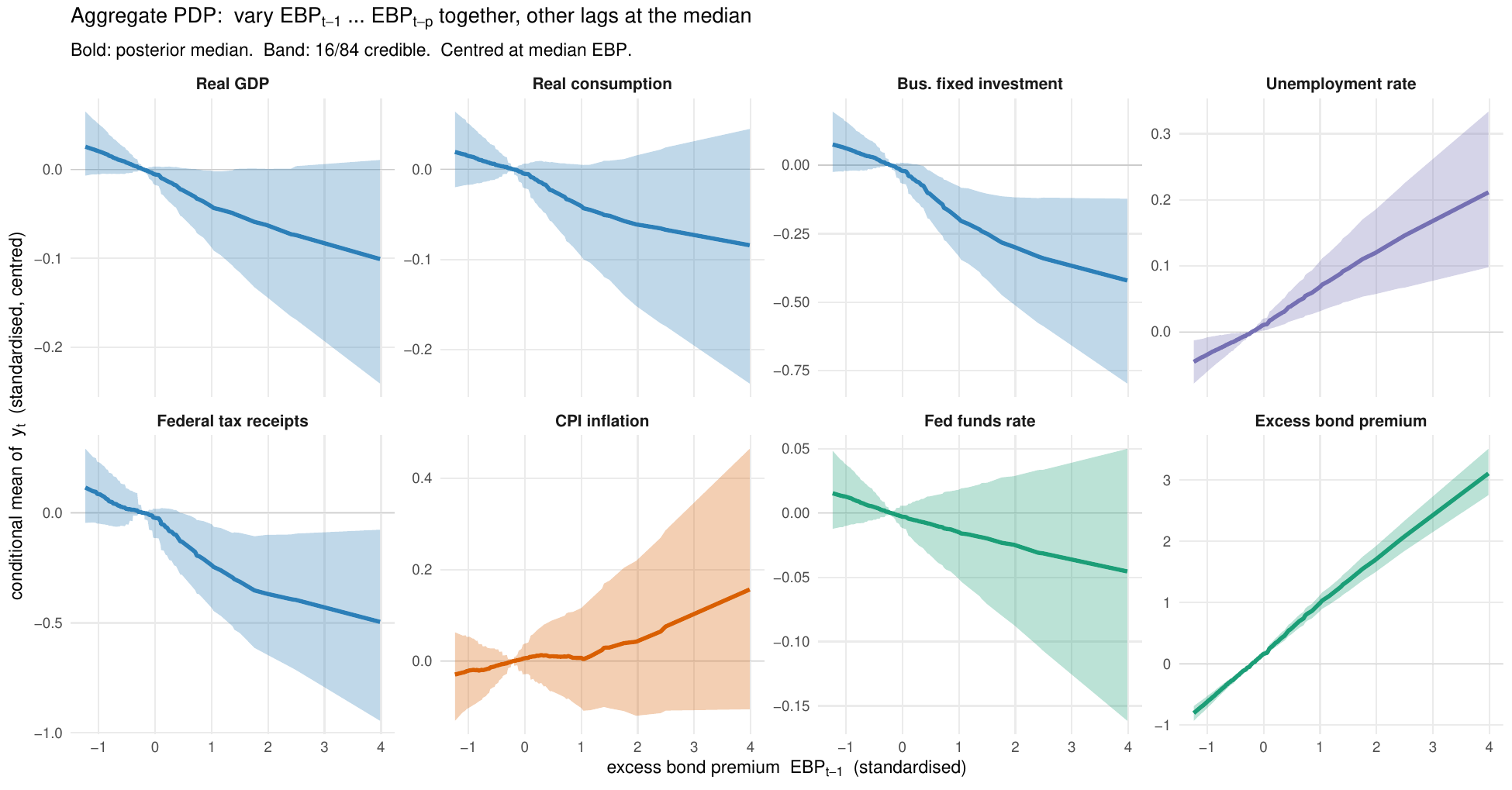}
    \caption*{\tiny \textbf{Notes:} Posterior median (solid line) and $16/84$ credible band of the model's conditional mean of each focus variable when the $p$ EBP lags are varied jointly along their empirical quantile grid and all other lags are pinned at the sample median.  Each curve is {centered} at the median EBP quantile.}
    \caption{Sensitivity of the focus variables to changes in the EBP (aggregate response)}
    \label{fig:PDPs_EBP_agg}
\end{figure}

To shed some light on the role of financial conditions in macroeconomic developments, \autoref{fig:PDPs_EBP_agg} shows posterior distributions of how the eight focus variables of the model change with movements in the excess bond premium (EBP).  All $p$ lags of EBP are varied jointly along the empirical EBP quantile grid, while the remaining elements of $\bm x_t$ are kept at the sample median.  Each per-variable curve is {centered} at the median EBP quantile so it passes through zero there, and the shaded band reports the 16$^{th}$ and 84$^{th}$ posterior percentiles.  Movements in the EBP impact the focus variables through both the linear BVAR component $\bm X \bm A^\ast$ and the nonlinear function component $\bm \Lambda_\mu \bm \mu(\bm x_t)$; the figure plots the aggregate conditional mean implied by the sum of the two.  Real activity (GDP, consumption, business fixed investment), the fiscal block (federal tax receipts) and the policy rate all decline monotonically with the EBP, while unemployment rises and the inflation response is mildly positive.  The curves steepen visibly in the right tail of EBP, consistent with stronger nonlinearity at high financial stress.

\FloatBarrier

\subsection{Additional generalized impulse responses}\label{app:fullirf}
%%   This appendix complements \autoref{fig:irfs_financial_large} and
%%   \autoref{fig:irfs_monetary_large} in the main text by reporting:
%%   (i) the GIRFs of the 14 non-focus response variables to large
%%   financial and monetary shocks, and (ii) the GIRFs of the eight
%%   focus and the 14 non-focus response variables to the corresponding
%%   small shocks ($\pm 5$).

\subsubsection{GIRFs of the remaining response variables to large shocks}

\begin{figure}[htbp]
    \centering
    \includegraphics[width=\linewidth]{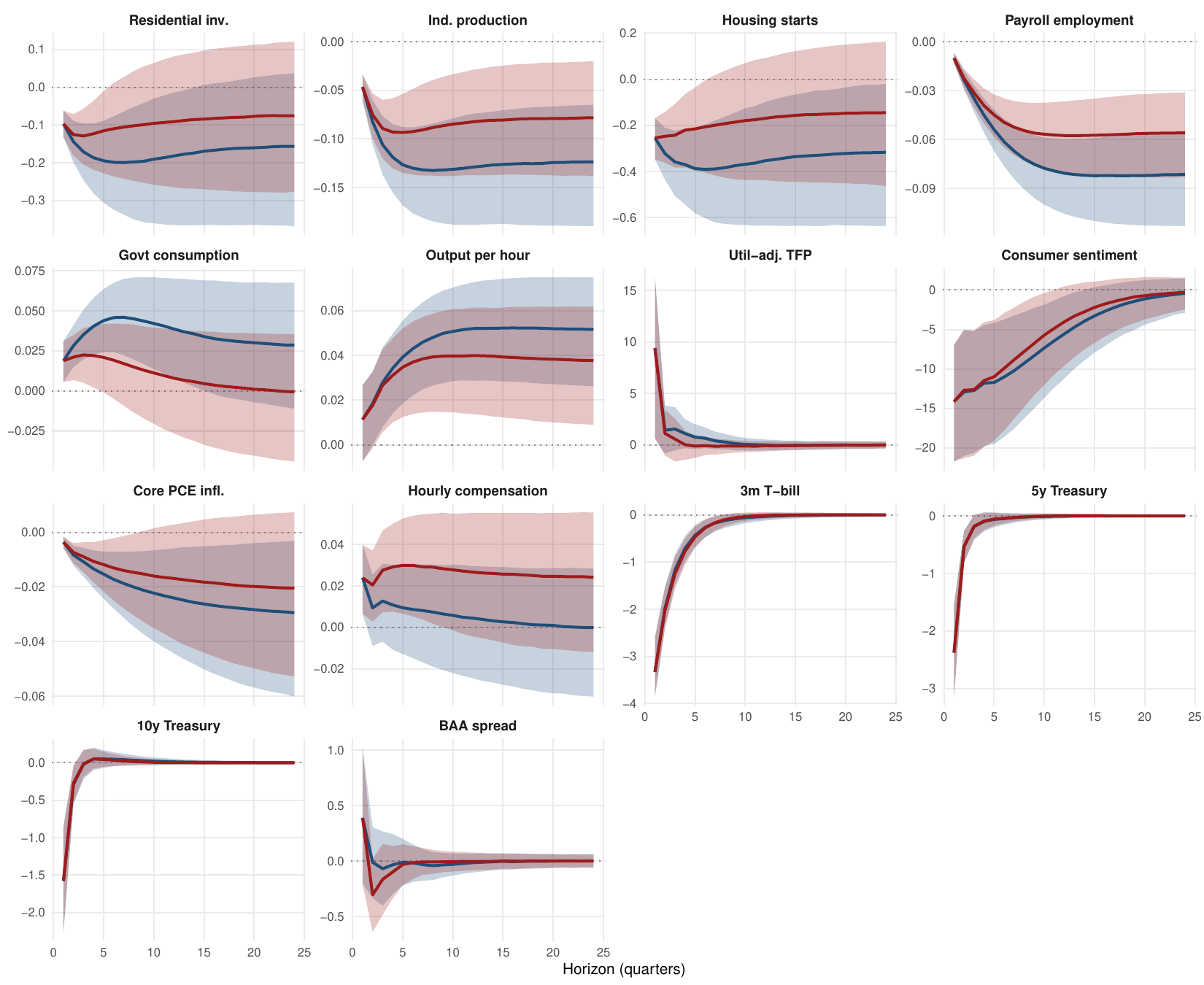}
    \caption{GIRF of the remaining response variables to a \emph{large} financial-conditions shock.  Conventions as in \autoref{fig:irfs_financial_large}; the 14 non-focus response variables are shown.}
    \label{fig:nonfocus_financial_large}
    \caption*{\tiny\textbf{Notes:} Blue solid line and shaded blue band: posterior median and $16/84$ band of the response to the contractionary ($+$) shock.  Red solid line and shaded red band: posterior median and $16/84$ band of the response to the expansionary ($-$) shock, mirrored to the positive axis.  Grey dotted line marks zero.}
\end{figure}

\begin{figure}[htbp]
    \centering
    \includegraphics[width=\linewidth]{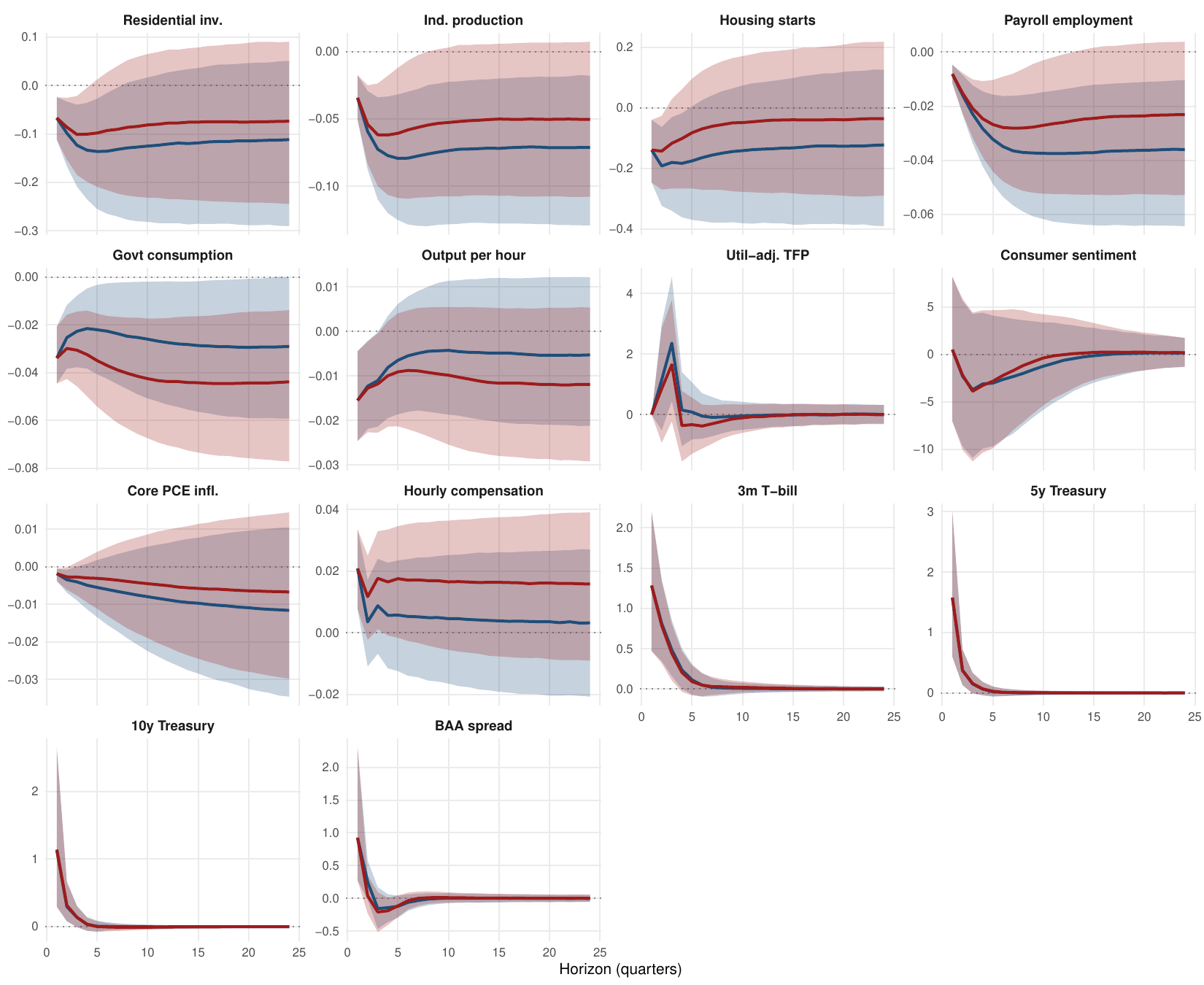}
    \caption{GIRF of the remaining response variables to a \emph{large} monetary policy shock.  Conventions as in \autoref{fig:irfs_monetary_large}; the 14 non-focus response variables are shown.}
    \label{fig:nonfocus_monetary_large}
    \caption*{\tiny\textbf{Notes:} Blue solid line and shaded blue band: posterior median and $16/84$ band of the response to the contractionary ($+$) shock.  Red solid line and shaded red band: posterior median and $16/84$ band of the response to the expansionary ($-$) shock, mirrored to the positive axis.  Grey dotted line marks zero.}
\end{figure}

\FloatBarrier

\subsubsection{GIRFs to small financial and monetary policy shocks}

\begin{figure}[htbp]
    \centering
    \includegraphics[width=\linewidth]{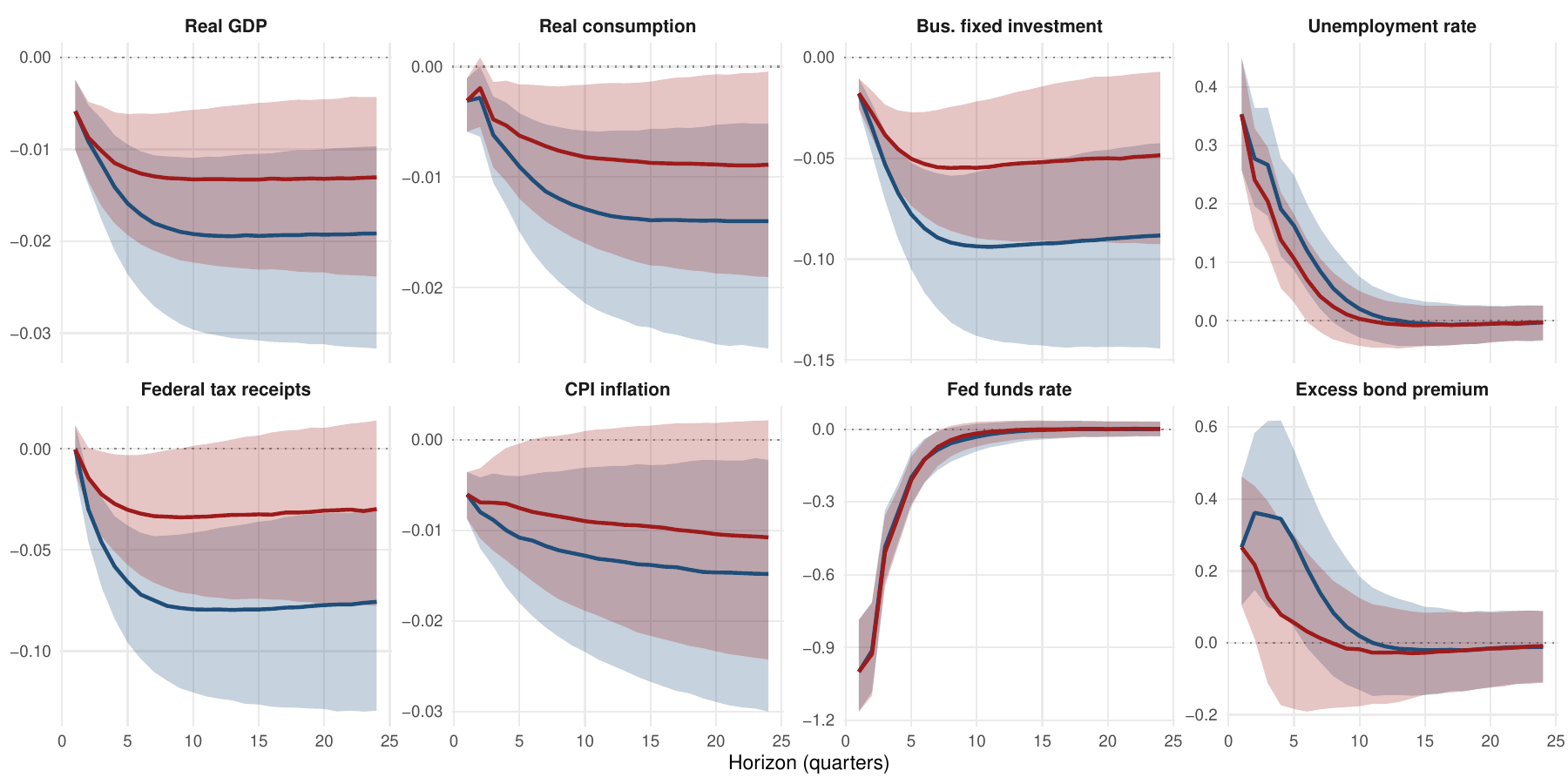}
    \caption{GIRF of the focus variables to a \emph{small} financial-conditions shock.  Conventions as in \autoref{fig:irfs_financial_large}; the eight focus response variables are shown.}
    \label{fig:focus_financial_small}
    \caption*{\tiny\textbf{Notes:} Blue solid line and shaded blue band: posterior median and $16/84$ band of the response to the contractionary ($+$) shock.  Red solid line and shaded red band: posterior median and $16/84$ band of the response to the expansionary ($-$) shock, mirrored to the positive axis.  Grey dotted line marks zero.}
\end{figure}

\begin{figure}[htbp]
    \centering
    \includegraphics[width=\linewidth]{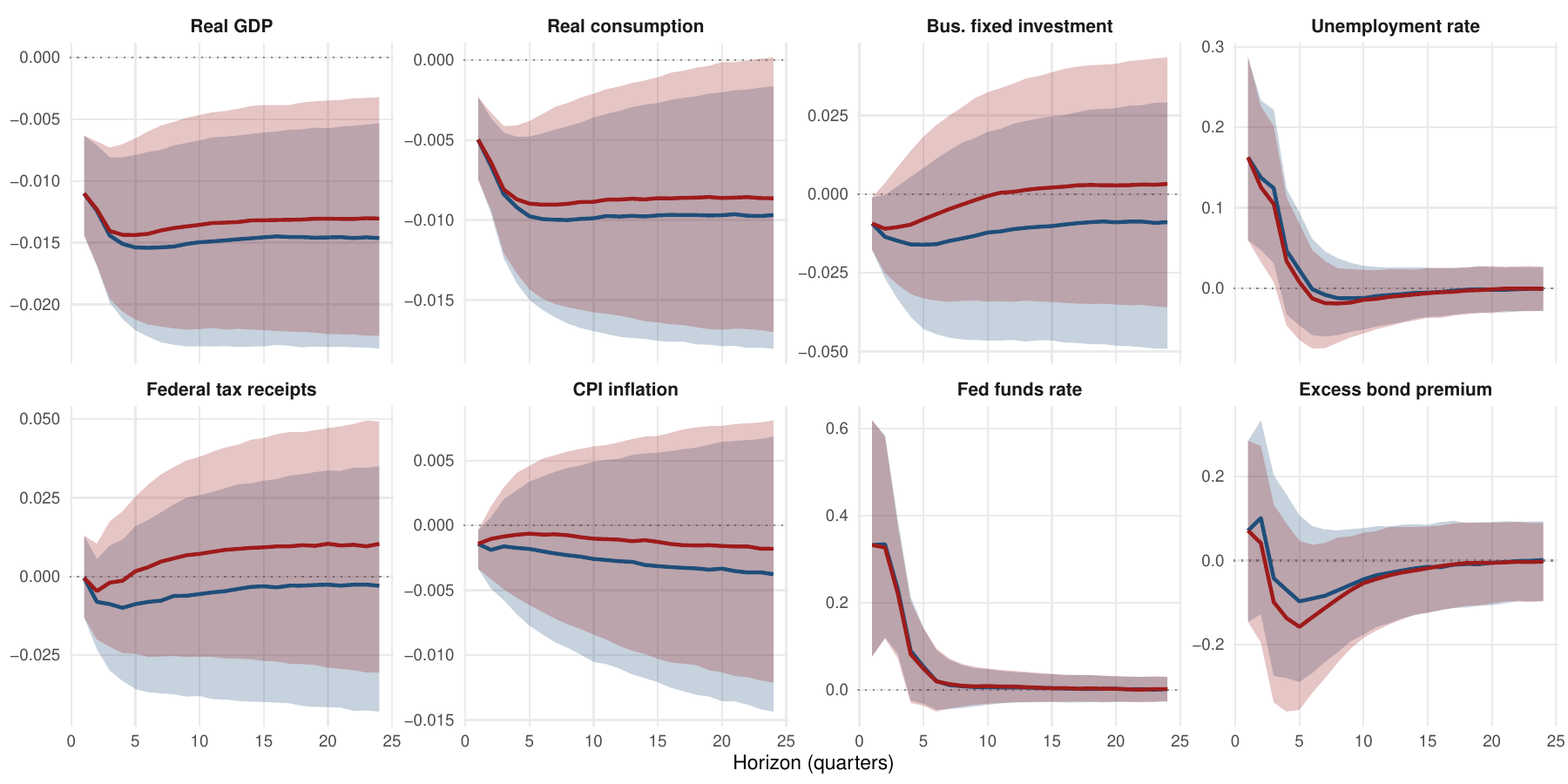}
    \caption{GIRF of the focus variables to a \emph{small} monetary policy shock.  Conventions as in \autoref{fig:irfs_monetary_large}; the eight focus response variables are shown.}
    \label{fig:focus_monetary_small}
    \caption*{\tiny\textbf{Notes:} Blue solid line and shaded blue band: posterior median and $16/84$ band of the response to the contractionary ($+$) shock.  Red solid line and shaded red band: posterior median and $16/84$ band of the response to the expansionary ($-$) shock, mirrored to the positive axis.  Grey dotted line marks zero.}
\end{figure}

\begin{figure}[htbp]
    \centering
    \includegraphics[width=\linewidth]{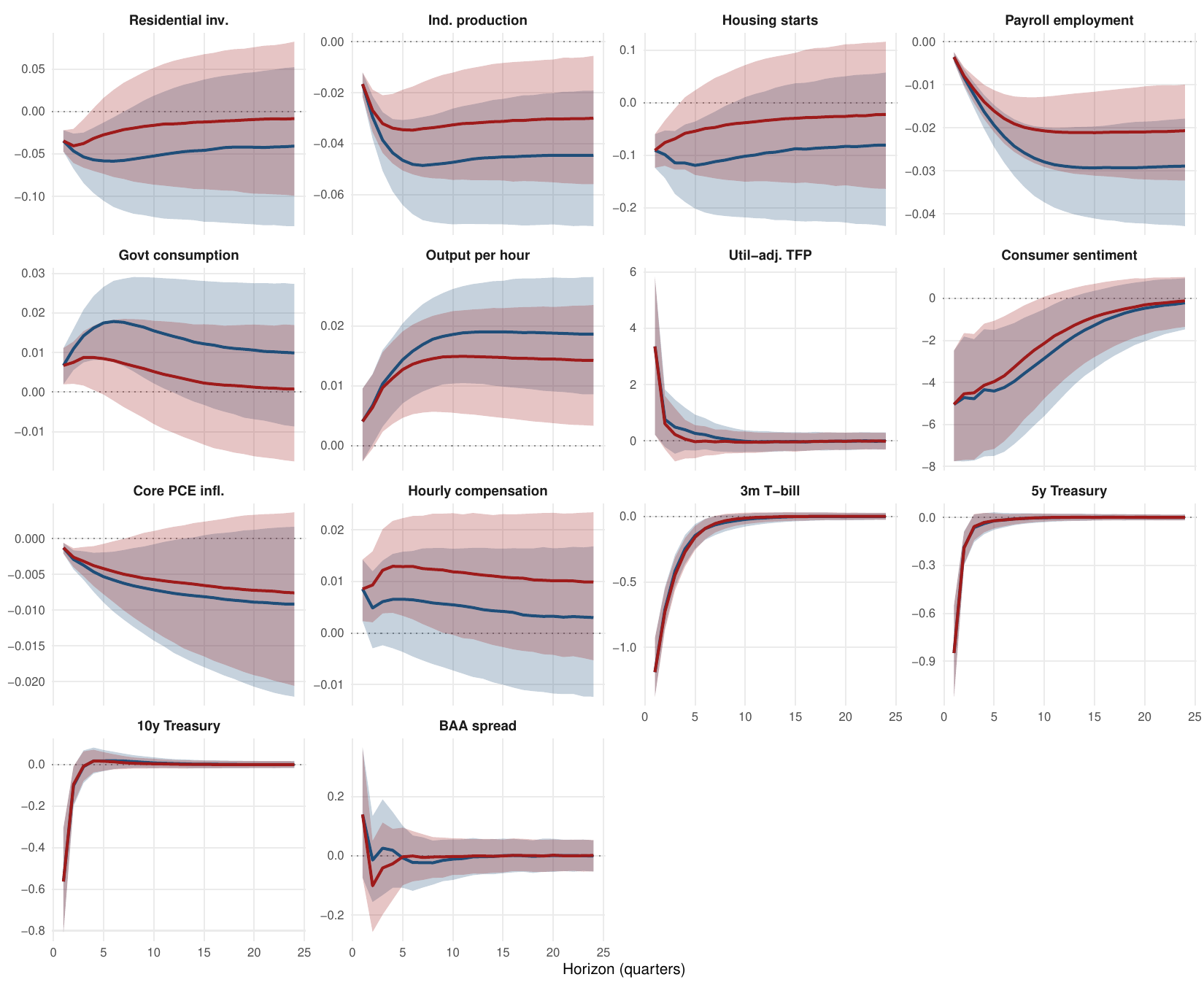}
    \caption{GIRF of the remaining response variables to a \emph{small} financial-conditions shock.  Conventions as in \autoref{fig:irfs_financial_large}; the 14 non-focus response variables are shown.}
    \label{fig:nonfocus_financial_small}
    \caption*{\tiny\textbf{Notes:} Blue solid line and shaded blue band: posterior median and $16/84$ band of the response to the contractionary ($+$) shock.  Red solid line and shaded red band: posterior median and $16/84$ band of the response to the expansionary ($-$) shock, mirrored to the positive axis.  Grey dotted line marks zero.}
\end{figure}

\begin{figure}[htbp]
    \centering
    \includegraphics[width=\linewidth]{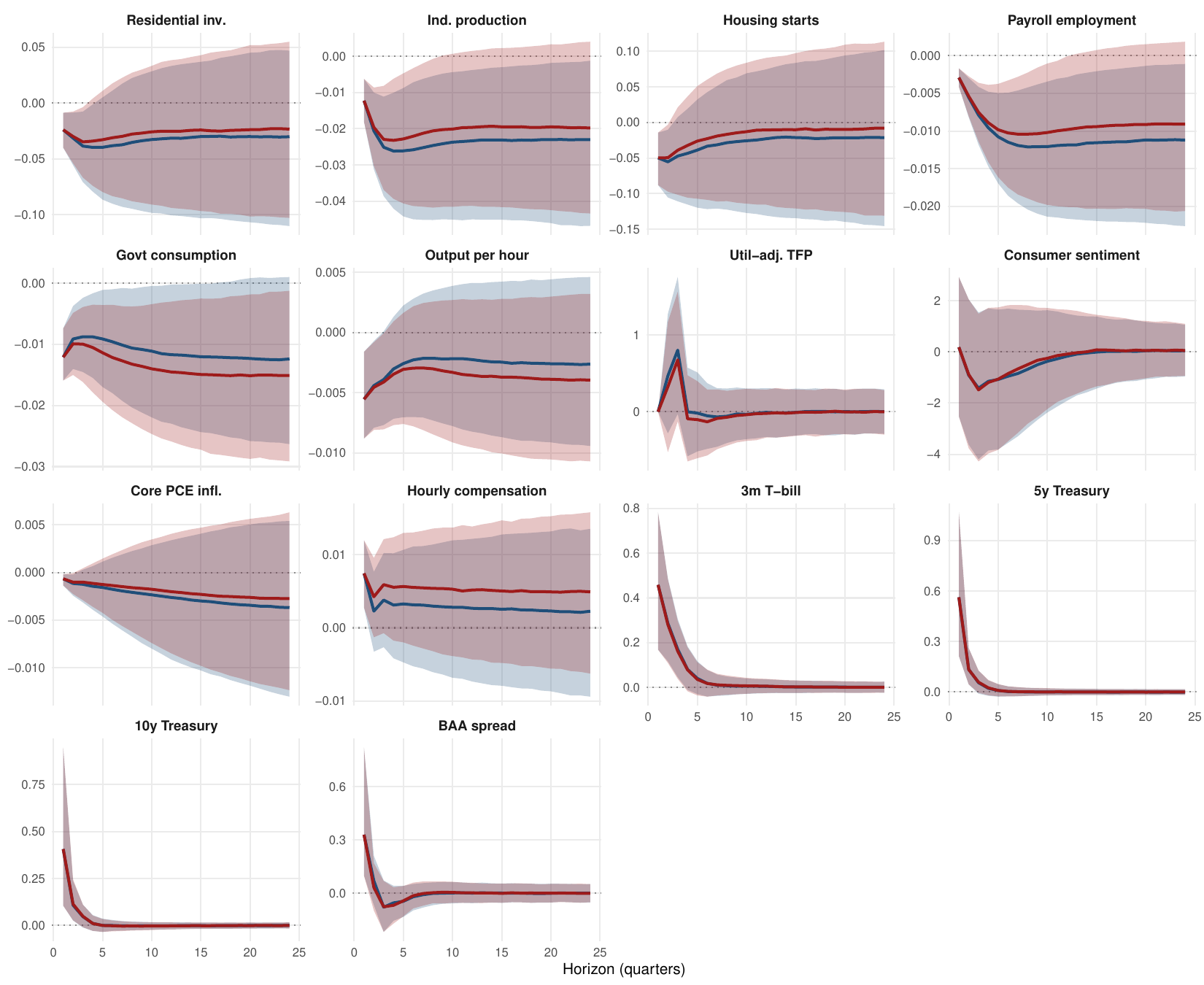}
    \caption{GIRF of the remaining response variables to a \emph{small} monetary policy shock.  Conventions as in \autoref{fig:irfs_monetary_large}; the 14 non-focus response variables are shown.}
    \label{fig:nonfocus_monetary_small}
    \caption*{\tiny\textbf{Notes:} Blue solid line and shaded blue band: posterior median and $16/84$ band of the response to the contractionary ($+$) shock.  Red solid line and shaded red band: posterior median and $16/84$ band of the response to the expansionary ($-$) shock, mirrored to the positive axis.  Grey dotted line marks zero.}
\end{figure}

\FloatBarrier

\end{appendices}
\end{document}